\documentclass[aps,pre,twocolumn,superscriptaddress,groupedaddress]{revtex4}

\usepackage{latexsym}
\usepackage[dvips]{graphics}
\usepackage{amssymb,amstext,amsfonts,amsmath}

\usepackage[T1]{fontenc} 
\usepackage[utf8x]{inputenc}
\usepackage[usenames,dvipsnames]{color}

\usepackage{natbib} 
\date{\today}

\begin{document}

\title{Generalization of Powell's results to population out of steady state}

\author{Jakub Jędrak}
\email[]{jjedrak@ichf.edu.pl}

\author{Marcin Rubin}

\affiliation{Institute of Physical Chemistry, Polish Academy of Sciences, ul. Kasprzaka 44/52, 01-224 Warsaw, Poland}

\author{Anna Ochab-Marcinek}
\affiliation{Institute of Physical Chemistry, Polish Academy of Sciences, ul. Kasprzaka 44/52, 01-224 Warsaw, Poland}

\date{\today}

\begin{abstract}
Since the seminal work of Powell, the relationships between the population growth rate, the probability distributions of generation time, and the distribution of cell age have been known for the bacterial population in a steady state of exponential growth. Here, we generalize these relationships to include an unsteady (transient) state for both the batch culture and the mother machine experiment. In particular, we derive a time-dependent Euler-Lotka equation (relating the generation time distributions to the population growth rate) and a generalization of the inequality between the mean generation time and the population doubling time. To do this, we use a model proposed by Lebowitz and Rubinow, in which each cell is described by its age and generation time. We show that our results remain valid for a class of more complex models that use other state variables in addition to cell age and generation time, as long as the integration of these additional variables reduces the model to Lebowitz-Rubinow form. As an application of this formalism, we calculate the fitness landscapes for phenotypic traits (cell age, generation time) in a population that is not growing exponentially. We clarify that the known fitness landscape formula for the cell age as a phenotypic trait is an approximation to the exact time-dependent formula.

\end{abstract}
 
\maketitle

\section{Introduction}  

The properties of microbial populations, including their temporal evolution, usually depend on the variability of quantities describing individual cells. For example, the growth rate of the population (the rate at which the total number of cells increases) depends on how the values of generation time (the duration of the cell cycle) are distributed among the cells. 

Due to the complexity of population dynamics, it seems reasonable to focus on populations in steady state. For such populations many theoretical results have been obtained \cite{powell1956growth, powell1964note, lebowitz1974theory, hashimoto2016noise, lin2017effects, quedeville2019critical, levien2020interplay}. By ``steady state'' we mean either steady exponential growth in batch culture (where the total number of cells in a population increases exponentially with time), or a true steady state in mother-machine experiments or continuous culture. In the latter two cases, the increase in cell number is compensated for by the removal of cells from the system, and the population size remains constant.  

But the opportunity for the study of many interesting phenomena is lost if the focus is only on the steady state. For this reason, cell populations that are out of the steady state have been studied experimentally \cite{Maruyama1956,Newton1959,Shehata1970} and theoretically \cite{Brown1940,Kendall1948, Hirsch1966,Burnett-Hall1967,bell1967cell,anderson1967cell,anderson1969cell,lebowitz1974theory, Hagander1980,chiorino2001desynchronization} for decades. The desynchronization of an initially nearly synchronous population (i.e., all cells are in nearly the same phase of their cell cycle) of bacteria or cancer cells and its evolution to the steady state can serve as an example \cite{Hirsch1966, Burnett-Hall1967,Hagander1980, chiorino2001desynchronization,gavagnin2021synchronized}.

In recent years, there has been a renewed interest in population dynamics out of the steady state. In particular, the oscillatory behavior of cell number and population growth rate in a desynchronizing bacterial culture has been studied \cite{jafarpour2018bridging, jafarpour2019cell, gavagnin2021synchronized, hein2022asymptotic} in the context of cell division timing strategies and noise in the intrinsic parameters of individual cells. 

Thus, there seems to be a need for generalizing the results obtained so far for the steady state to the unsteady case, that is, to transient population dynamics. These results include the relationships between the generation time distributions of mother cells (those just dividing), newborn cells (those just after cell division), and extant cells (all cells present in the population at a given time). These three generation-time distributions should not be confused with each other, as is still sometimes done; see discussion in \cite{quedeville2019critical}. Only the mother generation time can be observed experimentally. This is because the generation time (also called ``cell cycle time'' or ``interdivision time'') is a ``hidden variable'': Its value is not known until the end of the cell cycle. 

Relationships between various generation time distributions, as well as between generation time distributions, cell age distribution (age structure), and population growth rate, such as the Euler-Lotka equation, were found by Powell \cite{powell1956growth, powell1964note}, Lebowitz and Rubinow \cite{lebowitz1974theory}, and more recently by others \cite{lin2017effects, thomas2017making,levien2020interplay}, but only for bacterial cultures in steady state, for which all probability distributions are time-independent. To the best of our knowledge, the results derived decades ago by Powell and others have not yet been generalized to the case of unsteady state. 

In order to make such a generalization, we use the model proposed by Lebowitz and Rubinow \cite{lebowitz1974theory}. Within this approach, each cell is described by its age $a$ and the generation time $\tau$. As a consequence, one can obtain both the age structure and all of the generation-time probability distributions mentioned above from a single quantity -- the joint probability distribution of $a$ and $\tau$. Using this formalism, we find the relationships between various time-dependent probability distributions of interest, as well as the generalization of the Euler-Lotka equation, for both the batch culture and the mother machine experiment. We also derive inequalities linking averaged rates of appearance and disappearance of cells of generation time $\tau$, which generalize known relations between mean generation time and the population doubling time.

Most of our results for the population in the unsteady state reduce to the known results for the steady exponential growth in batch culture. However, we have also obtained some new results for the steady state: The inequalities concerning the moments of the two generation time distributions: of the mother cells and of the extant cells. These inequalities follow from the stationary limit of the equations relating the time evolution of the moments of different probability distributions. Solutions of these equations are also given.  

Although some probability distributions in Powell's approach and the Lebowitz-Rubinow model are unobservable, the present theoretical framework allows us to express them in terms of experimentally observable quantities.

The results presented here may be useful in a variety of contexts. For example, some of our results are needed to quantify the strength of natural selection by the phenotypic fitness landscape (here we use this term as defined in ref. \cite{nozoe2017inferring}) or by the growth rate of the subpopulation carrying a given phenotypic trait \cite{nozoe2017inferring, genthon2020fluctuation, genthon2021universal}. Therefore, we show how to use our results to calculate fitness landscapes for cell age and generation time. By ``calculate''  here we mean ``express in terms of observable quantities''.

Finally, we discuss possible generalizations of our approach. In particular, we show that any extension of the Lebowitz-Rubinow model that includes additional variables besides cell age and generation time (cell volume, individual growth rate, or concentrations of different proteins) reduces to the effective Lebowitz-Rubinow model when these additional variables are integrated out. Thus, results derived within the Lebowitz-Rubinow model are valid for a much broader class of population models.

We have relegated some of our findings to appendices. Appendix \ref{App_Tab} contains a table of the most important quantities that are used in the text.

\section{Theory \label{Theory}}

\subsection{McKendrick-von Foerster model \label{McKendrick_model_NA}}

We begin with a brief reminder of the McKendrick-von Foerster model \cite{MKendrick1925,von1959kinetics,trucco1965mathematical1, trucco1965mathematical2, rudnicki2014modele}. We need it here as a point of reference for a more general formalism of Lebowitz and Rubinow \cite{lebowitz1974theory}, which is analyzed in the next section. Equations of the McKendrick-von Foerster model read
%
%
\begin{eqnarray}
\label{NAPBE_1}
\left(\frac{\partial}{\partial t} + \frac{\partial}{\partial a} + \gamma(t, a) + D(t)\right) n(t, a) = 0, \\
\label{NAPBE_2}
n(t, 0) = 2^{\sigma} \int_{0}^{\tau_l} \gamma(t, a) n(t, a) d a, \\
\label{NAPBE_3}
n(0, a) = n_0(a).
\end{eqnarray}
Eq. (\ref{NAPBE_1}) describes the time evolution of the number density $n(t, a)$ of cells whose age is $a$; the boundary condition (\ref{NAPBE_2}) describes the influx of newborn cells, while (\ref{NAPBE_3}) is the initial condition. We neglect cell death. $\gamma(t, a)$ is the cell division rate and $\tau_l$ is the maximum possible cell age: $n(t, a) = 0$ for $a > \tau_l$. 

The fermenter dilution rate $D(t)$ in (\ref{NAPBE_1}) may vary with time, but it does not depend on $a$. Eqs. (\ref{NAPBE_1})--(\ref{NAPBE_3}) describe a batch culture ($D(t)=0$, $\sigma = 1$), a continuous culture ($D(t) > 0$, $\sigma = 1$), or a mother machine experiment ($D(t)=0$, $\sigma = 0$). (The mother machine device can be seen as an experimental realization of the ensemble of cell lineages, see Appendix \ref{Forward_vs_Backward}). 

The combination of (\ref{NAPBE_2}) and (\ref{NAPBE_3}) gives us the consistency condition at $t = a = 0$:
\begin{eqnarray}
\label{NAPBE_4}
n_0(0) = 2^{\sigma} \int_{0}^{\tau_l} \gamma(0, a) n_0(a) d a.
\end{eqnarray}
For both $\sigma = 0$ and $\sigma = 1$ 
\begin{equation}
N(t) = \int_{0}^{\tau_l} n(t, a) d a,
\label{N_definition_of_NA}
\end{equation}
is the total number of cells in the population at time $t$. Within the framework of the McKendrick-von Foerster model, it is possible to define both the age distribution of all cells in a population
\begin{equation}
\phi(t, a) = \frac{n(t, a)}{N(t)},
\label{phi_NA_definition}
\end{equation}
and the age distribution of currently dividing cells (mothers), for which age equals generation time \cite{quedeville2019critical}
\begin{eqnarray}
f_1(t, a) &=& \frac{\gamma(t, a) n(t, a)}{\int_{0}^{\tau_l} \gamma(t, a) n(t, a) d a}  = \frac{\gamma(t, a) \phi(t, a)}{\Lambda(t)}. 
\label{carrier_PDF_time_dependent_NA}
\end{eqnarray}
$\Lambda(t)$, which appears in (\ref{carrier_PDF_time_dependent_NA}), is given by 
\begin{equation}
\Lambda(t) = \int_{0}^{\tau_l} \gamma(t, a) \phi(t, a) d a.
\label{definition_of_Lambda_generalized_exponential_growth}
\end{equation}
From (\ref{NAPBE_2}), (\ref{phi_NA_definition}) and (\ref{definition_of_Lambda_generalized_exponential_growth}) we get 
\begin{equation}
n(t, 0) = 2^{\sigma}\Lambda(t)N(t),
\label{boundary_condition_again}
\end{equation}
and therefore $\Lambda(t)$ is proportional to the cell's birth rate. 

For the batch culture ($\sigma = 1$ and $D(t)=0$), $\Lambda(t)$ defined by (\ref{definition_of_Lambda_generalized_exponential_growth}) is the instantaneous population growth rate. In a general case, the population growth rate is given by $\sigma[\Lambda(t)-D(t)]$. This quantity vanishes both for the mother machine ($\sigma = 0$) and for the continuous culture if only $\Lambda(t) = D(t)$, i.e. if the dilution compensates for the increase in cell number.

However, the solution to (\ref{NAPBE_1}) with $\sigma = 1$ and $D(t) \neq 0$ can easily be obtained from the solution to the case where $D(t) = 0$ (see Appendix \ref{CC_vs_BC_McK_vF}), so from now on we will put $D(t) = 0$ in (\ref{NAPBE_1}). In this case, integrating (\ref{NAPBE_1}) with respect to $a$ and using (\ref{NAPBE_2}), we obtain 
\begin{equation}
\label{N_dot_sigma_Lambda_N}
\frac{d{N}(t)}{dt} = {\sigma} N(t)\Lambda(t),
\end{equation}
and therefore 
%
%
\begin{eqnarray}
\label{Number_of_cells_solution}
N(t) &=&  N_0 e^{ \sigma \int_0^{t} \Lambda(t^{\prime}) d t^{\prime}},
\end{eqnarray}
where $N_0 \equiv N(0)$. 

\subsection{Lebowitz-Rubinow model \label{a_tau_model_A}} 

Most of the results presented in this paper were obtained using the theoretical framework of the Lebowitz-Rubinow model. Therefore, we will now briefly discuss this model.

Starting with the McKendrick-von Foerster model (\ref{NAPBE_1})--(\ref{NAPBE_3}), one can derive the relationship between the two age distributions: $\phi(t, a)$ (\ref{phi_NA_definition}) for all cells in the population and $f_1(t, a)$  (\ref{carrier_PDF_time_dependent_NA}) for mothers, see Appendix \ref{McKendrick_model_NA_Appendix}. 

However, the generation time $\tau$ (duration of the cell cycle), and thus its inheritance, is not explicitly present in (\ref{NAPBE_1})--(\ref{NAPBE_3}). Therefore, the McKendrick-von Foerster model is not the preferred choice if one wants to obtain similar relationships involving the remaining generation time distributions: $f_0(t, \tau)$ (newborns) and $f_2(t, \tau)$ (extant cells, i.e., those present in a population at any given time), or the joint probability distribution of cell age and generation time, $\chi(t, a, \tau)$. For this reason, we use the model proposed by Lebowitz and Rubinow \cite{lebowitz1974theory}. In this approach, the generation time $\tau$ becomes an additional, non-dynamic variable ($d{\tau}/dt=0$) besides the cell age.

$n(t, a)$ in Eqs. (\ref{NAPBE_1}--\ref{NAPBE_3}) is now replaced by the number density $n(t, a, \tau)$ of the cells whose age is $a$ and whose generation time is $\tau$. These two densities are related by 
\begin{equation}
\label{n_generalized_reduced_to_n_standard}
n(t, a) = \int_{\underline{a}}^{\tau_l} n(t, a, \tau) d\tau,
\end{equation}
so the number of cells in the population is given by 
\begin{equation}
\label{Number_of_cells_definition_LR_model}
N(t) = \int_{0}^{\tau_l} \int_{\underline{a}}^{\tau_l} n(t, a, \tau) d \tau d a = \int_{\tau_s}^{\tau_l} \int_{0}^{\tau} n(t, a, \tau) d a d \tau.
\end{equation}
We assume that $\tau$ is bounded,  $0<\tau_s \leq \tau \leq \tau_l <\infty$; hence $\tau_l$ is the longest possible generation time, and $\tau_s$ is the shortest possible generation time: 
\begin{equation}
\label{a_tau_domain_definition_of}
0 \leq a \leq \tau, ~~~~ \underline{a} \equiv \max(a, \tau_s) \leq \tau \leq \tau_l,
\end{equation}
%
%
because $\tau$ must be greater than both $a$ and $\tau_s$. For \textit{E. coli} growing under optimal conditions $\tau_s \approx $ 20 min \cite{taheri2015cell}.

\subsubsection{Model equation, its boundary and initial conditions \label{Equations_of_the_LR_model}}

Within the approach of ref. \cite{lebowitz1974theory}, Eq. (\ref{NAPBE_1}) of the McKendrick-von Foerster model is replaced by
\begin{eqnarray}
\label{APBE_1}
\frac{\partial}{\partial t}n(t, a, \tau) + \frac{\partial}{\partial a} n(t, a, \tau) = 0, 
\end{eqnarray}
while the boundary and initial conditions are now 
\begin{equation}
\label{APBE_2}
n(t, 0, \tau) = 2^{\sigma} \int_{\tau_s}^{\tau_l} h(\tau | \tau^{\prime}) n(t, \tau^{\prime}, \tau^{\prime}) d \tau^{\prime} \equiv \Psi(t, \tau), 
\end{equation} 
and
\begin{eqnarray}
\label{APBE_3}
n(0, a, \tau) = n_0(a, \tau) \equiv \Phi(a, \tau). 
\end{eqnarray}
 The  consistency condition must be satisfied at $a = t = 0$:
\begin{equation}
\label{consistency_at_zero_zero_boundary_initial}
\Phi(0, \tau) = \Psi(0, \tau). 
\end{equation}
$h(\tau | \tau^{\prime}) d\tau$   in Eq. (\ref{APBE_2}) is the probability that the generation time of the two daughters is $\tau$, provided that the generation time of their mother was $\tau^{\prime}$. (We assume  that both daughter cells inherit the same generation time but this assumption can be relaxed, see Appendix \ref{Sisters}).

When the mother-daughter generation time correlations vanish, $h(\tau | \tau^{\prime}) = f(\tau)$. In such a case, the inherited generation time no longer depends on $\tau^{\prime}$. The opposite extreme is when daughters inherit the value of their mother's generation time at cell division: $h(\tau | \tau^{\prime}) = \delta(\tau - \tau^{\prime})$ \cite{lebowitz1974theory}. 

For continuous culture, the term $D(t) n(t, a, \tau)$ responsible for cell dilution should also be added to the l.h.s. of Eq. (\ref{APBE_1}). However, we omit this term because the solution to (\ref{APBE_1}) with $D(t) \neq 0$ can easily be obtained from the solution to the case $D(t) = 0$, just as in the McKendrick-von Foerster model (see Appendix \ref{CC_vs_BC_McK_vF}). The same remark applies to the term describing cell death, included in the original formulation of the Lebowitz-Rubinow model \cite{lebowitz1974theory}.

As in the case of Eq. (\ref{NAPBE_2}), in Eq. (\ref{APBE_2}) we have $\sigma = 0$ for a single cell lineage or the mother machine experiment (which, as we will show in Appendix \ref{Forward_vs_Backward}, can be treated as a realization of the ensemble of single cell lineages) and $\sigma = 1$ for the batch culture. Note that most of the quantities of interest are different for $\sigma = 0$ and $\sigma = 1$. Only the probability distribution of the inherited generation times $h(\tau | \tau^{\prime})$, which appears in (\ref{APBE_2}), and the initial condition $\Phi(a,\tau)$ (\ref{APBE_3}) are the same for both cases. (If the Lebowitz-Rubinow model was derived as an effective description from a more general model, then $h(\tau | \tau^{\prime})$ can also depend on both $\sigma$ and the observation time $t$, see subsection \ref{GLRM} and Appendix \ref{correlations_and_inheritance}). 

To distinguish between $\sigma = 0$ and $\sigma = 1$, we introduce an index: 
\begin{equation}
\ell(\sigma) =\begin{cases}
c ~~~  \text{for}~~~\sigma = 0, \\ \\
r ~~~ \text{for}~~~\sigma = 1. 
\label{F_or_r_ell_od_sigma}
\end{cases}
\end{equation}
%
%
In the above, $c$ refers to `chronological' or `forward', and $r$ to `retrospective' or `backward'; these terms are related to the two ways the population lineage tree can be sampled, see \cite{nozoe2017inferring, genthon2020fluctuation, genthon2021universal}. That $\sigma = 0$ corresponds to chronological  and $\sigma = 1$ to retrospective sampling of the lineage tree will be shown in the Appendix \ref{Forward_vs_Backward}. However, we explicitly distinguish between $\sigma = 0$ and $\sigma = 1$ only when the quantities with different values of $\sigma$ appear in the same formula (as e.g. in section  \ref{Fitness_landscapes} and Appendix \ref{Forward_vs_Backward}) or when we consider the steady state situation.


\subsubsection{Formal solution of the Lebowitz-Rubinow model}

A solution to (\ref{APBE_1}) has the form $n(t, a, \tau) = \mathcal{F}(t - a, \tau)$, where $\mathcal{F}(x_1, x_2)$ is a function of two real variables. Taking into account the initial and boundary conditions, we obtain \cite{lebowitz1974theory}
\begin{eqnarray}
n(t, a, \tau) &=&
\begin{cases}
\Phi(a - t, \tau) = n(0, a - t, \tau)~ \text{if}~a \geq t, \\ \\
\Psi(t - a, \tau) = n(t - a, 0, \tau)~ \text{if}~a \leq t, \label{solution_for_free_propagation_t_a_tau_characteristics}
\end{cases}
\end{eqnarray}
where the initial condition $\Phi(a, \tau)$ is defined by (\ref{APBE_3}) and the boundary condition $\Psi(t, \tau)$ is defined by (\ref{APBE_2}). The condition (\ref{consistency_at_zero_zero_boundary_initial}) ensures that $\Phi(a-t, \tau) = \Psi(t - a, \tau)$ for $a = t$. 

From (\ref{APBE_2}) and (\ref{solution_for_free_propagation_t_a_tau_characteristics}) we get the renewal equation: 
\begin{eqnarray}
\label{solution_for_free_propagation_t_a_tau_self_consistent_a_la_RR}
 \Psi(t, \tau) &=& 2^{\sigma} \Theta(\tau_l - t) \int_{\underline{t}}^{\tau_l} h(\tau | \tau^{\prime}) \Phi(\tau^{\prime} - t, \tau^{\prime}) d \tau^{\prime} \nonumber \\ & + & 2^{\sigma}\Theta(t-\tau_s) \int_{\tau_s}^{\overline{t}} h(\tau | \tau^{\prime}) \Psi(t-\tau^{\prime}, \tau^{\prime}) d \tau^{\prime}, \nonumber \\
\end{eqnarray}
where $\Theta (x)$ is the Heaviside step function and
\begin{equation}
\label{t_bar_definition}
\overline{t} \equiv \min(t, \tau_l), ~~~ \underline{t} \equiv \max(t, \tau_s).
\end{equation}
We can now rewrite (\ref{consistency_at_zero_zero_boundary_initial}) as 
\begin{equation}
\label{consistency_at_zero_LR}
\Phi(0, \tau) = \Psi(0, \tau) = 2^{\sigma} \int_{\tau_s}^{\tau_l} h(\tau | \tau^{\prime}) \Phi(\tau^{\prime}, \tau^{\prime}) d \tau^{\prime}.
\end{equation}
By integrating (\ref{consistency_at_zero_LR}) with respect to $\tau$, we get (\ref{NAPBE_4}).

\subsection{Reduction of the Lebowitz-Rubinow model to the McKendrick-von Foerster model \label{Reduction_of_LR_to_McKvF}}

Before proceeding, we want to clarify the connection between the Lebowitz-Rubinow and the McKendrick-von Foerster models. When the generation time $\tau$ is integrated out, eqs. (\ref{APBE_1})--(\ref{APBE_3}) of the former model should reduce to eqs. (\ref{NAPBE_1})--(\ref{NAPBE_3}) of the latter model. This should be the case not only for the special form of the initial condition considered in ref. \cite{lebowitz1974theory}, i.e. 
\begin{equation}
\label{Initial_condition_LR}
\Phi(a, \tau) = n_0(a) \frac{f(\tau)}{\int_a^{\tau_l}f(\tau^{\prime})d \tau^{\prime}}
\end{equation}
but in a general situation. 

If $a \geq \tau_s$, then if one integrates (\ref{APBE_1}) with respect to $\tau$ from $a$ to $\tau_l$ and uses the Leibniz integral rule, one gets 
\begin{eqnarray}
\label{APBE_1_integrated_with_respect_to_tau}
\frac{\partial}{\partial t}n(t, a) + \frac{\partial}{\partial a} n(t, a) + n(t, a, a) = 0, 
\end{eqnarray}
where $n(t, a)$ is defined by (\ref{n_generalized_reduced_to_n_standard}).
Similarly, integrate (\ref{APBE_2}) with respect to $\tau$ from $\tau_s$ to $\tau_l$. Keeping in mind that for any $\tau^{\prime}$ we have $\int_{\tau_s}^{\tau_l} h(\tau | \tau^{\prime}) d \tau = 1$, we get
\begin{equation}
\label{APBE_2_integrated_with_respect_to_tau}
n(t, 0) = \int_{\tau_{s}}^{\tau_{l}} \Psi(t, \tau) d \tau = 2^{\sigma} \int_{\tau_s}^{\tau_l} n(t, \tau^{\prime}, \tau^{\prime}) d \tau^{\prime}. 
\end{equation}
Comparing (\ref{APBE_1_integrated_with_respect_to_tau}) with (\ref{NAPBE_1}) and (\ref{APBE_2_integrated_with_respect_to_tau}) with (\ref{NAPBE_2}), we see that if
\begin{equation}
n(t, a, a) = \gamma(t, a)n(t, a) = f_1(t, a) N(t)\Lambda(t),
\label{carrier_NA_vs_anticipating_correspondence}
\end{equation}
then $n(t, a)$ defined by (\ref{n_generalized_reduced_to_n_standard}) satisfies the equations (\ref{NAPBE_1})--(\ref{NAPBE_3}). $f_1(t, \tau)$ in Eq. (\ref{carrier_NA_vs_anticipating_correspondence}) is the mother age distribution, which is defined by Eq. (\ref{carrier_PDF_time_dependent_NA}) in the McKendrick-von Foerster model, while for $\sigma=1$, $\Lambda(t)$ (\ref{definition_of_Lambda_generalized_exponential_growth}) is the instantaneous growth rate of the population. For $a < \tau_s$, $\underline{a} = \tau_s$ and $n(t, a, a) = \gamma(t, a) = 0$; thus the Lebowitz-Rubinow model reduces again to the McKendrick-von Foerster model. If the condition (\ref{carrier_NA_vs_anticipating_correspondence}) is satisfied, all the results of section \ref{McKendrick_model_NA} remain valid. In particular, the time dependence of the total number of cells is given by  (\ref{Number_of_cells_solution}).

Finally, note that from the Lebowitz-Rubinow model one can derive not only the McKendrick-von Foerster model, but also the model formally identical to that proposed by Rubinow in 1968 \cite{rubinow1968maturity}, see Appendix \ref{Rubinow}.

\subsection{Definitions of cell age and generation time distributions\label{Definitions_of_f_i_s} }

In this subsection, we introduce the probability distributions that will be used throughout the rest of this paper.

One can obtain all the generation time probability distributions: $f_i(t, \tau)$, $i=0$, $1$, $2$ and the cell age distribution $\phi(t, a)$ as either conditional or marginal probabilities from a single quantity -- the joint probability  distribution $\chi(t, a, \tau)$ of age and generation time. The latter distribution is the cell number density $n(t, a, \tau)$ normalized by the total number of cells $N(t)$:
%
%
\begin{equation}
\chi(t, a, \tau) \equiv \frac{n(t, a, \tau)}{N(t)}.
\label{chi_definition_of}
\end{equation}
%
We also define 
\begin{equation}
\chi_0(a, \tau) \equiv \chi(0, a, \tau)  = \frac{\Phi(a, \tau)}{N_0}.
\label{chi_0_definition_of}
\end{equation}
With $\chi(t, a, \tau)$ we get the age distribution of all cells in the population (extant cells)
\begin{equation}
\label{phi_derived_from_chi_definition}
\phi(t, a) = \int_{\underline{a}}^{\tau_l} \chi(t, a, \tau) d \tau = \frac{n(t, a)}{N(t)},
\end{equation}
as well as the generation time distribution of extant cells, 
\begin{equation}
f_2(t, \tau) = \int_{0}^{\tau} \chi(t, a, \tau) d a.
\label{t_dependent_f_2_definition}
\end{equation}
Next, define the following conditional distribution:
\begin{equation}
\chi(t, \tau |a) \equiv \frac{\chi(t, a, \tau)}{\phi(t, a)}.
\label{conditional_chi_definition}
\end{equation}
The generation time distribution of the newborns is then given by
\begin{equation}
f_0(t, \tau) \equiv \chi(t, \tau |0) \equiv \frac{\chi(t, 0, \tau )}{\phi(t, 0)} = \frac{\chi(t, 0, \tau)}{2^{\sigma} \Lambda(t)}.
\label{f_0_definition}
\end{equation}
The identity
\begin{equation}
\label{phi_at_zero}
{\phi(t, 0)} = 2^{\sigma} \Lambda(t)
\end{equation}
used in (\ref{f_0_definition}) follows from (\ref{boundary_condition_again}) and (\ref{phi_derived_from_chi_definition}).

The mother age distribution $f_1(t, \tau)$ (called `carrier distribution' by Powell \cite{powell1956growth,powell1964note}) has already been defined by (\ref{carrier_NA_vs_anticipating_correspondence}),
%
%
\begin{equation}
f_1(t, a) \equiv \frac{\chi(t, a, a)}{\int^{\tau_l}_{\tau_s} \chi(t, a, a) da} = \frac{\chi(t, a, a)}{\Lambda(t)}.
\label{t_dependent_f_1_definition}
\end{equation}
Note that the above definition of $f_1(t, \tau)$ is consistent with its definition (\ref{carrier_PDF_time_dependent_NA}) within the McKendrick von Foerster model.

Not only $f_{0}(t, \tau)$, but also $f_{1}(t, \tau)$ is a conditional distribution. To see it, define 
\begin{eqnarray}
\overline{\chi}(t, a, \theta) &\equiv & \chi(t, a, a + \theta), ~~~ \theta \equiv \tau - a,
\label{chi_tilde_solution}
\end{eqnarray}
and the corresponding conditional distribution
\begin{equation}
\overline{\chi}(t, a |\theta) = \frac{\overline{\chi}(t, a, \theta)}{\int_{a_s(\theta)}^{a_l(\theta)} \overline{\chi}(t, a, \theta) d a},
\label{mothers_rederrived_both_sigma_t_dep_a}
\end{equation}
where $a_s(\theta) = \max(0, \tau_s - \theta)$,  $a_l(\theta) = \tau_l - \theta$. For mothers, $\theta = 0$, so $f_1(t, a) = \overline{\chi}(t, a|0)$. Note that $f_{1}(t, a) \neq \chi(t, a |a)$ since the latter is not a probability distribution of $a$. From (\ref{carrier_NA_vs_anticipating_correspondence}), (\ref{chi_definition_of}) and (\ref{conditional_chi_definition}) it follows that $\chi(t, a |a) = \gamma(t, a)$.

As mentioned in the introduction, $f_{1}(t, \tau)$ is an experimentally observable quantity. Therefore, it makes sense to express other probability distributions using $f_{1}(t, \tau)$. 

\subsection{Definition of steady state}  

In this subsection, we give the definition of a steady state.

For mother-machine experiments ($\sigma = 0$), ``steady state'' means that all quantities are independent of time (stationary), i.e., $n(t, a, \tau)=n_c(a, \tau)$, $n(t, a)=n_c(a)$, $N(t) = N_0$, $\chi(t, a, \tau)=\chi_c(a, \tau)$, $\Lambda(t)=\Lambda_c$, and so on. This is therefore a true steady state. Note that for the mother  machine we always have $N(t) = N_0$, not only in the steady state limit.

For batch culture ($\sigma = 1$), by ``steady state'' we mean steady exponential growth, for which we have
\begin{equation}
n(t, a, \tau)= \chi_r(a, \tau) N_0 e^{\Lambda_r t}.
\label{n_LR_ss_def}
\end{equation}
So we also have  
\begin{equation}
N(t) = N_0 e^{\Lambda_r t}.
\label{N_LR_ss_def}
\end{equation}
Eq. (\ref{n_LR_ss_def}) explains why steady exponential growth is sometimes called ``self-similar growth.'' This term derives from the fact that the shape of the plot of the cell number density as a function of its variables other than the observation time $t$ (here: $a$ and $\tau$) does not depend on $t$. From (\ref{chi_0_definition_of}), (\ref{n_LR_ss_def}), and (\ref{N_LR_ss_def}) it is clear that not only $\chi(t, a, \tau)$ (\ref{chi_definition_of}) but also the remaining probability distributions do not depend on $t$.

\section{Results \label{Results}}

\subsection{Relationships between probability distributions of cell age and generation time\label{GTD_relationships_between}}

In this subsection, we present our main results: The relations between the probability distributions defined in the previous subsection: $\chi(t, a, \tau)$, $\phi(t, a)$, $f_0(t, \tau)$, $f_1(t, \tau)$, and $f_2(t, \tau)$. We also derive two forms of the generalized Euler-Lotka equation for a transient population dynamics.

We will present here the relationships between different probability distributions for the same values of $\sigma$. In other words, we consider two different experimental situations separately: the mother machine experiments ($\sigma = 0$) and the batch culture ($\sigma = 1$). The answer to another question: how is a given distribution for batch culture expressed in terms of the same distribution for the mother machine experiment, is given in Appendix \ref{Forward_vs_Backward}.

\subsubsection{Relationships between generation time probability distributions of mother and daughter cells} 

Using (\ref{solution_for_free_propagation_t_a_tau_characteristics}), (\ref{carrier_NA_vs_anticipating_correspondence}) and (\ref{f_0_definition}), we rewrite (\ref{APBE_2}) as
\begin{eqnarray}
n(t, 0, \tau) &=& 2^{\sigma} N(t)\Lambda(t) \int_{\tau_s}^{\tau_l} h(\tau | \tau^{\prime}) f_1(t, \tau^{\prime}) d \tau^{\prime} \nonumber \\ &=& 2^{\sigma} N(t)\Lambda(t) f_0(t, \tau) = \Psi(t, \tau), 
\label{Psi_expressed_via_f1_f0}
\end{eqnarray}
where the generation time distribution of newborn cells $f_0(t, \tau)$ is defined by (\ref{f_0_definition}). In this way, we obtain 
\begin{eqnarray}
f_0(t, \tau) &\equiv & \int_{\tau_s}^{\tau_l} h(\tau |\tau^{\prime}) f_1(t, \tau^{\prime}) d \tau^{\prime}. 
\label{f_0_definition_2}
\end{eqnarray}
One can think of Eq. (\ref{f_0_definition_2}) as another definition of $f_0(t, \tau)$, alternative to (\ref{f_0_definition}). If there are no mother-daughter generation time correlations, we have $h(\tau |\tau^{\prime}, t) = f(\tau) = f_0(t, \tau)$. (Following refs. \cite{powell1956growth, powell1964note, lin2017effects}, we denote the uncorrelated generation time distribution of newborns as $f(\tau)$ instead of $f_0(t, \tau)$). At the opposite extreme, where daughters inherit exactly the same generation time as their mother had, i.e., $h(\tau | \tau^{\prime}) = \delta(\tau - \tau^{\prime})$ we have $f_0(t, \tau) = f_1(t, \tau)$. For $t \geq \tau$, from (\ref{APBE_2}), (\ref{solution_for_free_propagation_t_a_tau_characteristics}), (\ref{carrier_NA_vs_anticipating_correspondence}), (\ref{t_dependent_f_1_definition}) and (\ref{Psi_expressed_via_f1_f0}), we get
\begin{equation}
\label{relation_between_f_1_and_f_0_time_dependent_derivation}
N(t)\Lambda(t) f_1(t, \tau) 
= 2^{\sigma} N(t-\tau)\Lambda(t-\tau) f_0(t-\tau, \tau). 
\end{equation}
$N(t)\Lambda(t)dt$ is the total number of cell divisions in the population at time $t$, and the factor $2^{\sigma}$ accounts for the number of daughter cells remaining in the population after each cell division. Therefore, the interpretation of the identity (\ref{relation_between_f_1_and_f_0_time_dependent_derivation}) is simple: The cells that divide at the time $t$ are those that were born at the time $t-\tau$, and which inherited the generation time $\tau$ at birth to divide when they reach the age $a=\tau$. Using (\ref{Number_of_cells_solution}), we rewrite (\ref{relation_between_f_1_and_f_0_time_dependent_derivation}) as 
%
%
\begin{equation}
\label{relation_between_f_1_and_f_0_time_dependent_final_form}
f_1(t, \tau) = \frac{2^{\sigma} \Lambda(t-\tau)e^{- \int_{t - \tau}^{t} \sigma \Lambda(t^{\prime}) d t^{\prime} }}{\Lambda(t)} f_0(t-\tau, \tau). 
\end{equation}
For $\sigma = 1$ (batch culture, corresponding to the retrospective probabilities and retrospective sampling \cite{nozoe2017inferring})  in the steady-state limit, we obtain from (\ref{relation_between_f_1_and_f_0_time_dependent_final_form}) the well-known relationship between generation time distributions of newborn and mother cells \cite{powell1956growth,rubinow1968maturity,lin2017effects,thomas2017making}
\begin{equation}
f_{1r}(\tau) = 2e^{-\Lambda_r \tau} f_{0r}(\tau).
\label{f_1_definition_by_f_0_ss}
\end{equation}
For $\sigma = 0$, i.e. for the mother machine experiment (corresponding to chronological probabilities and chronological sampling), in the steady state limit we get:
\begin{equation}
f_{1c}(\tau) = f_{0c}(\tau).
\label{relationship_between_f_0_f_1_c_ss}
\end{equation}
Now let us return to the case of the transient state. For $t \geq \tau_l$, we get the generalization of Eq. (18) of ref. \cite{powell1956growth} (Eq. (32) of ref. \cite{lebowitz1974theory}) by combining (\ref{f_0_definition_2}) and (\ref{relation_between_f_1_and_f_0_time_dependent_final_form}) 
\begin{equation}
\label{time_dependent_Euler_Lotka_wyrazenie_podcalkowe_0}
f_0(t, \tau) 
= 2^{\sigma} \int_{\tau_s}^{\tau_l} h(\tau |\tau^{\prime}) \mathcal{L}(t, \tau^{\prime})  f_0(t-\tau^{\prime}, \tau^{\prime}) d \tau^{\prime},
\end{equation}
where we introduced the shorthand notation
\begin{eqnarray}
\label{time_dependent_Euler_Lotka_definition_of_L_mathcal}
\mathcal{L}(t, \tau) &\equiv & \frac{\Lambda(t-\tau)e^{\sigma \int_{0}^{t - \tau}  \Lambda(t^{\prime}) d t^{\prime} }}{\Lambda(t)e^{\sigma \int_{0}^{t}  \Lambda(t^{\prime}) d t^{\prime} }}.  
\end{eqnarray}
Similarly, from (\ref{f_0_definition_2}) and (\ref{relation_between_f_1_and_f_0_time_dependent_final_form}) we get the analogous equation for $f_1(t, \tau)$, valid for $t \geq \tau$,
\begin{equation}
\label{time_dependent_Euler_Lotka_wyrazenie_podcalkowe_1}
f_1(t, \tau) 
= 2^{\sigma} \mathcal{L}(t, \tau) \int_{\tau_s}^{\tau_l} h(\tau |\tau^{\prime}) f_1(t-\tau, \tau^{\prime}) d \tau^{\prime}.
\end{equation}
Equation (\ref{time_dependent_Euler_Lotka_wyrazenie_podcalkowe_1}) can help find the functional forms of $h(\tau |\tau^{\prime})$ that are consistent with the measured values of observable quantities: instantaneous population growth rate $\Lambda(t)$ and generation time distributions of mothers, $f_1(t, \tau)$, determined from the experiment. 

\subsubsection{The first form of the Euler-Lotka equation \label{EL_1st}}

For the population at the state of steady exponential growth, the Euler-Lotka equation \cite{powell1956growth, lebowitz1974theory, lin2017effects} is the normalization condition for $f_{1r}(\tau)$ (\ref{f_1_definition_by_f_0_ss}), 
\begin{equation}
1 = 2 \int_{\tau_s}^{\tau_l} e^{-\Lambda_{r} \tau} f_{0r}(\tau) d\tau.
\label{Euler_Lotka_ss}
\end{equation}
In order to generalize (\ref{Euler_Lotka_ss}) to the case of transient population dynamics we can use 
(\ref{relation_between_f_1_and_f_0_time_dependent_final_form}),  (\ref{time_dependent_Euler_Lotka_wyrazenie_podcalkowe_0}) or  (\ref{time_dependent_Euler_Lotka_wyrazenie_podcalkowe_1}). Integrating both sides of (\ref{relation_between_f_1_and_f_0_time_dependent_final_form}) and remembering that $\tau \leq t$ we get
\begin{eqnarray}
\label{time_dependent_Euler_Lotka_simplest}
\int_{\tau_s}^{\overline{t}} f_1(t, \tau) d \tau &=& 2^{\sigma} \int_{\tau_s}^{\overline{t}} \mathcal{L}(t, \tau) f_0(t-\tau, \tau) d \tau, \nonumber \\ 
\end{eqnarray}
where $\mathcal{L}(t, \tau)$ has been defined by (\ref{time_dependent_Euler_Lotka_definition_of_L_mathcal})
and $\overline{t} \equiv \min(t, \tau_l)$ by (\ref{t_bar_definition}). For $t \geq \tau_l$ we have $\overline{t} = \tau_l$. The r.h.s. of (\ref{time_dependent_Euler_Lotka_simplest}) is then equal to one, and we get
\begin{eqnarray}
\label{time_dependent_Euler_Lotka}
1 &=& 2^{\sigma} \int_{\tau_s}^{\tau_l} \mathcal{L}(t, \tau) f_0(t-\tau, \tau) d \tau \nonumber \\ &=& 2^{\sigma} \int_{\tau_s}^{\tau_l} \int_{\tau_s}^{\tau_l} \mathcal{L}(t, \tau) h(\tau |\tau^{\prime}) f_1(t-\tau, \tau^{\prime}) d \tau^{\prime} d \tau. \nonumber \\ 
\end{eqnarray}
By using the normalization of $f_0(t, \tau)$ given by (\ref{time_dependent_Euler_Lotka_wyrazenie_podcalkowe_0}) we do not get a new form of the Euler-Lotka equation, but we come back to Eq. (\ref{time_dependent_Euler_Lotka}).
Strictly speaking, if $\tau_l=\infty$, then (\ref{time_dependent_Euler_Lotka}) is only satisfied in the limit $t \to \infty$, where we get Eq. (\ref{Euler_Lotka_ss}) for $\sigma=1$ and the normalization of $f_{0c}(\tau)$ for $\sigma=0$. However, it is reasonable to expect that there is an intermediate time scale for which replacing $\overline{t}$ by $\infty$ in (\ref{time_dependent_Euler_Lotka_simplest}) gives a satisfactory approximation, even though the system is still far enough from the steady state. 

Both $\Lambda(t)$ and $f_1(t, \tau)$ can be determined directly from the experiment. But since neither $h(\tau |\tau^{\prime})$ nor $f_0(t, \tau)$ is experimentally measurable, the equation (\ref{time_dependent_Euler_Lotka}) does not provide an alternative way to determine $\Lambda(t)$ from other measurable quantities. Nevertheless, with experimentally determined values of $\Lambda(t)$ and $f_1(t, \tau)$ one can check whether the form $h(\tau |\tau^{\prime})$ postulated by a given theoretical model is not excluded by (\ref{time_dependent_Euler_Lotka}).


In subsection \ref{EL_2nd} we will derive another form of the Euler-Lotka equation which is valid for all values of the observation time $t$ and is expressed only in terms of experimentally observable quantities.

\subsubsection{Joint distribution of cell age and generation time, expressed by the generation time distributions of the mothers or of the newborns. \label{n_expressed_via_f_i}}

Our task now is to express the most important probability distribution: $\chi(t, a, \tau)$ in terms of observable quantities:  $f_1(t, \tau)$ and $\Lambda(t)$. For the sake of completeness, we will also give formulas expressing $\chi(t, a, \tau)$ in terms $f_0(t, \tau)$.

First, we will express the cell number density $n(t, a, \tau)$ using the generation time distribution $f_1(t, \tau)$ of mothers or $f_0(t, \tau)$ of newborns. In what follows, we assume that $t\geq a$. From (\ref{solution_for_free_propagation_t_a_tau_characteristics}), (\ref{Psi_expressed_via_f1_f0}), (\ref{relation_between_f_1_and_f_0_time_dependent_derivation}), and (\ref{relation_between_f_1_and_f_0_time_dependent_final_form}), we obtain 
\begin{eqnarray}
n(t, a, \tau) &=& \Lambda(t-a + \tau) N_0 e^{\sigma\Omega(t-a + \tau)} {f}_1(t-a + \tau, \tau)\nonumber  \\ &=& 2^\sigma \Lambda(t-a) N_0 e^{\sigma\Omega(t-a)} {f}_0(t-a, \tau)  
\label{n_solution_through_f_0_f1}
\end{eqnarray}
and therefore
\begin{eqnarray}
\chi(t, a, \tau) &=&  \Lambda(t-a + \tau) e^{\sigma\Omega(t-a + \tau) -\sigma\Omega(t)} {f}_1(t-a + \tau, \tau), \nonumber \\  &=& 2^\sigma \Lambda(t-a) e^{\sigma\Omega(t-a)} e^{-\sigma\Omega(t)} {f}_0(t-a, \tau),
\label{chi_solution}
\end{eqnarray}
where we define
\begin{equation}
\Omega(t) \equiv \int_0^{t} \Lambda_r(t^{\prime}) d t^{\prime}.
\label{Omega_definition}
\end{equation}
Note that although the r.h.s. of both (\ref{n_solution_through_f_0_f1}) and (\ref{chi_solution}) is defined for all $\tau \in [\tau_s, \tau_l]$, the l.h.s. of each of these two formulas only makes sense for $\tau > a$.
For $\sigma = 1$ in the steady-state limit ($t \to \infty$), we have $\Lambda(t) = \Lambda_r$, ${f}_0(t-a, \tau) = {f}_{0r}(\tau)$, $\Omega(t) = \Lambda_r t$ and (\ref{chi_solution}) reduces to
\begin{equation}
\label{solution_for_chi_sigma_1}
\chi_r(a, \tau) = 2 \Lambda_r f_{0r}(\tau) e^{-\Lambda_r a} = \Lambda_r f_{1r}(\tau) e^{\Lambda_{r}(\tau - a)}. 
\end{equation}
In this limit, $n(t, a, \tau)$ (\ref{n_solution_through_f_0_f1}) is indeed of the form (\ref{n_LR_ss_def}), as it should be. For $\sigma = 0$ ($\ell(\sigma) = c$) we get \cite{thomas2017making}
\begin{equation}
\label{chi_MM}
\chi_c(a, \tau) = \Lambda_c f_{0c}(\tau) = \frac{f_{0c}(\tau)}{\int_{\tau_s}^{\tau_l} \tau^{\prime} f_{0c}(\tau^{\prime}) d \tau^{\prime}}.
\end{equation}

\subsubsection{Generation time distribution of extant cells expressed in terms of generation time distributions for mothers and newborns}

First, we want to express the unobservable generation time distribution of extant cells, $f_2(t, \tau)$, by the observable generation time distributions for mothers, $f_1(t, \tau)$. Using (\ref{t_dependent_f_2_definition}) and the properties of the solution of the Lebowitz-Rubinow model, it can be shown that for both $t \leq \tau$ and $t \geq \tau$ we have
\begin{eqnarray}
\label{f_2_expressed_using_f_1}
f_2(t, \tau) &=& \int_{t}^{t+\tau} e^{\sigma \int_{t}^{t^{\prime}} \Lambda(\tilde{t}) d \tilde{t}} \Lambda(t^{\prime}) f_1(t^{\prime}, \tau) d t^{\prime}, 
\end{eqnarray}
see Appendix \ref{phi_and_f1_R_App}. Note that the value of $f_2(t, \tau)$ at time $t$ is expressed by $f_1(t^{\prime}, \tau)$ at later times: $t^{\prime} \in [t, t + \tau]$. This means that we cannot determine $f_2(t, \tau)$ using only observations made at time $t$. Nevertheless, (\ref{f_2_expressed_using_f_1}) has a simple and intuitive interpretation, which is best seen when it is rewritten in terms of the number density
\begin{equation}
\nu(t,\tau) \equiv \int_{0}^{\tau} n(t, a, \tau) d a = N(t)f_2(t, \tau).
\label{nu_definition}
\end{equation}
%
of the cells whose generation time is $ \tau $ at time $ t $. Now multiply both sides of (\ref{f_2_expressed_using_f_1}) by $N(t)$ to get
\begin{equation}
\label{nu_expressed_using_f_1}
\nu(t, \tau) = \int_{t}^{t+\tau} \Lambda(t^{\prime}) N(t^{\prime}) f_1(t^{\prime}, \tau) d t^{\prime} . 
\end{equation}
Eq. (\ref{nu_expressed_using_f_1}) expresses the fact that all cells assigned with a given generation time $\tau$ that were present in the population at the observation time $t$ (and only such cells) will divide during the time interval $[{t}, {t+\tau}]$. In contrast, the cells born within this time interval and inheriting the generation time $\tau$ will divide at $t^{\prime \prime } > {t+\tau}$.

Using (\ref{t_dependent_f_2_definition}) and (\ref{chi_solution}) we can also get the expression that connects $f_2(t, \tau)$ and the generation time distribution for newborns  $f_0(t, \tau)$:
\begin{equation}
f_2(t, \tau) = 2^{\sigma} e^{-\sigma\Omega(t)} \int^{t}_{t - \tau} \Lambda(\tilde{t}) e^{\sigma\Omega(\tilde{t})} f_0(\tilde{t}, \tau) d \tilde{t}. 
\label{t_dependent_f_2_derived_from_chi}
\end{equation}
Eq. (\ref{t_dependent_f_2_derived_from_chi}) tells us that all those cells present in the population at observation time $t_{\text{obs}} = t$, whose generation time is $\tau$, must have been born between $t - \tau$ and $t$, because the cells assigned to generation time $\tau$ and born earlier have already divided.

For $t \geq \tau$ we can deduce another relationship between $f_2(t, \tau)$, $f_1(t, \tau)$ and $f_0(t, \tau)$. Our starting point now is the equation for the time evolution of $\nu(t,\tau)$ (\ref{nu_definition}). To obtain it, we integrate Eq. (\ref{APBE_1}) with respect to $a$, as in (\ref{nu_definition}), and we use the boundary condition (\ref{APBE_2}). Then, using Eq. (\ref{carrier_NA_vs_anticipating_correspondence}), we obtain 
\begin{eqnarray}
\frac{d{\nu}(t,\tau)}{dt} & = & n(t, 0, \tau) - n(t, \tau, \tau) \nonumber \\ 
&=& \left[2^{\sigma} f_0(t, \tau) - f_1(t, \tau) \right] N(t)\Lambda(t). 
\label{time_evolution_of_nu}
\end{eqnarray}
It is easy to identify the gain and loss terms in equation (\ref{time_evolution_of_nu}): The influx of newborns with generation time $\tau$ and the loss of such cells through division. (\ref{time_evolution_of_nu}) can be written down simply by counting the number of cells whose generation time is $\tau$ that enter and leave the population at any given time. Finally, from (\ref{Number_of_cells_solution}), (\ref{nu_definition}), and (\ref{time_evolution_of_nu}), we obtain the time evolution equation for $f_2(t, \tau)$ that we are looking for: 
\begin{eqnarray}
\frac{d f_2(t,\tau)}{dt} + \sigma\Lambda(t) f_2(t, \tau) = \Lambda(t)\left[2^{\sigma} f_0(t, \tau) - f_1(t, \tau) \right]. \nonumber \\
\label{time_evolution_of_f_2_equation}
\end{eqnarray}
Eq. (\ref{time_evolution_of_f_2_equation}) is easy to solve; we get
\begin{eqnarray}
f_2(t, \tau) &=& e^{-\sigma\int_{t_0}^{t} \Lambda(\eta)d \eta} \Big[f_2(t_0, \tau) + \nonumber \\ &+&\int_{t_0}^{t}\left[2^{\sigma} f_0(\tilde{t}, \tau) - f_1(\tilde{t}, \tau) \right] \Lambda(\tilde{t}) e^{\sigma \int_{t_0}^{\tilde{t}} \Lambda(\eta)d \eta} d \tilde{t} \Big]. \nonumber \\
\label{time_evolution_of_f_2_solution}
\end{eqnarray}
It may seem that only the cells born not earlier than at $t-\tau $ can affect the value of $f_2(t, \tau)$, and therefore $0 \leq t-\tau \leq t_0 \leq t$. However, since (\ref{time_evolution_of_f_2_solution}) is derived from accounting for all cells entering and leaving the population (\ref{time_evolution_of_nu}), the value of $t_0$ is arbitrary.

For $t \geq \tau$ we now have three seemingly different equations relating $f_2(t, \tau)$ to the remaining two generation time distributions ($f_1(t, \tau)$ or $f_0(t, \tau)$): (\ref{f_2_expressed_using_f_1}), (\ref{t_dependent_f_2_derived_from_chi}), and  (\ref{time_evolution_of_f_2_solution}). But with (\ref{relation_between_f_1_and_f_0_time_dependent_final_form}) one can show that for $t \geq \tau$ all these three expressions are equivalent. However, only (\ref{f_2_expressed_using_f_1}) is defined for both $t \geq \tau$ and $t \leq \tau$.

Now let us consider the case of steady exponential growth. First, for $\sigma = 1$ we get from (\ref{time_evolution_of_f_2_equation}) the well known formula \cite{powell1964note, lebowitz1974theory,lin2017effects,thomas2017making}:

\begin{equation}
f_{1r}(\tau) + f_{2r}(\tau) = 2 f_{0r}(\tau).
\label{relationship_between_f_0_f_1_f_2_ss}
\end{equation}
For $\sigma = 0$, Eq. (\ref{time_evolution_of_f_2_equation}) in the steady state limit yields only the condition $f_{1c}(\tau) = f_{0c}(\tau)$ (\ref{relationship_between_f_0_f_1_c_ss}), but not the explicit form of $f_{2c}(\tau)$. We can get the latter by using (\ref{f_2_expressed_using_f_1}) or (\ref{t_dependent_f_2_derived_from_chi}):  
\begin{equation}
\label{f_2_MM_SS_first}
f_{2c}(\tau) = \Lambda_{c} \tau f_{1c}(\tau) = \frac{\tau f_{1c}(\tau)}{\int_{\tau_s}^{\tau_l} \tau^{\prime} f_{1c}(\tau^{\prime}) d \tau^{\prime}}.
\end{equation}
The second equality follows from the normalization of $f_{2c}(\tau)$. Note also that in equation (\ref{f_2_MM_SS_first}) $f_{1c}(\tau)$ can be replaced by $f_{0c}(\tau)$ because of (\ref{relationship_between_f_0_f_1_c_ss}). An alternative way to get (\ref{f_2_MM_SS_first}) was given in ref. \cite{thomas2017making}.

\subsubsection{The second form of the Euler-Lotka equation \label{EL_2nd}}

The time-dependent generalization of the Euler-Lotka equation can also be formulated as a normalization condition for $f_2(t, \tau)$. Integrating both sides of (\ref{f_2_expressed_using_f_1}) with respect to $\tau$, we get
\begin{eqnarray}
\label{time_dependent_Euler_Lotka_f_2}
1 &=& \int_{\tau_s}^{\tau_s} \int_{t}^{t+\tau} \Lambda(t^{\prime}) e^{\sigma \int_{t}^{t^{\prime}} \Lambda(\tilde{t}) d \tilde{t}} f_1(t^{\prime}, \tau) d t^{\prime} d \tau.
\end{eqnarray}
In contrast to (\ref{time_dependent_Euler_Lotka}), now there are no restrictions for $t$, as (\ref{f_2_expressed_using_f_1}) is valid for both $\tau \leq t$ and $\tau \geq t$. Moreover, (\ref{f_2_expressed_using_f_1}) is expressed only by quantities that can be measured experimentally.

The time-independent Euler-Lotka equation for the exponentially growing population in batch culture (\ref{f_1_definition_by_f_0_ss}) usually has many solutions for $\Lambda_r$. In the simplest case, we have a finite number of complex roots, but only one of them - the one with the largest real part and an imaginary part equal to zero - has an interpretation of the population growth rate $\Lambda_r$. Here the solution(s) $\Lambda(t)$ of the time-dependent generalizations of Euler-Lotka equation: (\ref{time_dependent_Euler_Lotka_f_2}) or (\ref{time_dependent_Euler_Lotka}) are not numbers, but functions of the observation time $t$. Our interest here is in this $\Lambda(t)$, which for $\sigma=1$ can be determined from observations using the equation (\ref{N_dot_sigma_Lambda_N}): $\Lambda(t) = \dot{N}(t)/N(t)$, and for both values of $\sigma$ can be determined using the equation (\ref{boundary_condition_again}). However, we are not able to say whether this is the unique solution to equation  (\ref{time_dependent_Euler_Lotka_f_2_SS_sigma_1}), or whether there are multiple solutions. The equation (\ref{time_dependent_Euler_Lotka_f_2}) can be treated not so much as an equation to determine $\Lambda(t)$, but rather as a consistency condition that must be satisfied by both $\Lambda(t)$ and $f_1(t, \tau)$.

For the steady exponential growth in batch culture we get from (\ref{time_dependent_Euler_Lotka_f_2})
\begin{eqnarray}
\label{time_dependent_Euler_Lotka_f_2_SS_sigma_1}
1 &=& \int_{\tau_s}^{\tau_s} \left(e^{\Lambda_r \tau} - 1 \right) f_{1r}(\tau) d \tau,
\end{eqnarray}
which is equivalent to (\ref{Euler_Lotka_ss}) due to (\ref{f_1_definition_by_f_0_ss}).

On the other hand, for the mother machine experiment, it follows from (\ref{f_2_MM_SS_first}) that in the stationary limit $\Lambda_{c}$ is inversely proportional to the average generation time of the mothers (or, equivalently, to the average generation time of the newborns, since $f_{1c}(\tau) = f_{0c}(\tau)$), 
\begin{equation}
\label{f_2_MM_SS_Lambda_c_average_tau}
\Lambda_{c} = \frac{1}{\int_{\tau_s}^{\tau_l} \tau^{\prime} f_{1c}(\tau^{\prime}) d \tau^{\prime}}  = 1/\langle \tau \rangle_{1c}.
\end{equation}
Thus, we can treat (\ref{f_2_MM_SS_Lambda_c_average_tau}) as the Euler-Lotka equation for the mother machine setup at the steady state.

\subsubsection{Age structure expressed in terms of generation times for mothers or newborns \label{phi_and_f1_R}} 

As in the case of the generation time distribution of extant cells $f_2(t, \tau)$, the age distribution $\phi(t, a)$ can also be expressed by generation time distributions for mothers $f_1(t, \tau)$ or newborns $f_0(t, \tau)$.

If $a \geq t$ then using (\ref{Number_of_cells_solution}), (\ref{solution_for_free_propagation_t_a_tau_characteristics}), (\ref{chi_definition_of}), (\ref{chi_0_definition_of}),  (\ref{phi_derived_from_chi_definition}) and (\ref{Omega_definition}), we obtain
\begin{eqnarray}
\label{psi_from_chi_1}
\phi(t, a) &=& \phi_0(a-t)e^{-\sigma\Omega(t)} \nonumber \\ &-& \int_{t-a +\underline{a-t}}^{t-a + \underline{a}}  \Lambda(\tilde{t}) e^{\sigma\Omega(\tilde{t})-\sigma\Omega(t)} f_1(\tilde{t}, \tilde{t} - t + a) d \tilde{t}. \nonumber \\ 
\end{eqnarray}
For $a \leq t$, we have
\begin{eqnarray}
\label{psi_from_chi_2}
\phi(t, a) &=&  2^{\sigma} \Lambda(t - a)e^{\sigma\Omega(t-a)}e^{-\sigma\Omega(t)} \nonumber \\ &-& \int_{t-a + \tau_s}^{t-a + \underline{a}} \Lambda(\tilde{t}) e^{\sigma\Omega(\tilde{t})-\sigma\Omega(t)} f_1(\tilde{t}, \tilde{t} - t + a) d \tilde{t}. \nonumber \\
\end{eqnarray}
The details of the derivations of (\ref{psi_from_chi_1}) and (\ref{psi_from_chi_2}) are given in the Appendix \ref{phi_and_f1_R_App}. The second term on the r.h.s. of both (\ref{psi_from_chi_1}) and (\ref{psi_from_chi_2}) vanishes for $a < \tau_s$, and we get the age distribution of cells that are too young to divide: 
\begin{eqnarray}
\label{tilde_phi_definition_earlier}
\tilde{\phi}(t, a) &=&
\begin{cases}
\phi_0(a-t)e^{-\sigma\Omega(t)}~~~ \text{for}~~~a \geq t, \\ \\
2^\sigma e^{\sigma \Omega(t-a)} e^{-\sigma\Omega(t)} \Lambda(t-a)~~~ \text{for}~~~a \leq t. 
\end{cases}
\end{eqnarray}
With the help of (\ref{phi_derived_from_chi_definition}) and (\ref{chi_solution}) we can also derive the following expression
%
%
\begin{equation}
\label{phi_as_derived_from_chi_definition_f_0}
\phi(t, a) = 2^\sigma \Lambda(t-a) e^{\sigma\Omega(t-a)} e^{-\sigma\Omega(t)} \bar{F}_0(t-a, a), 
\end{equation}
where
\begin{equation}
\bar{F}_0(t, a) \equiv \int_{\underline{a}}^{\tau_l} f_0(t, \tau) d \tau.
\label{definition_c_0_minus}
\end{equation}
If there are no mother-daughter generation time correlations, we have $f_0(t, \tau) = f(\tau)$, $\bar{F}_0(t, a) = \bar{F}(a)$, and (\ref{phi_as_derived_from_chi_definition_f_0}) takes the simple form
\begin{eqnarray}
\label{phi_as_derived_from_chi_definition_f_no_correlations}
\phi(t, a) & = & 2^\sigma \Lambda(t-a) e^{\sigma\Omega(t-a)} e^{-\sigma\Omega(t)} \bar{F}(a). 
\end{eqnarray}
At steady state (\ref{phi_as_derived_from_chi_definition_f_0}) for $\sigma = 1$ equals
\begin{eqnarray}
\phi(a) &=& \Lambda e^{-\Lambda a} \left(2 - \int_{0}^{\underline{a}} f_1(\tilde{a}) e^{\Lambda \tilde{a}} d \tilde{a} \right) \label{age_structure_expressed_via_f_1_f_0_ss_1st_line} \\ &= & 2\Lambda e^{-\Lambda a} \left(1 - \int_{0}^{\underline{a}} f_0(\tilde{a}) d \tilde{a} \right) \label{age_structure_expressed_via_f_1_f_0_ss_2nd_line} \\ &=& 2\Lambda e^{-\Lambda a} \bar{F}_{0}(a). 
\label{age_structure_expressed_via_f_1_f_0_ss}
\end{eqnarray} 
Eq. (\ref{age_structure_expressed_via_f_1_f_0_ss_1st_line}) is Eq. (16) of ref. \cite{quedeville2019critical}, (\ref{age_structure_expressed_via_f_1_f_0_ss_2nd_line}) is Eq. (9) of \cite{powell1956growth}, whereas in (\ref{age_structure_expressed_via_f_1_f_0_ss}), following Powell \cite{powell1956growth}, we have introduced the quantity
\begin{equation}
\label{F_minus_definition}
\bar{F}_{0}(a) \equiv 1 - \int_{0}^{\underline{a}} f_0(\tilde{a}) d \tilde{a} = \int_{\underline{a}}^{\tau_l} f_0(\tilde{a}) d \tilde{a}
\end{equation}
which is the stationary conterpart of (\ref{definition_c_0_minus}). For $\sigma = 0$, at the steady state we get from (\ref{phi_as_derived_from_chi_definition_f_0}) Eq. (D6) of ref. \cite{thomas2017making}, 
\begin{equation}
\phi_c(a) = {\Lambda_c \bar{F}_{0c}(a)} = \frac{\bar{F}_{0c}(a)}{\int_{\tau_s}^{\tau_l} \tau^{\prime} f_{0c}(\tau^{\prime}) d \tau^{\prime}}.
\label{phi_MM}
\end{equation}

\subsection{Time-dependent generalization of inequalities between mean generation time and population doubling time} 

For a population in the steady state of exponential growth, it can be shown  \cite{hashimoto2016noise, garcia2019linking, quedeville2019critical} that
%
%
\begin{equation}
\left\langle \tau \right \rangle_1 \leq \frac{\ln 2}{\Lambda_r} \leq \left\langle \tau \right \rangle_0,
\label{MGT_mothers_daughters_inequality}
\end{equation}
%
%
where $\ln 2/\Lambda_r $ is the population doubling time, $\left\langle \tau \right \rangle_0  \equiv \int_{\tau_s}^{\tau_l} \tau f_{0}(\tau) d \tau$ and similarly for $\left\langle \tau \right \rangle_1$. There is no corresponding inequality for $\sigma = 0$; in this case we have $f_{0c}(\tau) = f_{1c}(\tau)$, see Eq. (\ref{f_2_MM_SS_first}).

The double inequality (\ref{MGT_mothers_daughters_inequality}) can be derived by using the fact that for any two probability distributions $p(t, x)$, $q(t, x)$ we have
\begin{equation}
\mathcal{D}[p(t, x)||q(t, x)] \equiv \int p(t, x) \ln \left[ \frac{p(t, x)}{q(t, x)} \right] dx \geq 0.
\label{KL_divergence_definition}
\end{equation}
The $\mathcal{D}[p(t, x)||q(t, x)]$ appearing in the above formula is a Kullback-Leibler divergence (or relative entropy). In particular, (\ref{MGT_mothers_daughters_inequality}) follows from non-negativity of both $\mathcal{D}[f_{1r}(\tau)||f_{0r}(\tau)]$ and $\mathcal{D}[f_{0r}(\tau)||f_{1r}(\tau)]$ \cite{hashimoto2016noise, garcia2019linking}. 

What is the time-dependent counterpart of (\ref{MGT_mothers_daughters_inequality})? First, consider the case of batch culture ($\sigma=1$). Using
(\ref{KL_divergence_definition}), (\ref{relation_between_f_1_and_f_0_time_dependent_final_form}) and (\ref{Psi_expressed_via_f1_f0}) one can show that the condition $\mathcal{D}[f_{1r}(t, \tau)||f_{0r}(t, \tau)] \geq 0$ implies that 
\begin{equation}
\left\langle  \ln \left[ \frac{\Psi(t, \tau)}{\Psi(t-\tau, \tau)} \right] \right \rangle_{1r} \leq \ln 2, 
\label{KL_D_inequality_t_dep_left_1} 
\end{equation}
where now $\left\langle  (\ldots) \right \rangle_{1r}  \equiv \int_{\tau_s}^{\tau_l} (\ldots) f_{1r}(t, \tau) d \tau$. In a similar way, from the inequality $\mathcal{D}[f_{0r}(t, \tau)||f_{1r}(t, \tau)] \geq 0$ one gets
\begin{equation}
\ln 2 \leq \left\langle  \ln \left[ \frac{\Psi(t, \tau)}{\Psi(t-\tau, \tau)} \right] \right \rangle_{0r},
\label{KL_D_inequality_t_dep_right_0} 
\end{equation}
where $\left\langle (\ldots) \right \rangle_{0r}  \equiv \int_{\tau_s}^{\tau_l} (\ldots) f_{0r}(t, \tau) d \tau$. Combining (\ref{KL_D_inequality_t_dep_left_1}) and (\ref{KL_D_inequality_t_dep_right_0}) we obtain double inequality generalizing (\ref{MGT_mothers_daughters_inequality}).

From (\ref{N_dot_sigma_Lambda_N}) and (\ref{Psi_expressed_via_f1_f0}) we get
\begin{equation}
\Psi(t, \tau) = 2 \frac{d{N}(t)}{dt}  f_0(t, \tau).
\end{equation}
Therefore, the numerators of the expressions under the logarithm in (\ref{KL_D_inequality_t_dep_left_1}) and (\ref{KL_D_inequality_t_dep_right_0}) are proportional to the number of cells born at observation time $t$ and inheriting generation time $\tau$ (since this number is equal to $\Psi(t, \tau)dt d\tau$), while the denominators are proportional to the number of cells born at time $t - \tau$ and inheriting generation time $\tau$. The latter cells divide at time $t$. Thus, in both (\ref{KL_D_inequality_t_dep_left_1}) and (\ref{KL_D_inequality_t_dep_right_0}) we have the average of the logarithm of the ratio of the number of cells born at time $t$ and inheriting the generation time $\tau$ to the number of the mother cells dividing at time $t$ at the age $\tau$. In other words, we average the logarithms of the ratio of the number of cells appearing in the population  at time $t$ with the inherited generation time $\tau$ to the number of such cells disappearing from the population. At steady state, from (\ref{KL_D_inequality_t_dep_left_1}) and (\ref{KL_D_inequality_t_dep_right_0}) we recover (\ref{MGT_mothers_daughters_inequality}).


Both (\ref{KL_D_inequality_t_dep_left_1}) and (\ref{KL_D_inequality_t_dep_right_0}) become equations for $h(\tau | \tau^{\prime}) = \delta(\tau - \tau^{\prime})$ as then $f_0(t, \tau) = f_1(t, \tau)$. In this case, each cell division increases the number of cells with generation time $\tau$ by one: one such cell disappears and two are born.

We can repeat the same reasoning for $\sigma = 0$, that is, for the mother machine experiments. We get 
\begin{eqnarray}
\left\langle  \ln \left[ \frac{\Psi(t, \tau)}{\Psi(t-\tau, \tau)} \right] \right \rangle_{1c} \leq  0 \leq \left\langle  \ln \left[ \frac{\Psi(t, \tau)}{\Psi(t-\tau, \tau)} \right] \right \rangle_{0c} \nonumber \\
\label{KL_D_inequality_t_dep_double_c} 
\end{eqnarray}
Again, $\Psi(t, \tau)dt d\tau$ is the number of cells born between $t$ and $t+dt$ that inherit generation time $\tau$, but now we have $\Psi(t, \tau) = N_0 \Lambda(t) f_{0}(t, \tau)$. The mean values that appear in (\ref{KL_D_inequality_t_dep_double_c}) are defined as $\left\langle  (\ldots) \right \rangle_{0c}  \equiv \int_{\tau_s}^{\tau_l} (\ldots) f_{0c}(t, \tau) d \tau$, $\left\langle  (\ldots) \right \rangle_{1c}  \equiv \int_{\tau_s}^{\tau_l} (\ldots) f_{1c}(t, \tau) d \tau$.

The stationary limit of (\ref{KL_D_inequality_t_dep_double_c}) is trivial: $0 \leq 0 \leq 0$. As with the batch culture, the inequality (\ref{KL_D_inequality_t_dep_double_c}) also becomes equality for transient dynamics if $h(\tau | \tau^{\prime}) = \delta(\tau - \tau^{\prime})$.

\subsection{The time evolution of the moments of $\chi(t, a, \tau)$} 

From the equations (\ref{N_dot_sigma_Lambda_N}), (\ref{APBE_1}), (\ref{APBE_2}), and (\ref{chi_definition_of}), we obtain the time evolution equation for $\chi(t, a, \tau)$,
\begin{eqnarray}
\label{time_evolution_of_chi}
\frac{\partial \chi(t, a, \tau)}{\partial t} + \frac{\partial \chi(t, a, \tau)}{\partial a} + {\sigma} \Lambda(t) \chi(t, a, \tau) = 0, 
\end{eqnarray}
and the boundary condition,  
\begin{equation}
\label{APBE_2_via_chi}
\chi(t, 0, \tau) = 2^{\sigma} \int_{\tau_s}^{\tau_l} h(\tau | \tau^{\prime}) \chi(t, \tau^{\prime}, \tau^{\prime}) d \tau^{\prime}  = 2^{\sigma} \Lambda(t) f_0(t, \tau).
\end{equation}
The initial condition follows from (\ref{APBE_3}) and (\ref{chi_definition_of}). By integrating (\ref{time_evolution_of_chi}) with respect to $a$ one obtains (\ref{time_evolution_of_f_2_equation}). Similarly, by integrating (\ref{time_evolution_of_chi}) with respect to $\tau$ one obtains the time evolution equation for $\phi(t, a)$ as given by (\ref{dynamics_of_phi_I_1}). From (\ref{time_evolution_of_chi}) we can also get the time evolution equations for the moments of $\chi(t, a, \tau)$:
\begin{eqnarray}
\label{moments_chi_ODE}
\frac{d {U}_{km}(t)}{dt} &=& - \sigma \Lambda(t) {U}_{km}(t) +  k {U}_{k-1 m}(t) \nonumber \\ &-&  \Lambda(t) \mathcal{T}_{k+m}(t) +  \delta_{k0} 2^{\sigma} \Lambda(t)  \mathcal{Z}_{m}(t),
\end{eqnarray}
where $\delta_{kl}$ is Kronecker delta and 
\begin{eqnarray}
\label{moments_chi_f0_f1_definitions}
\mathcal{Z}_k(t) &=& \int_{\tau_s}^{\tau_l} \tau^k f_0(t, \tau) d \tau, \nonumber \\ 
\mathcal{T}_k(t) &=& \int_{\tau_s}^{\tau_l} \tau^k f_1(t, \tau) d \tau, \nonumber \\ 
{U}_{km}(t) &=& \int_{\tau_s}^{\tau_l} \int_{0}^{\tau} a^k \tau^m \chi(t, a, \tau) d a d \tau.
\end{eqnarray}
Before solving (\ref{moments_chi_ODE}), we first consider its steady state solution, $d{U}_{km}(t)/dt = 0$. Then (\ref{moments_chi_ODE}) becomes a system of algebraic equations. For $\sigma = 0 $ the solution is simple: 
\begin{eqnarray}
\label{moments_chi_SS_r_solutions}
{U}_{km} &=& \frac{\Lambda_{c}}{k+1}  \mathcal{T}_{k+m+1}, ~~~ k \geq 1.
\end{eqnarray}
The same result can be obtained by using the explicit form of $\chi_c(a, \tau)$ given by (\ref{chi_MM}). For $\sigma = 1$ we get
\begin{eqnarray}
\label{moments_chi_SS_r_algebraic_equations}
{U}_{km} &=& \Lambda_r^{-1}  k {U}_{k-1 m} - \mathcal{T}_{k+m}  + 2   \delta_{k0}  \mathcal{Z}_{m}.
\end{eqnarray}
For $k \geq 1$ the solution of (\ref{moments_chi_SS_r_algebraic_equations}) is given by
\begin{eqnarray}
\label{moments_chi_SS_r_solutions}
{U}_{km} &=& \frac{k!}{\Lambda_r^k}\left({U}_{0m} - \sum_{j=1}^{k} \frac{\Lambda_r^j}{j!}  \mathcal{T}_{j+m} \right),
\end{eqnarray}
where ${U}_{0m} \equiv \mathcal{W}_{m}$ is the steady-state value of the $m$-th moment of $f_2(t, \tau)$: 
\begin{eqnarray}
\label{moments_f2_definitions}
\mathcal{W}_{m}(t) &=&  \int_{\tau_s}^{\tau_l}  \tau^m f_2(t, \tau)d \tau = {U}_{0 m}(t). 
\end{eqnarray}
In particular, for $m=0$ we get Eq. (\ref{moment_kth_phi_f1_ODE_ss}) derived in a different way in the Appendix \ref{McKendrick_model_NA_Appendix}:
\begin{eqnarray}
\label{moment_kth_phi_f1_ODE_ss_main}
\mathcal{A}_k =  \frac{k!}{\Lambda_r^k}\left(1 - \sum_{j=1}^{k} \frac{\Lambda_r^j}{j!}  \mathcal{T}_{j} \right).
\end{eqnarray}
$\mathcal{A}_k =  \mathcal{U}_{k 0}$ is the steady-state value of the $k$-th moment of the cell age distribution $\phi(t, a)$ (\ref{phi_derived_from_chi_definition}), 
%
%
\begin{equation}
\label{moments_phi_f1_definitions_main}
\mathcal{A}_k(t) = \int_{0}^{\tau_l} a^k \phi(t, a) d a  = {U}_{k 0}(t).
\end{equation}
The equation (\ref{moments_chi_SS_r_algebraic_equations}) can also be obtained directly from $\chi_r(a, \tau)$ (\ref{solution_for_chi_sigma_1}), but solving the moment equations seems to be a more convenient way to get it.

Since ${U}_{km} \geq 0$, from (\ref{moments_chi_SS_r_solutions}) one obtains series of inequalities involving moments of $f_2(\tau)$ and those of $f_1(\tau)$. In particular, for $k=1, 2, 3$ we have for any $m$
\begin{eqnarray}
\label{moments_chi_SS_r_inequality_1} 
\mathcal{W}_{m} - \Lambda_{r} \mathcal{T}_{m+1} \geq 0,
\end{eqnarray}
\begin{eqnarray}
\label{moments_chi_SS_r_inequality_2} 
2 \mathcal{W}_{m} - 2 \Lambda_{r} \mathcal{T}_{m+1} - \Lambda^2_{r} \mathcal{T}_{m+2} \geq 0,
\end{eqnarray}
\begin{eqnarray}
\label{moments_chi_SS_r_inequality_3} 
6 \mathcal{W}_{m} - 6 \Lambda_{r} \mathcal{T}_{m+1} -  3 \Lambda^2_{r} \mathcal{T}_{m+2} - \Lambda^3_{r} \mathcal{T}_{m+3} \geq 0.
\end{eqnarray}
Now let us return to the case of unsteady state and to the equation (\ref{moments_chi_ODE}). To solve this system of equations, we first consider its two special cases. For $k=0$, Eq. (\ref{moments_chi_ODE}) reduces to the time evolution equation for the moments of $f_2(t, \tau)$: 
\begin{equation}
\label{moments_f_2_ODE}
\frac{d {W}_{m}(t)}{dt} + \sigma \Lambda(t) {W}_{m}(t) + \Lambda(t) \mathcal{T}_{m}(t) - 2^{\sigma} \Lambda(t) \mathcal{Z}_{m}(t) = 0,
\end{equation}
where ${W}_{m}(t)$ is defined by (\ref{moments_f2_definitions}), while $\mathcal{T}_{m}(t)$ and $\mathcal{Z}_{m}(t)$ are defined by (\ref{moments_chi_f0_f1_definitions}).
Eq. (\ref{moments_f_2_ODE}), which can also be derived from Eq. (\ref{time_evolution_of_f_2_equation}), is easy to solve, one gets
\begin{eqnarray}
\label{moments_f_2_ODE_solution}
\mathcal{W}_{m}(t) &=&    e^{-\sigma \Omega(t)} \Big \{  \mathcal{W}_{m}(0) + \nonumber \\ &+& \int_{0}^{t}\left[2^{\sigma}\mathcal{Z}_{m}(\tilde{t}) - \mathcal{T}_{m}(\tilde{t}) \right] \Lambda(\tilde{t}) e^{\sigma \Omega(\tilde{t})} d \tilde{t}\Big \}. \nonumber \\
\end{eqnarray}
Similarly, to get the time evolution equation for the moments of $\phi(t, a)$, we substitute $m=0$ in (\ref{moments_chi_ODE}). Assuming $k \geq 1$, we get 
\begin{eqnarray}
\label{moments_phi_f1_ODE_main}
\frac{d {\mathcal{A}}_k(t)}{dt} + \sigma \Lambda(t) \mathcal{A}_k(t) - k \mathcal{A}_{k-1}(t) = - \Lambda(t) \mathcal{T}_k(t). \nonumber \\
\end{eqnarray}
where $\mathcal{A}_k(t)$ is defined by (\ref{moments_phi_f1_definitions_main}). Eqs. (\ref{moments_phi_f1_ODE_main}) are solved in the Appendix \ref{McKendrick_model_NA_Appendix} (see Eq. (\ref{moments_phi_f1_solutions})), the solution is:
\begin{eqnarray}
\label{moments_phi_f1_solutions_main}
\mathcal{A}_k(t) &=& e^{-\sigma\Omega(t)} \Bigg \{ \sum_{l=0}^{k}\binom{k}{l} t^{k-l} \mathcal{A}_l(0) \nonumber \\ &-& \sum_{l=0}^{k}\binom{k}{l} \int_{0}^{t} \Lambda(t^{\prime}) e^{\sigma\Omega(t^{\prime})} (t-t^{\prime})^{k-l} \mathcal{T}_l(t^{\prime}) d t^{\prime} \nonumber \\ &+& 2^{\sigma} \int_{0}^{t} \Lambda(t^{\prime}) e^{\sigma\Omega(t^{\prime})} (t-t^{\prime})^k d t^{\prime}\bigg \}. 
\end{eqnarray}
Finally, to find the solution of (\ref{moments_chi_ODE}), we observe that for $k \geq 1$ the equation (\ref{moments_chi_ODE}) has the same form as the equation (\ref{moments_phi_f1_ODE_main}) if we identify ${U}_{km}(t) \longleftrightarrow \mathcal{A}_k(t)$, $\mathcal{T}_{k+m}(t) \longleftrightarrow \mathcal{T}_{k}(t)$. So (\ref{moments_phi_f1_solutions_main}) immediately gives us
\begin{eqnarray}
\label{moments_chi_f1_solutions}
\mathcal{U}_{km}(t) &=& e^{-\sigma\Omega(t)}\Bigg\{ \sum_{l=0}^{k}\binom{k}{l} t^{k-l} \mathcal{U}_{lm}(0) \nonumber \\ &-& \sum_{l=0}^{k}\binom{k}{l} \int_{0}^{t} \Lambda(\theta) e^{\sigma\Omega(\theta)} (t-\theta)^{k-l} \mathcal{T}_{l+m}(\theta) d \theta \nonumber \\ &+& 2^{\sigma} \int_{0}^{t} \Lambda(\theta) e^{\sigma\Omega(\theta)} (t-\theta)^k d \theta \bigg \}. 
\end{eqnarray}
%

\subsection{Application of our formalism: Fitness landscapes for phenotypic traits \label{Fitness_landscapes}}

Nozoe et al. \cite{nozoe2017inferring} proposed a formalism to quantify the fitness of the phenotype $ \mathbf{s} $ within a growing population composed of multiple phenotypes. These authors introduced the concept of  the fitness landscape $ H(t,\mathbf{s}) $, defined as \cite{nozoe2017inferring, genthon2020fluctuation, genthon2021universal}
\begin{equation}
\label{fitness_landscape_definition}
{H}(t,\mathbf{s}) \equiv \overline{\Lambda_{r}(t)} + \frac{1}{t} \ln \left[ \frac{P_{r}(t, \mathbf{s})}{P_{c}(t, \mathbf{s})} \right].
\end{equation}
$ \overline{\Lambda_{r}(t)} $ (denoted as $ \Lambda_{t}$ in ref. \cite{genthon2021universal}) is the time-averaged instantaneous population growth rate $\Lambda_r(t)$:
\begin{equation}
\overline{\Lambda_{r}(t)} = \frac{1}{t} \int_0^{t} \Lambda_r(t^{\prime}) d t^{\prime} = \frac{1}{t} \Omega(t).
\label{Lambda_t_definition}
\end{equation}
In Eq. (\ref{fitness_landscape_definition}), $P_{r}(t, \mathbf{s})$ is the retrospective (backward) probability for the phenotype $\mathbf{s}$, 
\begin{equation}
\label{P_r_FL_definition}
	P_r(t,\mathbf{s}) = \frac{n(t,\mathbf{s})}{N(t)},
\end{equation}
where $ n(t,\mathbf{s}) $ is the number of cells carrying $ \mathbf{s} $ and $ N(t) $ is the total cell number. $P_{r}(t, \mathbf{s})$ is the share of the phenotype $ \mathbf{s} $ at time $ t $ in the batch culture experiment ($\sigma=1$) that has been initiated from $N(0) = N_0$ cells at time $ t=0 $. 
$P_{c}(t, \mathbf{s})$, called the chronological or forward probability, is the probability that we get to the cell carrying the phenotype $ \mathbf{s} $ at time $ t $ if we start a random walk from one of the $ N_0 $ lineage tree roots at time $ t=0 $ (the random walk along the lineage involves some number $ m $ of random choices between two branches):
\begin{equation}
\label{P_c_FL_definition}
	P_c(t,\mathbf{s}) = \frac{1}{N(0)} \sum_{m} \frac{n(t,\mathbf{s};m)}{2^m}.
\end{equation}
Here, $ n(t,\mathbf{s};m) $ denotes the number of cells carrying the phenotype $ \mathbf{s} $, which have divided $ m $ times since $ t=0 $. $P_{c}(t,\mathbf{s})$ corresponds to the share of the phenotype $ \mathbf{s} $ at time $ t $ in the mother machine experiment ($\sigma=0$) that have started at $ t=0$  with $ N_0 $ cells.


We see that, in order to calculate ${H}(t,\mathbf{s})$ (\ref{fitness_landscape_definition}), we have to know both the chronological (forward) and retrospective (backward) time-dependent probability distributions of the phenotypic traits and the instantaneous population growth rate $\Lambda_r(t)$. The fitness landscape is flat in the long-time limit: ${H}(t, \mathbf{s})$ (\ref{fitness_landscape_definition}) approaches a constant, equal to the steady-state population growth rate,
\begin{equation}
\label{fitness_landscape_definition_SS_limit}
\lim_{t \to \infty}{H}(t, \mathbf{s}) = \Lambda_{r} = \lim_{t \to \infty} \Lambda_r(t).
\end{equation}

\subsubsection{Cell age as a phenotypic trait}

The concept of the fitness landscape (in the sense of ref. \cite{nozoe2017inferring}) for  cell age $ a $ as a phenotypic trait may be slightly counter-intuitive, because $ a $ increases linearly in time and does not directly carry the information about the cell's fitness based on its generation time. However, $ a $ is correlated with the generation time: If we find a cell at time $ t $ whose age is $ a $, then we can be sure that the cell's generation time will be larger than $ a $. Therefore, the fitness landscape ${H_{\phi}}(t, a)$ measures how much the current phenotypic state of the cell, being its age $ a $, affects the relative difference in the statistics of finding cells of age $ a $ in the batch culture compared to the statistics of finding cells of age $ a $ in the mother machine experiment.

In ref. \cite{genthon2020fluctuation}, the following formula has been given for ${H}(t, a)$ in the case of vanishing mother-daughter correlations of generation time (Eq. (45) of that reference, rewritten in our notation):
\begin{eqnarray}
\label{fitness_landscape_a_semi_SS}
H_{\phi}(t, a) &=& \Lambda_{r} + \frac{1}{t} \ln \left[\frac{\phi_{r}(a)}{\phi_{c}(a)} \right] \\ \nonumber &=& \frac{1}{t} \left[(t-a)\Lambda_{r} + \ln \left({2 \Lambda_r}{\Lambda}_c^{-1} \right)\right].
\end{eqnarray}
The above expression can only be treated as an approximation of the true fitness landscape, since it includes the time-independent probability distributions and steady-state values of $\Lambda_r(t)$ and $\Lambda_c(t)$: $\Lambda_r $ and $\Lambda_c = \lim_{t \to \infty} \Lambda_c(t)$, i.e., $\Lambda_r =  \int_{0}^{\tau_l} \gamma_r(a) \phi_r(a) d a$, $\Lambda_c = \int_{0}^{\tau_l} \gamma_c(a) \phi_c(a) d a$, see equation (\ref{definition_of_Lambda_generalized_exponential_growth}).
However, if one calculates the fitness landscape using the time-dependent quantities, one obtains
\begin{widetext}
\begin{eqnarray}
\label{fitness_landscape_a}
{H}_{\phi}(t, a) &=& \overline{\Lambda_{r}(t)} + \frac{1}{t} \ln \left[\frac{\phi_{r}(t, a)}{\phi_{c}(t, a)} \right] = \overline{\Lambda_{r}(t)} + \frac{1}{t} \ln \left[ \frac{\tilde{\phi}_{r}(t, a)}{\tilde{\phi}_{c}(t, a)} \right] + \frac{1}{t} \ln \left[ \frac{\bar{F}_{0r} (t-a, a)}{\bar{F}_{0c} (t-a, a)} \right] \nonumber \\ &=& \frac{1}{t} \int_0^{t-a} \Lambda_r(t^{\prime}) d t^{\prime} + \frac{1}{t} \ln \left[ \frac{2 \Lambda_r(t -a)}{\Lambda_c(t-a)} \right] + \frac{1}{t} \ln \left[ \frac{\bar{F}_{0r} (t-a, a)}{\bar{F}_{0c} (t-a, a)} \right], 
\end{eqnarray}
\end{widetext}
where $\bar{F}_{0}(t, a)=\int_{\underline{a}}^{\tau_l} f_0(t, \tau) d \tau$ is defined by (\ref{definition_c_0_minus}) and the age distribution $\tilde{\phi}_{}(t, a)$ of cells which are too young to divide is defined by (\ref{tilde_phi_definition_earlier}).

If there are no mother-daughter generation time correlations (this is the case that should be compared with (\ref{fitness_landscape_a_semi_SS})) then $h(\tau |\tau^{\prime}) = f(\tau) = f_0(t, \tau)$. In consequence, $f_{0r}(t, \tau) = f_{0c}(t, \tau) = f(\tau)$, $\bar{F}_{0r}(t, a) = \bar{F}_{0c}(t, a) = \bar{F}_{}(a)$, and the last term in (\ref{fitness_landscape_a}) vanishes. (Note that $h_r(\tau |\tau^{\prime}) = h_c(\tau |\tau^{\prime})$.) The resulting expression is similar but not identical to (\ref{fitness_landscape_a_semi_SS}). It seems, therefore, that the formula (\ref{fitness_landscape_a_semi_SS}) obtained by Genthon and Lacoste \cite{genthon2020fluctuation} is an approximation to the exact time-dependent formula (\ref{fitness_landscape_a}). 
{There may exist a time scale short enough that $H(t,\mathbf{s})$ is still not to  equal to $\Lambda_r$  (\ref{fitness_landscape_definition_SS_limit}) but long enough for $\ln \left[{\phi_{r}(t, a)}/{\phi_{c}(t, a)} \right]$ to be approximately constant in time. In such a case, Eq. (\ref{fitness_landscape_a}) will be equivalent to Eq. (\ref{fitness_landscape_a_semi_SS}) \cite{genthon2020fluctuation}. }
{(While revising the manuscript in the first round of peer review, we became aware of a new preprint by Genthon and Lacoste \cite{genthon2023cell} in which the authors use a fully time-dependent formula for the fitness landscape, such as postulated in the present paper.)}

\subsubsection{Generation time as a phenotypic trait}

One can also treat the second variable of the Lebowitz-Rubinow model, the generation time $\tau$, as a phenotypic trait. This choice seems more intuitive than the cell age $a$ because $\tau$ is more directly related to the volume growth rate of a single cell and the population growth rate. The distribution of $\tau$ that should be used in (\ref{fitness_landscape_definition}) is $f_2(t, \tau)$ because only this generation time distribution is defined for all cells. In addition, the fitness landscape for the generation time is not uniquely defined if we use $f_0(t, \tau)$ or $f_1(t, \tau)$ instead of $f_2(t, \tau)$, see the Appendix \ref{Forward_vs_Backward} for details.

 Using Eq. (\ref{f_2_expressed_using_f_1}), valid for both $t \leq \tau$ and $t \geq \tau$, we get
%
%
%
\begin{eqnarray}
\label{fitness_landscape_tau}
{H}_{f_2}(t, \tau) &=& \overline{\Lambda_{r}(t)} + \frac{1}{t} \ln \left[\frac{f_{2r}(t, \tau)}{f_{2c}(t, \tau)} \right] \nonumber \\ &=& \frac{1}{t} \ln \left[ \frac{\int_{t}^{t+\tau} e^{\Omega_r(t^{\prime})} \Lambda_r(t^{\prime}) f_{1r}(t^{\prime}, \tau) d t^{\prime}}{\int_{t}^{t+\tau} \Lambda_c(t^{\prime}) f_{1c}(t^{\prime}, \tau) d t^{\prime}} \right]. \nonumber \\
\end{eqnarray}
{The equation (\ref{fitness_landscape_tau}) contains only experimentally observable quantities. However, as in the case of equations (\ref{f_2_expressed_using_f_1}) and (\ref{time_dependent_Euler_Lotka_f_2}), to calculate the value of ${H}_{f_2}(t, \tau)$, we need to know $f_{1r}(t, \tau)$, $f_{1c}(t, \tau)$, $\Lambda_r(t)$ and $\Lambda_c(t)$ over the entire time interval $[t, t+\tau]$. }
 
As in the case of $\mathbf{s}=a$ and ${H}_{\phi}(t, a)$ (\ref{fitness_landscape_a}), one would need to use the results of Appendix \ref{mth_generation_LR_model} to exactly calculate (\ref{fitness_landscape_tau}) for arbitrary $t$ (i.e., to express $H_{f_2}(t, \tau)$ using only $h(\tau |\tau^{\prime})$ and the initial condition $\Phi(a, \tau)$ (\ref{APBE_3})). Such a calculation requires knowledge of the functional forms of $h(\tau |\tau^{\prime})$ and $\Phi(a, \tau)$ (both of which are unobservable) and evaluation of the integrals in the series solution of Section \ref{mth_generation_LR_model} (which is probably not analytically feasible, but could perhaps be done numerically for the specific systems).

\subsubsection{Cell age and generation time as a two dimensional phenotypic trait}

Besides one-dimensional phenotypic traits: $\mathbf{s}=a$ and $\mathbf{s}=\tau$, one can also consider two-dimensional trait: $\mathbf{s}= (a, \tau)$. Once again, our task is to express the fitness landscape in terms of observable quantities only. Using the equations (\ref{chi_solution}), (\ref{Omega_definition}), (\ref{fitness_landscape_definition}) and (\ref{Lambda_t_definition}) we get

\begin{eqnarray}
\label{fitness_landscape_a_tau}
{H}_{\chi}(t, a, \tau) &=& \overline{\Lambda_{r}(t)} + \frac{1}{t} \ln \left[\frac{\chi_r (t, a, \tau)}{\chi_c (t, a, \tau)} \right] \nonumber \\ &=& \frac{1}{t} \ln \left[ \frac{\Lambda_r(t-a+\tau)}{\Lambda_c(t-a+\tau)} \right] +  \frac{\Omega(t-a+\tau)}{t}    \nonumber \\ &+& \frac{1}{t}\ln \left[ \frac{f_{1r}(t-a+\tau, \tau)}{f_{1c}(t-a+\tau, \tau)} \right].
\end{eqnarray}

\subsection{Generalization of the Lebowitz-Rubinow model \label{GLRM}} 

Population dynamics are often described by the population balance models  \cite{bell1967cell, bell1968cell, thomas2017making, quedeville2019critical, garcia2019linking, genthon2020fluctuation}. In this approach, each cell is characterized by its age and possibly some additional variables. These could be, for example, the cell volume (mass, size), the volume growth rate, or the copy number (or concentration) of protein molecules of a particular type.

Population balance models are based on first-order partial differential equations describing the deterministic time evolution of the cell number density, supplemented by appropriate boundary and initial conditions. However, various probability distributions can be constructed from the cell number density. 

The word `balance', which refers to `accounting' for the number of cells of a given age, volume, etc., may be misleading here, since such models can just as well describe the system that is not in steady state. Other terms such as `structured population models' or `continuous rate models' are sometimes used, each referring to a different aspect of such a theoretical framework.

The simplest population balance model is the McKendrick-von Foerster model (\ref{NAPBE_1})-(\ref{NAPBE_3}), proposed almost a century ago by McKendrick \cite{MKendrick1925} and later independently by von Foerster \cite{von1959kinetics}, see also \cite{trucco1965mathematical1, trucco1965mathematical2, rudnicki2014modele}. However, from our point of view, the McKendrick-von Foerster model has a serious limitation: It does not contain explicit information about the generation time $\tau$ and its inheritance. Therefore, one cannot use this model to determine the dynamics of all generation time distributions of interest: $f_0(t, \tau)$ for cells whose age $a$ is zero (newborns), $f_1(t, \tau)$ for just dividing cells for which $a=\tau$ (mothers), and $f_2(t, \tau)$ for all cells present in the population at a given time (extant cells). 

The same remarks apply to any generalization of the McKendrick-von Foerster model that has a form of the population balance equation in which we have some variables in addition to cell age, but in which generation time $\tau$ does not explicitly appear as an independent variable. Such models can be found e.g. in \cite{bell1967cell, bell1968cell, garcia2019linking, quedeville2019critical, genthon2020fluctuation} and will be referred to here as ``generalized McKendrick-von Foerster models''. 

On the other hand, models that include $\tau$ as an independent variable belong to the same class as the Lebowitz-Rubinow model, and we will simply call them ``generalized Lebowitz-Rubinow models''. In the following, we will show how to define the most important probability distributions that appear in the generalized Lebowitz-Rubinow models. We will also show that each of these models reduces to the original Lebowitz-Rubinow model when the variables other than $a$ and $\tau$  are integrated out, so that all the results obtained in this paper remain valid within a rather broad class of population balance models. But before we move on to the generalizations of the Lebowitz-Rubinow model, let's stop for a moment at the generalized McKendrick-von Foerster models.

\subsubsection{Generalized McKendrick-von Foerster model \label{RPBE}}

The McKendrick-von Foerster model (\ref{NAPBE_1})--(\ref{NAPBE_3}) can be extended to the more complex population balance model, where each cell is characterized not only by its age but also by other state variables \cite{quedeville2019critical}:
\begin{equation}
(\xi_1, \xi_2, \ldots, \xi_d) \equiv \vec{\xi} \in \Xi \subset \mathbb{R}^d.
\end{equation}
Some of the $\xi_i$-s may have non-trivial dynamics, the most obvious example being cell volume $V$ or cell length. However, some variables may remain constant during the cell cycle, e.g. the cell volume growth rate $\lambda$. Nevertheless, the time evolution of $\vec{\xi}$ is assumed to be deterministic:
\begin{equation}
\label{RPBE_xi_time_evolution}
\dot{\vec{\xi}} = \vec{g}\left(t, a, \vec{\xi}\right), 
\end{equation}
or $\dot{\xi}_i = g_i(t, a, \xi_1, \xi_2, \ldots, \xi_d)$, $i = 1, 2, \ldots, d$. The dot denotes the derivative with respect to the observation time $t$ or the cell age $a$, depending on the sign of $a-t$. For example, the cell volume $V$ is often assumed to grow exponentially:
\begin{equation}
\dot{V} = \lambda V, 
\end{equation}
where $\lambda$ is a constant, $\dot{\lambda}=0$. (More generally, $g_i(t, a, \vec{\xi})=0$ for all non-dynamic variables). We also exclude state variables such as generation time $\tau$ or cell volume of dividing cells $V_d$, which are the values of the dynamic variables at cell division. The models in which such variables are present belong to the same class as the Lebowitz-Rubinow model and will be discussed in the next subsection.

The basic quantity is now the cell number density $n(t, a, \vec{\xi})$, which obeys the following time-evolution equation: \cite{quedeville2019critical}
%
\begin{eqnarray}
\label{RPBE_PBE_time_evolution}
\big[\partial_{t} ~+ ~\partial_{a} &+& \gamma(t, a, \vec{\xi})~ + ~D(t)\big] n(t, a, \vec{\xi})  \nonumber \\ & = & - \nabla_{\vec{\xi}} \left[\vec{g}(t, a, \vec{\xi})n(t, a, \vec{\xi}) \right],
\end{eqnarray}
where $\nabla_{\vec{\xi}} = (\partial_{\xi_1}, \partial_{\xi_2}, \dots, \partial_{\xi_d})$,  and  $\partial_{x} = \partial/\partial x$. The equation (\ref{RPBE_PBE_time_evolution}) must be supplemented with the boundary condition
\begin{eqnarray}
\label{RPBE_PBE_boundary_condition_1}
n(t, 0, \vec{\xi}) &=& 2^\sigma \int_{0}^{\tau_l} \int_{ \Xi} \mathcal{K}(\vec{\xi} | a, \vec{\zeta}) \gamma(t, a, \vec{\zeta}) n(t, a, \vec{\zeta}) d \vec{\zeta} da, \nonumber \\
\end{eqnarray}
and 
\begin{eqnarray}
\label{RPBE_PBE_boundary_condition_2}
\vec{0} &=& \vec{g}(t, a, \vec{\xi})n(t, a, \vec{\xi})\Big |_{\vec{\xi} \in \partial \Xi}
\end{eqnarray}
($\partial \Xi$ denotes the boundary of $ \Xi \subset \mathbb{R}^d$ \cite{quedeville2019critical}) as well as with the initial condition 
\begin{eqnarray}
\label{RPBE_PBE_intial_condition}
n_0(a, \vec{\xi}) &=& n(0, a, \vec{\xi}).
\end{eqnarray}
The kernel $\mathcal{K}(\vec{\xi} |a, \vec{\zeta})$ describing the inheritance of $\vec{\xi}$ is the probability distribution of $\vec{\xi}$ parameterized by $a$ and $ \vec{\zeta}$. And therefore, for arbitrary $t, a$ and $\vec{\zeta}$, we have
\begin{eqnarray}
\label{RPBE_normalization_of_K}
1 &=& \int_{ \Xi} \mathcal{K}(\vec{\xi} |a, \vec{\zeta}) d \vec{\xi}.
\end{eqnarray}
The cell number density of the general population balance model (\ref{RPBE_PBE_time_evolution})--(\ref{RPBE_PBE_intial_condition}) normalized by the total cell number, 
\begin{equation}
N(t) = \int_0^{\tau_l} \int_{ \Xi} n(t, a, \vec{\xi}) d \vec{\xi} d a,
\end{equation}
has a natural interpretation of the probability density: 
\begin{eqnarray}
\label{RPBE_definition_of_p_t_a}
\phi(t, a, \vec{\xi}) & \equiv & \frac{n(t, a, \vec{\xi})}{\int_0^{\tau_l} \int_{ \Xi} n(t, a, \vec{\xi}) d \vec{\xi} d a}.
\end{eqnarray}
The $\phi(t, a, \vec{\xi})$ defined above is a generalization of the cell age distribution $\phi(t, a)$ of the McKendrick-von Foerster model. However, following \cite{quedeville2019critical} we can also define another probability distribution,
\begin{eqnarray}
\label{RPBE_definition_of_f_1}
f_1(t, a, \vec{\xi}) = \frac{\gamma(t, a, \vec{\xi}) n(t, a, \vec{\xi})}{\int_{0}^{\tau_l}\int_{\Xi} \gamma(t, a, \vec{\xi}) n(t, a, \vec{\xi}) d \vec{\xi} d a},
\end{eqnarray}
which generalizes the mother age distribution $f_1(t, a)$ given by (\ref{carrier_PDF_time_dependent_NA}):
\begin{equation}
f_1(t, a) = \int_{ \Xi} f_1(t, a, \vec{\xi}) d \vec{\xi}.
\end{equation}
The denominator of Eq. (\ref{RPBE_definition_of_f_1}) is equal to $\Lambda(t) N(t)$, thus we can replace $\gamma(t, a, \vec{\xi}) n(t, a, \vec{\xi})$ with $ N(t)\Lambda(t) f_1(t, a, \vec{\xi})$ in (\ref{RPBE_PBE_time_evolution}) and (\ref{RPBE_PBE_boundary_condition_1}).

We are now ready to show that the general population balance model defined by (\ref{RPBE_PBE_time_evolution})-(\ref{RPBE_PBE_intial_condition}) can be reduced to the McKendrick-von Foerster model. Let 
\begin{eqnarray}
\label{RPBE_reduced_n_t_a}
n(t, a) &=& \int_{ \Xi} n(t, a, \vec{\xi}) d \vec{\xi},
\end{eqnarray}
where $n(t, a, \vec{\xi})$ is the solution of (\ref{RPBE_PBE_time_evolution}) satisfying  (\ref{RPBE_PBE_boundary_condition_1}), (\ref{RPBE_PBE_boundary_condition_2}), and (\ref{RPBE_PBE_intial_condition}). The term $-\nabla_{\vec{\xi}} \big[\vec{g}(t, a, \vec{\xi}) n(t, a, \vec{\xi}) \big]$,  appearing on the r.h.s. of (\ref{RPBE_PBE_time_evolution}) integrated with respect to $\vec{\xi}$ vanishes due to the Ostrogradsky-Gauss theorem and due to (\ref{RPBE_PBE_boundary_condition_2}). We obtain 
%
%
\begin{eqnarray}
\label{RPBE_PBE_time_evolution_reduced_n}
\big[\partial_{t} + \partial_{a} + \gamma_{e}(t, a) + D(t)\big] n(t, a) = 0,
\end{eqnarray}
where we define the effective rate of division to be
\begin{eqnarray}
\label{RPBE_definition_of_gamma_eff}
\gamma_{e}(t, a) = \frac{\int_{ \Xi} \gamma(t, a, \vec{\xi}) n(t, a, \vec{\xi}) d \vec{\xi}}{\int_{ \Xi} n(t, a, \vec{\xi}) d \vec{\xi}}.
\end{eqnarray}
Note that $\gamma_{e}(t, a)$ may depend on $\sigma$. That is, it may be different for the mother machine and batch culture scenarios.

Next, integrating (\ref{RPBE_PBE_boundary_condition_1}) with respect to $\vec{\xi}$ and using (\ref{RPBE_normalization_of_K}), (\ref{RPBE_reduced_n_t_a}), and (\ref{RPBE_definition_of_gamma_eff}), we get 
\begin{eqnarray}
\label{RPBE_PBE_boundary_condition_1_reduced_n}
n(t, 0) = 2^{\sigma} \int_{0}^{\tau_l} \gamma_e(t, a) n(t, a) d a.
\end{eqnarray}
Eq. (\ref{RPBE_PBE_time_evolution_reduced_n}) is identical to (\ref{NAPBE_1}), whereas (\ref{RPBE_PBE_boundary_condition_1_reduced_n}) is identical to (\ref{NAPBE_2}), provided we identify $\gamma_e(t, a)$ with $\gamma(t, a)$. The initial condition (\ref{RPBE_PBE_intial_condition}) reduces to (\ref{NAPBE_3}). In this way, we obtain the McKendrick-von Foerster model (\ref{NAPBE_1})--(\ref{NAPBE_3}) from the general population balance model given by (\ref{RPBE_PBE_time_evolution})--(\ref{RPBE_PBE_intial_condition}). Obviously, many models may yield the same effective McKendrick-von Foerster model.

\subsubsection{From the generalized McKendrick - von Foerster model to the generalized Lebowitz - Rubinow model \label{GLRM_proper}}

Any generalized McKendrick-von Foerster model as defined by (\ref{RPBE_PBE_time_evolution})--(\ref{RPBE_PBE_intial_condition}) can be extended in the way Lebowitz and Rubinow extended the original McKendrick-von Foerster model by adding generation time $\tau$ as an independent, non-dynamic variable. By integrating out $\tau$, such an extended model can be reduced to the original population balance equation. It can also be reduced to the original Lebowitz-Rubinow model by integrating the remaining variables, and to the McKendrick-von Foerster model by integrating all variables except cell age. This is discussed below.

As in the case of the general population balance model analyzed above, we assume that the time evolution of $\vec{\xi}$ is deterministic and can be described by a system of ordinary differential equations:
\begin{equation}
\label{GLRM_xi_time_evolution}
\dot{\vec{\xi}} = \vec{g}\left(t, a, \tau, \vec{\xi}\right). 
\end{equation}
The presence of $\tau$ in the above equation may be caused by the dependence of the dynamics of some quantities (e.g. cell volume) on the fraction of the cell cycle, $a/\tau$. 

Instead of (\ref{RPBE_PBE_time_evolution}), (\ref{RPBE_PBE_boundary_condition_1})  and (\ref{RPBE_PBE_boundary_condition_2}) we now have 
\begin{eqnarray}
\label{GLRM_PBE_time_evolution}
\partial_{t} n(t, a, \tau,  \vec{\xi}) &+& \partial_{a} n(t, a, \tau,  \vec{\xi})  \\ &+& \nabla_{\vec{\xi}} \big[\vec{g}(t, a, \tau, \vec{\xi})  n(t, a, \tau,  \vec{\xi})\big] = 0 \nonumber
\end{eqnarray}
%
%
\begin{eqnarray}
\label{GLRM_PBE_boundary_condition_1}
n(t, 0, \tau, \vec{\xi}) &=& 2^\sigma \int_{0}^{\tau_l} \int_{ \Xi} \mathcal{K}(\tau, \vec{\xi} | a, \vec{\zeta}) n(t, a, a, \vec{\zeta}) d \vec{\zeta} da, \nonumber \\
\end{eqnarray}
\begin{eqnarray}
\label{GLRM_PBE_boundary_condition_2}
\vec{0} &=& \vec{g}(t, a, \tau, \vec{\xi})n(t, a, \tau, \vec{\xi})\Big |_{\vec{\xi} \in \partial \Xi}
\end{eqnarray}
(we put $D(t)=0$ in (\ref{RPBE_PBE_time_evolution})), while the initial condition is
\begin{eqnarray}
\label{GLRM_PBE_intial_condition}
\Phi(a, \tau, \vec{\xi}) &=& n(0, a, \tau, \vec{\xi}).
\end{eqnarray}
Note that we have
\begin{eqnarray}
\label{GLRM_PBE_boundary_condition_1_integrated_K}
1 &=&  \int_{0}^{\tau_l} \int_{ \Xi} \mathcal{K}(\tau, \vec{\xi} | a, \vec{\zeta}) d \vec{\xi} d\tau.
\end{eqnarray}

\paragraph{Reduction of the generalized Lebowitz - Rubinow model to the generalized McKendrick-von Foerster model.}

Following essentially the same line of reasoning as in subsection \ref{Reduction_of_LR_to_McKvF}, in this paragraph we show how the generalized Lebowitz-Rubinow model can be reduced to the generalized McKendrick-von Foerster model.  

First, the cell number densities of these two models are related by an expression analogous to Eq. (\ref{n_generalized_reduced_to_n_standard}):
\begin{equation}
\label{n_generalized_reduced_to_n_standard_GLRM}
n(t, a, \vec{\xi}) = \int_{\underline{a}}^{\tau_l} n(t, a, \tau, \vec{\xi}) d\tau,
\end{equation}
Next, we integrate (\ref{GLRM_PBE_time_evolution}) with respect to $\tau$. We get
\begin{eqnarray}
\label{GLRM_PBE_time_evolution_integrated_wrt_tau}
{\partial_t}n(t, a, \vec{\xi})  &+& {\partial_a} n(t, a, \vec{\xi}) ~ + ~ n(t, a, a, \vec{\xi}) \nonumber \\  &+ & \nabla_{\vec{\xi}} \left[\vec{g}_e(t, a,  \vec{\xi}) n(t, a, \vec{\xi})\right]   = 0, 
\end{eqnarray}
with $n(t, a, \vec{\xi})$ defined by (\ref{n_generalized_reduced_to_n_standard_GLRM}) and $\vec{g}_e(t, a, \vec{\xi})$ by
\begin{equation}
\label{g_eff_GMcK_vF_GLRM}
\vec{g}_e(t, a,  \vec{\xi})   \equiv  \frac{\int_{\underline{a}}^{\tau_l} \vec{g}(t, a, \tau, \vec{\xi}) n(t, a, \tau, \vec{\xi}) d\tau}{\int_{\underline{a}}^{\tau_l} n(t, a, \tau, \vec{\xi}) d\tau}.
\end{equation}
If $\vec{g}(t, a, \tau, \vec{\xi})$ does not depend on $\tau$, then $\vec{g}_e = \vec{g}$. Note that $n(t, a, a, \vec{\xi})=0$ for $a < \tau_s$. Next, we integrate the boundary conditions (\ref{GLRM_PBE_boundary_condition_1}) and (\ref{GLRM_PBE_boundary_condition_2}) and obtain
\begin{eqnarray}
\label{GLRM_PBE_boundary_condition_1_integrated_wrt_tau}
n(t, 0, \vec{\xi}) &=& 2^\sigma \int_{0}^{\tau_l} \int_{ \Xi} \mathcal{K}_e(\vec{\xi} | a, \vec{\zeta}) n(t, a, a, \vec{\zeta}) d \vec{\zeta} da \nonumber \\
\end{eqnarray}
and
%
\begin{eqnarray}
\label{GLRM_PBE_boundary_condition_2_integrated_wrt_tau}
\vec{0} &=& \vec{g}_e(t, a, \vec{\xi})n(t, a, \vec{\xi})\Big |_{\vec{\xi} \in \partial \Xi},
\end{eqnarray}
where $\vec{g}_e(t, a, \vec{\xi})$ is defined by (\ref{g_eff_GMcK_vF_GLRM}), and where 
%
%
\begin{equation}
\label{K_eff_GMcK_vF_GLRM}
\mathcal{K}_e(\vec{\xi} | a, \vec{\zeta}) = \int_{\underline{a}}^{\tau_l} \mathcal{K}(\tau, \vec{\xi} | a, \vec{\xi})  d\tau.
\end{equation}
If the equations (\ref{GLRM_PBE_time_evolution_integrated_wrt_tau}), (\ref{GLRM_PBE_boundary_condition_1_integrated_wrt_tau}), and (\ref{GLRM_PBE_boundary_condition_2_integrated_wrt_tau}) are to be identical to the corresponding equations: (\ref{RPBE_PBE_time_evolution}), (\ref{RPBE_PBE_boundary_condition_1}), and (\ref{RPBE_PBE_boundary_condition_2}), then the following conditions must be satisfied:
\begin{eqnarray}
\label{GLRM_PBE_red_cons_cond_1}
n(t, a, a, \vec{\xi}) = \gamma(t, a, \vec{\xi}) n(t, a, \vec{\xi}),
\end{eqnarray}
\begin{eqnarray}
\label{GLRM_PBE_red_cons_cond_2}
\vec{g}_e(t, a, \vec{\xi}) = \vec{g}(t, a, \vec{\xi}),
\end{eqnarray}
\begin{eqnarray}
\label{GLRM_PBE_red_cons_cond_3}
\mathcal{K}_e(\vec{\xi} | a, \vec{\zeta}) = \mathcal{K}(\vec{\xi} | a, \vec{\zeta}).
\end{eqnarray}
In the above, $\vec{g}_e(t, a, \vec{\xi})$ is given by (\ref{g_eff_GMcK_vF_GLRM}) and $\mathcal{K}_e(\vec{\xi} | a, \vec{\zeta})$ by (\ref{K_eff_GMcK_vF_GLRM}). The first of these three conditions is an obvious generalization of equation (\ref{carrier_NA_vs_anticipating_correspondence}).

Finally, by integrating the boundary condition (\ref{GLRM_PBE_intial_condition}) with respect to $\tau$ we obtain (\ref{RPBE_PBE_intial_condition}). Thus, we have shown that the generalized Lebowitz - Rubinow model as defined by (\ref{RPBE_PBE_time_evolution})-(\ref{GLRM_PBE_intial_condition}) can be reduced to an effective model of the same form as the population balance model defined by the equations (\ref{RPBE_PBE_time_evolution})-(\ref{RPBE_PBE_intial_condition}).

\paragraph{Definitions of generalized cell age and generation time distributions in the generalized Lebowitz - Rubinow model.}

Analogous to the case of the original Lebowitz-Rubinow model, we define
\begin{equation}
\chi(t, a, \tau, \vec{\xi}) \equiv \frac{n(t, a, \tau, \vec{\xi})}{N(t)},
\label{chi_definition_of_GLRM}
\end{equation}
where 
\begin{eqnarray}
\label{Number_of_cells_definition_GLRM_model}
N(t) &=& \int_{0}^{\tau_l} \int_{\underline{a}}^{\tau_l} \int_{\Xi} n(t, a, \tau, \vec{\xi})  d \vec{\xi} d \tau d a\nonumber \\ &=& \int_{\tau_s}^{\tau_l} \int_{0}^{\tau} \int_{\Xi} n(t, a, \tau, \vec{\xi})  d \vec{\xi} d a d \tau
\end{eqnarray}
is the total number of cells in the population. In direct analogy to what was done in the subsection \ref{Definitions_of_f_i_s}, we can now define natural generalizations of all three generation time probability distributions: $f_{0}(t, \tau)$, $f_{1}(t, \tau)$, and $f_{2}(t, \tau)$, as well as the cell age distribution $\phi(t, a)$. These can be obtained from $\chi(t, a, \tau, \vec{\xi})$ (\ref{chi_definition_of_GLRM}) as either  conditional or marginal probabilities. 

First, we define the joint distribution of cell age $a$ and $\vec{\xi}$ of all cells in the population, 
\begin{equation}
\label{phi_derived_from_chi_definition_GLRM}
\phi(t, a, \vec{\xi}) \equiv \int_{\underline{a}}^{\tau_l} \chi(t, a, \tau, \vec{\xi}) d \tau = \frac{n(t, a, \vec{\xi})}{N(t)}.
\end{equation}
We also define the joint distribution of $\tau$ and $\vec{\xi}$ for extant cells, 
\begin{equation}
f_2(t, \tau, \vec{\xi}) \equiv \int_{0}^{\tau} \chi(t, a, \tau, \vec{\xi}) d a,
\label{t_dependent_f_2_definition_GLRM}
\end{equation}
and analogous distribution for mother cells 
\begin{equation}
f_1(t, \tau, \vec{\xi}) \equiv \frac{\chi(t, \tau, \tau, \vec{\xi})}{\int_{\Xi} \int^{\tau_l}_{\tau_s}  \chi(t, \tau, \tau, \vec{\xi}) d\tau d \vec{\xi}} = \frac{\chi(t, \tau, \tau, \vec{\xi})}{\Lambda(t)}.
\label{t_dependent_f_1_definition_GLRM}
\end{equation}
Similar to the case of $\phi(t, a, \vec{\xi})$, such a defined $f_1(t, \tau, \vec{\xi})$ agrees with (\ref{RPBE_definition_of_f_1}). Finally, the joint distribution of $\tau$ and $\vec{\xi}$ for daughters is
\begin{equation}
f_0(t, \tau, \vec{\xi}) \equiv \chi(t, \tau, \vec{\xi} | 0) \equiv \frac{\chi(t, 0, \tau, \vec{\xi} )}{\phi(t, 0)} = \frac{\chi(t, 0, \tau, \vec{\xi})}{2^{\sigma} \Lambda(t)},
\label{f_0_definition_GLRM}
\end{equation}
where is the age distribution defined by (\ref{phi_NA_definition}) or (\ref{phi_derived_from_chi_definition}) and $\chi(t, \tau, \vec{\xi} | a) \equiv {\chi(t, a, \tau, \vec{\xi})}/{\phi(t, a)}$. Note that we have
\begin{equation}
f_0(t, \tau, \vec{\xi}) = \int_{0}^{\tau_l} \int_{ \Xi} \mathcal{K}(\tau, \vec{\xi} | \tilde{\tau}, \vec{\zeta}) f_1(t,  \tilde{\tau}, \vec{\zeta}) d \vec{\zeta} d \tilde{\tau}.
\label{f_0_definition_GLRM_2}
\end{equation}
When additional variables $\vec{\xi}$ are integrated out, $\chi(t, a, \tau, \vec{\xi})$, $\phi(t, a, \vec{\xi})$, and $f_i(t, \tau, \vec{\xi})$, $i=0, 1, 2$ reduce to the corresponding distributions of the original Lebowitz-Rubinow model.

\paragraph{Reduction of the generalized Lebowitz-Rubinow model to the original Lebowitz-Rubinow model.}

In this paragraph we show that if $n(t, a, \tau, \vec{\xi})$ satisfies the equations of the generalized Lebowitz-Rubinow model (\ref{RPBE_PBE_time_evolution})-(\ref{GLRM_PBE_intial_condition}), then the reduced cell number density 
\begin{eqnarray}
\label{number_density_of_cells_red_GLRM_to_LRM}
n(t, a, \tau) &=&  \int_{\Xi} n(t, a, \tau, \vec{\xi})  d \vec{\xi}
\end{eqnarray}
is a solution of the equations (\ref{APBE_1})-(\ref{APBE_3}) of the original Lebowitz-Rubinow model with an appropriately chosen probability distribution of the inherited generation times, $h(t, \tau | \tau^{\prime})$.

To show this, we first integrate (\ref{RPBE_PBE_time_evolution}) with respect to $\vec{\xi}$. The term $\nabla_{\vec{\xi}}[\vec{g}(t, a, \tau, \vec{\xi}) n(t, a, \tau, \vec{\xi})]$ vanishes due to the boundary condition (\ref{GLRM_PBE_boundary_condition_2}), the derivatives of $n(t, a, \tau, \vec{\xi})$ with respect to $t$ and $a$ reduce to the corresponding terms in the Lebowitz-Rubinow equations and we indeed obtain the time evolution equation (\ref{APBE_1}). The initial condition (\ref{GLRM_PBE_intial_condition}) reduces to (\ref{APBE_3}).
The only non-trivial part is the reduction of the boundary condition (\ref{GLRM_PBE_boundary_condition_1}) to (\ref{APBE_2}). We have
\begin{eqnarray}
\label{GLRM_PBE_boundary_condition_1_red_GLRM_to_LRM}
n(t, 0, \tau) &=& \int_{\Xi} n(t, 0, \tau, \vec{\xi})  d \vec{\xi}   \nonumber \\ &=& 2^\sigma   \int_{0}^{\tau_l} \int_{\Xi} \int_{ \Xi} \mathcal{K}(\tau, \vec{\xi} | a, \vec{\zeta}) n(t, a, a, \vec{\zeta})  d \vec{\xi}  d \vec{\zeta}  da \nonumber \\  &=& 2^\sigma   \int_{0}^{\tau_l} \int_{\Xi}  \tilde{\mathcal{K}}(\tau | a, \vec{\zeta}) n(t, a, a, \vec{\zeta})  d \vec{\zeta}  da \nonumber \\  &=& 2^\sigma   \int_{0}^{\tau_l} h_e(t, \tau | a) n(t, a, a)   da, \nonumber \\
\end{eqnarray}
where 
\begin{eqnarray}
\label{h_e_def_red_GLRM_to_LRM}
 h_e(t, \tau | a)   &=& \int_{\Xi} \frac{ \tilde{\mathcal{K}}(\tau | a, \vec{\zeta}) n(t, a, a, \vec{\zeta})  d \vec{\zeta}}{\int_{\Xi} n(t, a, a, \vec{\zeta})  d \vec{\zeta}} \nonumber \\  &=& \int_{\Xi}\tilde{\mathcal{K}}(\tau | a, \vec{\zeta}) \mathcal{P}(t, \vec{\zeta} | a, a)  d \vec{\zeta}. 
\end{eqnarray} 
and where we have defined: 
\begin{eqnarray}
\label{Chi_xi_a_tau_def_red_GLRM_to_LRM}
\mathcal{P}(t, \vec{\zeta} | a, \tau) &\equiv & \frac{n(t, a, \tau, \vec{\zeta})}{ n(t, a, \tau)} = \frac{\chi(t, a, \tau, \vec{\zeta})}{ \chi(t, a, \tau)}.
\end{eqnarray}
Since $n(t, a, \tau)$ defined by (\ref{number_density_of_cells_red_GLRM_to_LRM}) satisfies the equations of the Lebowitz-Rubinow model, all results obtained within this model (in particular, the relationships between different probability distributions) remain valid. Note, however, that the effective $h_e(t, \tau | t, a)$ defined by (\ref{h_e_def_red_GLRM_to_LRM}) may explicitly depend on the observation time $t$. It may also be different for the mother machine experiment and for the batch culture, i.e., for different values of the $\sigma$ parameter. This is because $h_e(t, \tau | a)$ is determined by $n(t, a, \tau, \vec{\xi})$ or $\mathcal{P}(t, \vec{\zeta} | a, \tau)$, and these quantities are usually different for $\sigma = 0$ and $\sigma = 1$. In the full model (i.e., the generalized Lebowitz-Rubinow model given by (\ref{RPBE_PBE_time_evolution})-(\ref{GLRM_PBE_intial_condition})), the value of the generation time inherited by daughters usually depends not only on the generation time of the mother, but also on other variables characterizing the mother cell.

Therefore, the effective Lebowitz-Rubinow model with $h_e(t, \tau | a)$ given by (\ref{h_e_def_red_GLRM_to_LRM}) should not be used to obtain the relation (\ref{n_r_to_n_c_ratio}) between $n_r(t, a, \tau; m)$ and $n_c(t, a, \tau; m)$.  On the technical side, this is because if we try to repeat the step with the introduction of the effective $h_e(t, \tau | a)$ in (\ref{APBE_2_m}), we will in general get different $h_e(t, \tau |  a)$ for different values of the parameter $m$, i.e. for different generations. However, the relation between $n_r(t, a, \tau; m)$ and $n_c(t, a, \tau; m)$ can be obtained within a more complete description provided by the generalized Lebowitz-Rubinow model. Within the latter, we have 
\begin{equation}
\label{n_r_to_n_c_ratio_GLRM}
n_r(t, a, \tau, \vec{\xi}; m) = 2^m n_c(t, a, \tau, \vec{\xi}; m), 
\end{equation}
see discussion in subsection \ref{Forward_vs_Backward}. By integrating both sides of (\ref{n_r_to_n_c_ratio_GLRM}) with we get a desired result, i.e. Eq. (\ref{n_r_to_n_c_ratio}). 

\section{Conclusions \label{SandD}}

We have used the extension of the McKendrick-von Foerster model proposed by Lebowitz and Rubinow \cite{lebowitz1974theory} to generalize the seminal results of Powell \cite{powell1956growth, powell1964note} to the case of a population in a unsteady (transient) state: We have derived the exact relationships between cell age and generation time probability distributions. Such relationships were found decades ago for the steady state, but, to the best of our knowledge, they have not yet been derived for the transient population dynamics. In particular, we have derived a generalization of the Euler-Lotka equation that links the generation time distribution of just dividing cells (mothers) to the instantaneous population growth rate. We have also derived the inequalities linking the rates of appearance and disappearance of cells of generation time $\tau$. These inequalities generalize the known relationship between the mean generation time of mothers, the mean generation time of newborns, and the population doubling time.

Not all probability distributions in Powell's approach and the Lebowitz-Rubinow model are experimentally observable \cite{quedeville2019critical}. We have established the identities that link the unobservable generation time distributions of newborn and extant cells to the observable mother generation time distribution.

Our results can help to infer information about generation-time inheritance: The experimentally measured instantaneous population growth rate $\Lambda(t)$ and the mother generation time distribution $f_1(t, \tau)$ constrain the possible functional forms of the probability distribution $h(\tau |\tau^{\prime})$ of inherited generation times.

We have shown that the Lebowitz-Rubinow model can always be reduced to the McKendrick-von Foerster model. This finding extends the results of Lebowitz and Rubinow, who obtained the McKendrick-von Foerster equation from their model only for the initial condition, which is a product of two functions: One depending solely on cell age, the other solely on generation time \cite{lebowitz1974theory}.

We have also discussed the connection between the Lebowitz-Rubinow model and the model based on the `maturity representation' proposed by Rubinow in 1968 \cite{rubinow1968maturity}, see Appendix \ref{Rubinow}.


As an application of our formalism, we have calculated the fitness landscapes (as defined in refs. \cite{nozoe2017inferring, genthon2020fluctuation, genthon2021universal}) for certain phenotypic traits in a population out of the steady state. We have shown that the fitness landscape formula proposed in ref.  \cite{genthon2020fluctuation} for the cell age as a phenotypic trait is an approximation to the exact time-dependent formula derived in the present paper. We have also calculated the fitness landscape for the generation time as a phenotypic trait.

The results obtained in the framework of the Lebowitz-Rubinow model can be generalized in several ways. First, we discussed an extension of this model in which each cell is described not only by its age and generation time, but also by additional variables such as volume growth rate and current volume. Second, the original Lebowitz-Rubinow model explicitly included only the mother-daughter generation time correlations. Therefore, we have considered another generalization of this model that explicitly takes into account the non-vanishing correlations between the more distant generations (see Appendix \ref{correlations_and_inheritance}). Such an extended Lebowitz-Rubinow model uses the distribution $h(\tau | \tau_1, \tau_2, \ldots, \tau_G)$ of inherited generation times $\tau$, which is a function of the generation times $\tau_i$ of $G$ previous generations. It reduces to the original model after integrating the generation times of the grandmother, great-grandmother, etc. of a given cell. However, in such a case, the effective `Markovian' distribution of the inherited generation times $h_e(t, \tau | \tau_1)$, which appears in the standard Lebowitz-Rubinow equations, may depend on the observation time $t$, even though its counterpart $h(\tau | \tau_1, \tau_2, \ldots, \tau_G)$ of the extended Lebowitz-Rubinow model is time-independent.

\section{Author contributions}

JJ designed the study, performed the analytical calculations, and wrote the paper. MR, AOM, and JJ discussed the interpretation of the fitness landscape. MR and AOM took part in writing the paper and reviewed the manuscript. 

\section*{Acknowledgments}

We warmly acknowledge the discussions with Ryszard Rudnicki. 


AOM and MR were supported by the National Science Centre SONATA BIS 6 Grant No. 2016/22/E/ST2/00558. \\

$\ast$Corresponding author, e-mail: jjedrak@ichf.edu.pl


\appendix

\begin{widetext}

\section{Table: Key quantities used in this paper \label{App_Tab}}


%
\begin{center}
\begin{table}[h]
  \centering
  \begin{tabular}{| l | l | l| l | }
    \hline
{\textbf{Quantity}} & \textbf{Name/description} &   \textbf{Eq/Eqs.} \\ \hline 
$a \in [0, \tau]$  &  Cell age   &    \\ \hline
 $\tau \in [\tau_s, \tau_l]$  &   Generation time   &    \\ \hline
   $ n(t, a, \tau) $ &  Number density of cells whose age is $a$ and whose generation time is $\tau$  &  (\ref{n_solution_through_f_0_f1})   \\ \hline
    $\Phi(a, \tau) = n(0, a, \tau)$ &  Initial condition  &  
(\ref{APBE_3}), (\ref{solution_for_free_propagation_t_a_tau_characteristics})   \\ \hline
       $\Psi(t, \tau) = n(t, 0, \tau)$ &  Boundary condition  &  (\ref{APBE_2}), (\ref{solution_for_free_propagation_t_a_tau_characteristics}) \\ \hline
$ n(t, a) $ & Number density of cells whose age is $a$  &  (\ref{n_generalized_reduced_to_n_standard})  \\ \hline
$ N(t) $ &  Total number of cells in the population &  (\ref{Number_of_cells_solution}) (\ref{Number_of_cells_definition_LR_model})   \\ \hline 
 $\gamma(t, a)$  &  Cell division rate &  (\ref{carrier_NA_vs_anticipating_correspondence}), (\ref{gamma_carrier_NA_vs_anticipating_correspondence_all_a_t})  \\ \hline
$ \Lambda(t) $ & Age-averaged cell division rate $\gamma(t, a)$  &  (\ref{definition_of_Lambda_generalized_exponential_growth}), (\ref{boundary_condition_again}), (\ref{N_dot_sigma_Lambda_N}) \\ 
 & $N(t)\Lambda(t)dt$ is the total number of cell divisions in the population at time $t$  & (\ref{n_solution_through_f_0_f1}), (\ref{chi_solution}) \\ \hline
$ \Lambda_r(t) $ & $\Lambda(t)$ in batch culture: instantaneous growth rate of the population  &  (\ref{Omega_definition}) \\ \hline
$ \Omega(t) $ & $ \Omega(t) \equiv \int_0^{t} \Lambda_r(t^{\prime}) d t^{\prime} $  &   (\ref{Omega_definition}) \\ \hline
$h(\tau | \tau^{\prime})$  &  Probability distribution of the inherited generation times & (\ref{APBE_2}), (\ref{f_0_definition_2})  \\ \hline 
$ \chi(t, a, \tau) $ &  Joint distribution of cell age and generation time. &  (\ref{chi_definition_of}), (\ref{chi_solution})  \\ \hline  
$ \phi(t, a) $ &  Cell age distribution & (\ref{phi_NA_definition}), (\ref{phi_derived_from_chi_definition}), (\ref{psi_from_chi_1}), (\ref{psi_from_chi_2})  \\ \hline 
$ f_0(t, \tau) $ &  Generation time distribution of the newborn cells  & (\ref{f_0_definition}), (\ref{f_0_definition_2})  \\ \hline 
$ f_1(t, \tau), f_1(t, a) $ &   Generation time distribution of mother cells (age distribution of mothers)  &  (\ref{carrier_PDF_time_dependent_NA}),  (\ref{t_dependent_f_1_definition}), (\ref{relation_between_f_1_and_f_0_time_dependent_final_form}) \\ \hline 
$ f_2(t, \tau) $ &  Generation time distribution of extant cells  & (\ref{t_dependent_f_2_definition}), (\ref{f_2_expressed_using_f_1}), (\ref{t_dependent_f_2_derived_from_chi})   \\ \hline 
  \end{tabular}
  \begin{center}
    \caption{Notation: list of the most important quantities used in this paper. } 
  \end{center}  
  \label{tab:tabofetas}
\end{table}

\end{center}


\section{Relation between solutions of (\ref{NAPBE_1})-(\ref{NAPBE_3}) for batch and continuous culture\label{CC_vs_BC_McK_vF}}

One can easily get the solution to (\ref{NAPBE_1}) for continuous culture ($\sigma = 1$ and $D(t) \neq 0$) from the solution to the batch culture case where $D(t) = 0$. If $n(t, a)$ is the solution to (\ref{NAPBE_1}) with $D(t) \neq 0$ and ${\bar{n}}(t, a)$ is the solution to (\ref{NAPBE_1}) with $D(t) = 0$, then
\begin{equation}
\label{n_bar_vs_n_why_no_D}
{\bar{n}}(t, a) = n(t, a) e^{\int_0^t D(t^{\prime}) d t^{\prime}}.
\end{equation}
Integration of (\ref{n_bar_vs_n_why_no_D}) with respect to $a$ yields
\begin{equation}
\bar{N}(t) = N(t) e^{\int_0^t D(t^{\prime}) d t^{\prime}},
\end{equation}
and therefore ${\bar{\phi}}(t, a) = {\bar{n}}(t, a)/ {\bar{N}}(t) = n(t, a)/ N(t) = \phi(t, a)$. Cell division does not depend on the dilution rate $D(t)$: $\bar{\gamma}(t, a) = \gamma(t, a)$. As a consequence, we get $\bar{\Lambda}(t) = \Lambda(t)$ and ${\bar{f_1}}(t, a) = f_1(t, a)$ from (\ref{carrier_PDF_time_dependent_NA}) and (\ref{definition_of_Lambda_generalized_exponential_growth}). Since neither $\Lambda(t)$ nor the probability distributions depend on $D(t)$, we can put $D(t) = 0$ in (\ref{NAPBE_1}) without loss of generality.


\section{More on the relationships between the McKendrick-von Foerster and Lebowitz-Rubinow models \label{phi_tilde_and_gamma_expressed_via_f_0}}


The division rate $\gamma(t, a)$ of the McKendrick-von Foerster model (\ref{NAPBE_1})--(\ref{NAPBE_3}) can be expressed in terms of the quantities of the Lebowitz-Rubinow model (\ref{APBE_1})--(\ref{APBE_3}). By using (\ref{carrier_NA_vs_anticipating_correspondence}) and (\ref{conditional_chi_definition}) we get
\begin{equation}
\gamma(t, a) = \frac{n(t, a, a)}{n(t, a)} = \frac{\chi(t, a, a)}{\phi(t, a)} = \chi(t, a |a).
\label{gamma_carrier_NA_vs_anticipating_correspondence_all_a_t}
\end{equation}
Such a defined division rate is an effective quantity and may depend on $\sigma$, i.e. it may be different for the mother machine experiment ($\sigma = 0$) and the batch culture ($\sigma = 1$). This difference may occur because $\gamma(t, a)$ is no longer a known function that is treated as an input to the model. Instead, it is determined from the solutions of the Lebowitz-Rubinow equation (\ref{APBE_1}). This is in contrast to the situation when we base our description of population dynamics solely on the McKendrick-von Foerster model. If we then neglect cell-cell interactions and assume identical environmental conditions (temperature, pH, nutrients, etc.) for the batch culture and the mother machine experiment, then the McKendrick-von Foerster model's division rate $\gamma(t, a)$ should be identical in these two cases.

Therefore, if the McKendrick-von Foerster equation is derived from a model that contains more variables (e.g., the Lebowitz-Rubinow model), then in general there are two different cell division rate functions: one for the batch culture ($\gamma_r(t, a)$) and one for the mother machine experiment ($\gamma_c(t, a)$). In both cases, we can express $\gamma(t, a)$ for $t \geq a$ in terms of the daughter generation time distribution $f_0(t, \tau)$ using (\ref{carrier_NA_vs_anticipating_correspondence}) and (\ref{n_solution_through_f_0_f1})
%
%
\begin{equation}
\gamma(t, a) = \frac{f_0(t - a, a)}{\int_{a}^{\tau_l} f_0(t - a, \tau) d \tau} = \frac{f_0(t - a, a)}{ \bar{F}_0(t - a, a)}.
\label{gamma_carrier_NA_vs_anticipating_correspondence}
\end{equation}
where $\bar{F}_0(t, a)$ is defined by (\ref{definition_c_0_minus}). We have $f_0(t,a)=0$ for $a < \tau_s$ and then $\gamma(t, a)= 0$. In the steady state, we get from (\ref{gamma_carrier_NA_vs_anticipating_correspondence}) 
\begin{equation}
\gamma(a) = \frac{f_0(a)}{\int_{a}^{\tau_l} f_0(\tau) d \tau} = \frac{f_0(a)}{\bar{F}_{0}(a)}.
\label{gamma_carrier_NA_vs_anticipating_correspondence_t_independent}
\end{equation}

The fact that one can derive Eqs. (\ref{NAPBE_1})--(\ref{NAPBE_3}) from Eqs. (\ref{APBE_1})--(\ref{APBE_3}) shows that the McKendrick-von Foerster model does not exclude mother-daughter generation time correlations and generation time inheritance, it just does not allow their explicit description. This is because the distribution $h(\tau | \tau^{\prime})$ of inherited generation times, which appears in Eq. (\ref{APBE_2}), does not exist in the McKendrick-von Foerster model. Moreover, from Eq. (\ref{gamma_carrier_NA_vs_anticipating_correspondence}) we see that $\gamma(t, a)$ can explicitly depend on the observation time $t$ in the presence of such correlations, since $f_0(t, \tau)$ usually depends on $t$, see (\ref{f_0_definition_2}).

Now assume that there are no mother-daughter generation time correlations 
\begin{equation}
\label{definition_of_h_no_correlations_Appendix}
h(\tau | \tau^{\prime}) = f(\tau).
\end{equation} 
It follows from (\ref{f_0_definition_2}) that the distribution of inherited generation times no longer depends on $t$: $f_0(t, \tau) = f(\tau)$, so from (\ref{gamma_carrier_NA_vs_anticipating_correspondence}) we get $\gamma(t,a)=\gamma(a)$. In such a case we have 
\begin{equation}
\label{definition_of_f_through_gamma_a_la_chiorino}
f(a) = \gamma(a)e^{-\int_{0}^{a} \gamma(a^{\prime}) d a^{\prime}},
\end{equation}
and therefore
\begin{equation}
\gamma(a) = \frac{f(a)}{\int_{a}^{\tau_l} f(\tau) d \tau} = \frac{f(a)}{\bar{F}(a)}.
\label{gamma_carrier_NA_vs_anticipating_correspondence_t_independent_uncorrelated}
\end{equation}
The McKendrick-von Foerster model can be called the independent generation times (IGT) model only if $\gamma(a)$ is given by (\ref{gamma_carrier_NA_vs_anticipating_correspondence_t_independent_uncorrelated}). Equation (\ref{gamma_carrier_NA_vs_anticipating_correspondence}) is a generalization of the division rate of the IGT model (\ref{gamma_carrier_NA_vs_anticipating_correspondence_t_independent_uncorrelated}) to the situation where the mother-daughter generation time correlations are present.

\section{What relationships between probability distributions are derivable within the McKendrick-von Foerster model? \label{McKendrick_model_NA_Appendix}} 

Here we show how some of the results presented in the main text can be derived in an alternative way using only the framework of the McKendrick-von Foerster model. This model can be used to derive the relationship between the generation time distribution of mothers (equivalent to their age distribution), $f_1(t, a)$ (\ref{carrier_PDF_time_dependent_NA}) and the age distribution $\phi(t, a)$ (\ref{phi_NA_definition}) of all cells in a population. To do this, we first obtain from equations (\ref{NAPBE_1})--(\ref{phi_NA_definition}), (\ref{boundary_condition_again}), and (\ref{N_dot_sigma_Lambda_N}) the time evolution equation, the boundary condition, and the initial condition for the cell age distribution $\phi(t, a)$ defined by (\ref{phi_NA_definition}):
\begin{eqnarray}
\label{dynamics_of_phi_I_1}
\left[\partial_t + \partial_a + \gamma(t, a) + \sigma\Lambda(t)\right] \phi(t, a) = 0, \\
\label{dynamics_of_phi_I_2}
\phi(t, 0) = 2^{\sigma} \Lambda(t), \\
\label{dynamics_of_phi_I_3}
\phi(0, a) = \phi_0(a).
\end{eqnarray}
The equations (\ref{dynamics_of_phi_I_1})--(\ref{dynamics_of_phi_I_3}) are valid for both the batch culture ($\sigma = 1$) and the mother machine ($\sigma = 0$). 
%
%
%
Using (\ref{carrier_PDF_time_dependent_NA}), we rewrite (\ref{dynamics_of_phi_I_1}) as
%
\begin{equation}
\label{dynamics_of_phi_II_1}
\left[\partial_t + \partial_a + \sigma\Lambda(t)\right] \phi(t, a) + \Lambda(t)f_1(t, a) = 0.
\end{equation}
Next, we apply the Laplace transform to (\ref{dynamics_of_phi_II_1}) and get 
\begin{equation}
\label{dynamics_of_phi_II_1_Laplace}
\frac{d\hat{\phi}(t, s)}{dt} + [s+\sigma\Lambda(t)] \hat{\phi}(t, s) = \Lambda(t)[2^{\sigma} - \hat{f}_1(t, s)], 
\end{equation}
where 
%
\begin{eqnarray}
\hat{\phi}(t, s) &=& \int_0^{\infty} e^{-sa} \phi(t, a) da, \\ \hat{f}_1(t, s) &=& \int_0^{\infty} e^{-sa} f_1(t, a) da.
\label{definitions_of_the_Laplace_transforms_of_phi_and_g}
\end{eqnarray}
Solving (\ref{dynamics_of_phi_II_1_Laplace}) yields
%
%
\begin{eqnarray}
\label{dynamics_of_phi_II_1_Laplace_solution_of}
\hat{\phi}(t, s) &=& e^{-st}e^{-\sigma\Omega(t)} \left\lbrace \hat{\phi}(0, s) + \int_0^{t} e^{\sigma\Omega(t^{\prime})+st^{\prime}} \Lambda(t^{\prime})[2^{\sigma} - \hat{f}_1(t^{\prime}, s)] d t^{\prime} \right\rbrace, 
\end{eqnarray}
%
where $\Omega(t)$ is defined by (\ref{Omega_definition}).
If we invert (\ref{dynamics_of_phi_II_1_Laplace_solution_of}), we obtain the equations (\ref{psi_from_chi_1}) and (\ref{psi_from_chi_2}), which were derived in a different way in section \ref{phi_and_f1_R}: 
%
\begin{eqnarray}
\phi(t, a) =
\begin{cases}
e^{-\sigma\Omega(t)} \Big[\phi_0(a-t) - \int_0^{t} \Lambda(t^{\prime}) e^{\sigma\Omega(t^{\prime})} {f}_1(t^{\prime}, a-t+t^{\prime}) d t^{\prime} \Big], ~~~ a\geq t, \\ \\
e^{-\sigma\Omega(t)} \Big[2^\sigma e^{\sigma\Omega(t-a)} \Lambda(t-a) - \int_{t-a}^{t} \Lambda(t^{\prime}) e^{\sigma\Omega(t^{\prime})} {f}_1(t^{\prime}, a-t+t^{\prime}) d t^{\prime} \Big], ~~~ a \leq t.
\label{phi_as_a_function_of_g_solution_of_PDE}
\end{cases}
\label{phi_as_a_function_of_g_solution_of_PDE}
\end{eqnarray}
Eq. (\ref{phi_as_a_function_of_g_solution_of_PDE}) is a generalization of Eq. (14) of ref. \cite{quedeville2019critical} to the case of unsteady population growth. (In Eq. (14) of ref. \cite{quedeville2019critical}, there is a constant $D(t) = D$ instead of $\Lambda$, because $D = \Lambda$ in the steady state limit considered there.) 

We are also interested in the time evolution equations for the moments of $\phi(t, a)$. These can be obtained using (\ref{dynamics_of_phi_II_1}) or (\ref{dynamics_of_phi_II_1_Laplace}),
\begin{eqnarray}
\label{moments_phi_f1_ODE}
\frac{d {\mathcal{A}}_k(t)}{dt} + \sigma \Lambda(t) \mathcal{A}_k(t) - k \mathcal{A}_{k-1}(t) = - \Lambda(t) \mathcal{T}_k(t), ~~~ k =1, 2, \ldots,
\end{eqnarray}
where 
\begin{equation}
\label{moments_phi_f1_definitions}
\mathcal{A}_l(t) = \int_{0}^{\tau_l} a^k \phi(t, a) d a, ~~~ \mathcal{T}_k(t)= \int_{0}^{\tau_l} a^k f_1(t, a) d a. \nonumber
\end{equation}
For $\sigma = 1$ from (\ref{moments_phi_f1_ODE}) in the steady-state limit, we obtain Eqs. (18) and (20) of ref. \cite{quedeville2019critical}, 
\begin{equation}
\label{moments_1_and_2st_phi_f1_ODE_ss}
\mathcal{A}_1 + \mathcal{T}_1 = \Lambda^{-1},~~~ \mathcal{A}_2 = {2}\Lambda^{-1} \mathcal{A}_1 - \mathcal{T}_2.
\end{equation}
%
For $\sigma = 0$, we get Eq. (4) of ref. \cite{jafarpour2018bridging}:
\begin{eqnarray}
\label{moment_1st_phi_f1_ODE_ss_cJ}
\mathcal{A}_1 = \frac{1}{2} \mathcal{T}_1 \left(1 + \frac{\mathcal{T}_2 - \mathcal{T}_1^2}{\mathcal{T}_1^2}\right) = \frac{\mathcal{T}_2}{2 \mathcal{T}_1}.
\end{eqnarray}
In general, for $\sigma =1$ and any $k\geq 1$, we get 
\begin{eqnarray}
\label{moment_kth_phi_f1_ODE_ss}
\mathcal{A}_k =  \frac{k!}{\Lambda_r^k}\left(1 - \sum_{j=1}^{k} \frac{\Lambda_r^j}{j!}  \mathcal{T}_{j} \right).
\end{eqnarray}
Note that (\ref{moment_kth_phi_f1_ODE_ss}) is a special case of the equation (\ref{moments_chi_SS_r_solutions}).

Now let us return to the case of transient dynamics. For $k=1, 2, \ldots, K$, Eq. (\ref{moments_phi_f1_ODE}) gives a closed system of $K$ equations that can be solved recursively for any $K<\infty$. However, it is much more convenient to solve (\ref{dynamics_of_phi_II_1_Laplace}) instead and find the moments using the Laplace transform $\hat{\phi}(t, s)$ (\ref{dynamics_of_phi_II_1_Laplace_solution_of}), which is equivalent to using the generating functions for probability distributions. We get
%
%
\begin{eqnarray}
\label{moments_phi_f1_solutions}
\mathcal{A}_k(t) &=& e^{-\sigma\Omega(t)}\Bigg[\sum_{l=0}^{k}\binom{k}{l} t^{k-l} \mathcal{A}_l(0) + 2^{\sigma} \int_{0}^{t} \Lambda(t^{\prime}) e^{\sigma\Omega(t^{\prime})} (t-t^{\prime})^k d t^{\prime} - \sum_{l=0}^{k}\binom{k}{l} \int_{0}^{t} \Lambda(t^{\prime}) e^{\sigma\Omega(t^{\prime})} (t-t^{\prime})^{k-l} \mathcal{T}_l(t^{\prime}) d t^{\prime} \Big]. \nonumber \\
\end{eqnarray}

%


Eq. (\ref{moments_phi_f1_solutions}) is a generalization of Eqs. (18) and (20) of ref. \cite{quedeville2019critical}. To our knowledge, neither (\ref{phi_as_a_function_of_g_solution_of_PDE}) nor (\ref{moment_kth_phi_f1_ODE_ss}) and (\ref{moments_phi_f1_solutions}) has been shown in the literature to date.


\section{Derivation of equations (\ref{f_2_expressed_using_f_1}),  (\ref{psi_from_chi_1}), and (\ref{psi_from_chi_2}) \label{phi_and_f1_R_App}}

Eq. (\ref{phi_as_a_function_of_g_solution_of_PDE}), derived using the formalism of the McKendrick-von Foerster model, links the cell age distribution $\phi(t, a)$ with the generation time distribution $f_1(t, \tau)$ of mothers. However, the Lebowitz-Rubinow model provides an alternative way to obtain that equation. 

Consider first the case of $a \geq t$. Using (\ref{Number_of_cells_solution}), (\ref{solution_for_free_propagation_t_a_tau_characteristics}), (\ref{chi_definition_of}), (\ref{chi_0_definition_of}) and (\ref{phi_derived_from_chi_definition}), we obtain 

%
\begin{eqnarray}
\label{phi_as_derived_from_chi_definition_part_1_App}
\phi(t, a) &\equiv & \int_{\underline{a}}^{\tau_l} \chi(t, a, \tau) d \tau = \int_{\underline{a}}^{\tau_l} \chi_0(a-t, \tau) e^{-\sigma\Omega(t)} d \tau \\ &=& \int_{\underline{a-t}}^{\tau_l} \chi_0(a-t, \tau) e^{-\sigma\Omega(t)} d \tau - \int_{\underline{a-t}}^{\underline{a}} \chi_0(a-t, \tau) e^{-\sigma\Omega(t)} d \tau = \label{phi_as_derived_from_chi_definition_part_1_last_line_App} \\ &=& \phi_0(a-t)e^{-\sigma\Omega(t)} - \frac{1}{N(t)}\int_{\underline{a-t}}^{\underline{a}} n(0, a-t, \tau) d \tau, 
\end{eqnarray}
%
where $\phi_0(a) \equiv \phi(0, a)$, $\chi_0(a, \tau) \equiv \chi(0, a, \tau)$ is defined by (\ref{chi_0_definition_of}) and $\underline{a} \equiv \max(a, \tau_s)$. Now consider the last integral in (\ref{phi_as_derived_from_chi_definition_part_1_last_line_App}). We have

\begin{eqnarray}
\label{phi_as_derived_from_chi_definition_part_2_App}
\int_{\underline{a-t}}^{\underline{a}} n(0, a-t, \tau) d \tau & = & \int_{\underline{a-t}}^{\underline{a}} n(t- a + \tau, \tau , \tau) d \tau = \int_{t-a +\underline{a-t}}^{t-a + \underline{a}} n(t^{\prime}, t^{\prime} - t + a, t^{\prime} - t + a) d t^{\prime} \nonumber \\ &=& \int_{t-a +\underline{a-t}}^{t-a + \underline{a}} \Lambda(t^{\prime})N(t^{\prime})f_1(t^{\prime}, t^{\prime} - t + a) d t^{\prime}. 
\end{eqnarray}

In (\ref{phi_as_derived_from_chi_definition_part_2_App}), we have used (\ref{solution_for_free_propagation_t_a_tau_characteristics}) and (\ref{carrier_NA_vs_anticipating_correspondence}). Combining (\ref{phi_as_derived_from_chi_definition_part_1_App}) and (\ref{phi_as_derived_from_chi_definition_part_2_App}), we finally obtain 
\begin{eqnarray}
\label{phi_as_derived_from_chi_definition_final_App}
\phi(t, a) &=& \phi_0(a-t)e^{-\sigma\Omega(t)} - \int_{t-a +\underline{a-t}}^{t-a + \underline{a}} e^{-\sigma\Omega(t)} \Lambda(t^{\prime}) e^{\sigma\Omega(t^{\prime})} f_1(t^{\prime}, t^{\prime} - t + a) d t^{\prime}. 
\end{eqnarray}
$\phi(t, a)$ (\ref{phi_as_derived_from_chi_definition_final_App}) is identical to both Eq. (\ref{psi_from_chi_1}) and to the first line of (\ref{phi_as_a_function_of_g_solution_of_PDE}) if one puts $\tau_s=0$; hence $\underline{a-t} = a-t$, $\underline{a}= a$ (we have assumed $\tau_s=0$ when deriving (\ref{phi_as_a_function_of_g_solution_of_PDE})). 

Now, for $a \leq t$, we have
\begin{eqnarray}
\label{identity_r1_App}
N(t)\phi(t, a) &=& n(t, a) = \int_{\underline{a}}^{\tau_l} n(t, a, \tau) d \tau = \int_{\tau_s}^{\tau_l} n(t -a, 0, \tau) d \tau - \int_{\tau_s}^{\underline{a}} n(t -a, 0, \tau) d \tau \\ &=& n(t -a, 0) - \int_{\tau_s}^{\underline{a}} n(t - a + \tau, \tau, \tau) d \tau \\ &=& \label{identity_r1_alsmost_last_line_App} 2^{\sigma} \Lambda(t - a)N(t - a)- \int_{t-a + \tau_s}^{t-a + \underline{a}} n(t^{\prime}, t^{\prime} - t + a, t^{\prime} - t + a) d t^{\prime} \\
&=& 2^{\sigma} \Lambda(t - a)N(t - a) - \int_{t-a + \tau_s}^{t-a + \underline{a}} \Lambda(t^{\prime})N(t^{\prime})f_1(t^{\prime}, t^{\prime} - t + a) d t^{\prime}.
\label{identity_r1_last_line_App}
\end{eqnarray}
%
%
Dividing both sides of Eq. (\ref{identity_r1_last_line_App}) by $N(t)$ (\ref{Number_of_cells_solution}), we get (\ref{psi_from_chi_2}), which is equivalent to the second line of (\ref{phi_as_a_function_of_g_solution_of_PDE}) if only $\tau_s =0$.

%
%


In order to find the relationship between the generation time distribution $f_2(t, \tau)$ of extant cells and the generation time distribution $f_1(t, \tau)$ of mothers, we proceed in a similar way. Assume first that $t \geq \tau \geq a$. Using (\ref{t_dependent_f_2_definition}), we get
%
\begin{eqnarray}
\label{f_2_as_derived_from_chi_definition_part_1_App}
f_2(t, \tau) &=& \int_{0}^{\tau} \chi(t, a, \tau) d a = \frac{1}{N(t)} \int_{0}^{\tau} n(t -a, 0, \tau) da = \frac{1}{N(t)} \int_{0}^{\tau} n(t - a + \tau, \tau, \tau) da = \frac{1}{N(t)} \int_{t}^{t+\tau} n(t^{\prime}, \tau, \tau) d t^{\prime} \nonumber \\ &=& \frac{1}{N(t)} \int_{t}^{t+\tau} \Lambda(t^{\prime})N(t^{\prime})f_1(t^{\prime}, \tau) d t^{\prime} = e^{-\sigma\Omega(t)} \int_{t}^{t+\tau} \Lambda(t^{\prime}) e^{\sigma\Omega(t^{\prime})} f_1(t^{\prime}, \tau) d t^{\prime}.
\end{eqnarray}
%
%
%
The derivation is analogous for $t \leq \tau$ but we have to consider two cases: $t \leq a$ and $t \geq a$, 
\begin{eqnarray}
\label{f_2_as_derived_from_chi_definition_part_2_App}
f_2(t, \tau) &=& \int_{0}^{t} \chi(t, a, \tau) d a + \int_{t}^{\tau} \chi(t, a, \tau) d a = \frac{1}{N(t)} \int_{0}^{t} n(t -a, 0, \tau) da + \frac{1}{N(t)} \int_{t}^{\tau} n(0, a-t, \tau) da \\ &=& \frac{1}{N(t)} \int_{0}^{t} n(t - a + \tau, \tau, \tau) da + \frac{1}{N(t)} \int_{t}^{\tau} n(t - a + \tau, \tau, \tau) da = \frac{1}{N(t)} \int_{0}^{\tau} n(t - a + \tau, \tau, \tau) da \\ &=& \frac{1}{N(t)} \int_{t}^{t+\tau} n(t^{\prime}, \tau, \tau) d t^{\prime} = e^{-\sigma\Omega(t)} \int_{t}^{t+\tau} \Lambda(t^{\prime}) e^{\sigma\Omega(t^{\prime})} f_1(t^{\prime}, \tau) d t^{\prime}.\label{f_2_as_derived_from_chi_definition_part_2_last_line_App}
\end{eqnarray}
From (\ref{f_2_as_derived_from_chi_definition_part_1_App}) and (\ref{f_2_as_derived_from_chi_definition_part_2_App})-- (\ref{f_2_as_derived_from_chi_definition_part_2_last_line_App}), we obtain a single formula for $f_2(t, \tau)$, valid both for $t \leq \tau$ and for $t \geq \tau$:
\begin{eqnarray}
\label{f_2_expressed_using_f_1_App}
f_2(t, \tau) &=& \int_{t}^{t+\tau} e^{-\sigma\Omega(t)} e^{\sigma\Omega(t^{\prime})} \Lambda(t^{\prime}) f_1(t^{\prime}, \tau) d t^{\prime} = \int_{t}^{t+\tau} e^{\sigma \int_{t}^{t^{\prime}} \Lambda(\tilde{t}) d \tilde{t}} \Lambda(t^{\prime}) f_1(t^{\prime}, \tau) d t^{\prime}.
\end{eqnarray}
This is the equation (\ref{f_2_expressed_using_f_1}) in the main text.

\section{Lebowitz-Rubinow equations for the m-th generation of cells and their solution in the form of a series \label{mth_generation_LR_model}}

The equations (\ref{APBE_1})-(\ref{APBE_3}) of the Lebowitz-Rubinow model can be rewritten in a form that explicitly includes the number of cell divisions in a given cell lineage. Let us call the cells existing at $t=0$ the zeroth generation. The $m$-th generation is the result of the $m$-th cell division (counted from $t=0$). The cell number density is $n(t, a, \tau; m)$. The boundary condition (\ref{APBE_2}) is now
\begin{equation}
\label{APBE_2_m}
n(t, 0, \tau; m) = 2^{\sigma} \int_{\tau_s}^{\tau_l} h(\tau | \tau^{\prime}) n(t, \tau^{\prime}, \tau^{\prime}; m-1) d \tau^{\prime}, 
\end{equation} 
while the initial condition is
\begin{eqnarray}
\label{APBE_3_m}
n_0(a, \tau, 0) = n(0, a, \tau; 0). 
\end{eqnarray}
We have $n(t, a, \tau; 0) = \Phi(a-t, \tau)$ and 
\begin{equation}
n(t, a, \tau) = \sum_{m = M_\text{min}}^{M_\text{max}} n(t, a, \tau; m),
\label{n_LR_as_a_sum_of_n_ms}
\end{equation}
%
%
where $M_\text{min}$ and $M_{\text{max}}$ are the minimum and maximum possible number of cell divisions in a lineage. Both $M_\text{min}$ and $M_{\text{max}}$ depend on $t, a, \tau_s$ and $\tau_l$:
\begin{equation}
\label{M_s_M_l_definitions}
M_\text{min} \equiv \lfloor (t-a)/\tau_l  \rfloor, ~~~ M_{\text{max}} \equiv \lfloor (t-a)/\tau_s \rfloor + 1,
\end{equation}
where $\lfloor x \rfloor$ is the floor function or the integer part of $x$. Between $t_{\text{obs}}=0$ and $t_{\text{obs}}=t$ there are not less than $M_\text{min}$ and not more than $M_{\text{max}}$ complete cell cycles.

The time evolution of $n(t, a, \tau; m)$ does not depend on $m$ and is given by Eq. (\ref{APBE_1}):
\begin{eqnarray}
\label{APBE_1_m}
\frac{\partial}{\partial t}n(t, a, \tau; m) + \frac{\partial}{\partial a} n(t, a, \tau; m) = 0. 
\end{eqnarray}
As a consequence, each $n(t, a, \tau; m)$ obeys Eq. (\ref{solution_for_free_propagation_t_a_tau_characteristics}). The renewal equation (\ref{solution_for_free_propagation_t_a_tau_self_consistent_a_la_RR}) can be rewritten as
%
%
\begin{eqnarray}
\label{solution_for_free_propagation_t_a_tau_self_consistent_a_la_RR_m}
 \Psi_m(t, \tau) &=& 2^{\sigma} \delta_{m1} \Theta(\tau_l - t) \int_{\underline{t}}^{\tau_l} h(\tau | \tau^{\prime}) \Phi(\tau^{\prime} - t, \tau^{\prime}) d \tau^{\prime} \nonumber \\ & + & 2^{\sigma} (1-\delta_{m1}) \Theta(t-\tau_s)  \int_{\tau_s}^{\overline{t}} h(\tau | \tau^{\prime}) \Psi_{m-1}(t-\tau^{\prime}, \tau^{\prime}) d \tau^{\prime}, 
\end{eqnarray}
where $\Psi_m(t, \tau) = n(t, 0, \tau; m)$ and $\overline{t}$ and $\underline{t}$ are defined by 
(\ref{t_bar_definition}). So we have 
\begin{eqnarray}
n(t, a, \tau) &=&
\begin{cases}
\Phi(a - t, \tau)~~~ \text{for}~~~a \geq t, \\ \\
\sum_{m = M_\text{min}}^{M_{\text{max}}}\Psi_m(t - a, \tau) ~~~ \text{for}~~~a \leq t, \label{solution_for_free_propagation_t_a_tau_characteristics_m}
\end{cases}
\end{eqnarray}
where 
%
\begin{equation}
\Psi_1(t_1, \tau_1) = 2^{\sigma} \Theta(\tau_l - t_1) \int_{\underline{t}_1}^{\tau_l} h(\tau_1 | \tau_0) \Phi(\tau_0 - t_1, \tau_0) d \tau_{0}, 
\label{Psi_explicit_form_starting_1}
\end{equation}%
and for $i\geq 2$ 
\begin{equation}
\Psi_i(t_i, \tau_i) = 2^{\sigma} \int_{{\tau_s}}^{\overline{t}_i} h(\tau_i | \tau_{i-1}) \Psi_{i-1}(t_{i-1}, \tau_{i-1}) d \tau_{i-1}.  
\label{Psi_explicit_form_starting_m}
\end{equation}
$\tau_j$ is the duration of the ${j}$th cell cycle: $\tau_j = t_{j+1} - t_{j}$ and $t_i = t_1 + \sum_{j=1}^{i-1} \tau_j$.

Equations (\ref{solution_for_free_propagation_t_a_tau_characteristics_m}--\ref{Psi_explicit_form_starting_m}) can be used to construct a formal solution of the Lebowitz-Rubinow model equations in the form of a series that depends only on the distribution of the inherited generation time $h(\tau |\tau^{\prime})$ and the initial condition $\Phi(a, \tau)$ \cite{lebowitz1974theory}. 

Equations (\ref{solution_for_free_propagation_t_a_tau_characteristics_m}--\ref{Psi_explicit_form_starting_m}) can also be used to derive the relationship
\begin{equation}
\label{n_r_to_n_c_ratio}
n_r(t, a, \tau; m) = 2^m n_c(t, a, \tau; m). 
\end{equation}
Equation (\ref{n_r_to_n_c_ratio}) follows from (\ref{Psi_explicit_form_starting_1})-(\ref{Psi_explicit_form_starting_m}) and from the fact that both the initial state and the distribution of inherited generation times are identical for the mother machine ($\sigma = 0$, $\ell(\sigma) = c$) and the batch culture ($\sigma = 1$, $\ell(\sigma) = r$): $\Phi_c(a, \tau) = \Phi_r(a, \tau)$, $h_c(\tau | \tau^{\prime}) = h_r(\tau | \tau^{\prime})$.

\section{Relationship between the form of probability distributions for mother machine experiments ($\sigma=0$) and batch culture ($\sigma=1$) \label{Forward_vs_Backward}}

In this Appendix, we show how to express a given probability distribution for the batch culture in terms of the same distribution for the mother machine experiment. It turns out that the relationship in question is different for the conditional distributions ($f_0(t, \tau)$ and $f_1(t, \tau)$) than for $\chi(t, a, \tau)$ and the marginal distributions that can be obtained from $\chi(t, a, \tau)$: $\phi(t, a)$ or $f_{2}(t, \tau)$. We also show how the instantaneous population growth rate $\Lambda_r(t)$ can be expressed in terms of quantities obtained from cell lineage statistics.

In Subsection \ref{GTD_relationships_between} we have considered relationships between probability distributions for the same values of $\sigma = 0$. That is, we looked separately at mother machine experiments ($\sigma = 0$) and separately at batch culture ($\sigma = 1$). This is justified in the sense that these are two different experimental situations. But as we have already pointed out, and as we will show in this appendix, $\sigma =0$ is equivalent to chronological sampling, and $\sigma =1$ is equivalent to retrospective sampling for the same batch culture population.

The difference between chronological and retrospective sampling (and their respective probabilities) was explained in subsection \ref{Fitness_landscapes}. Briefly, in retrospective sampling, we assign to each cell the statistical weight $1/N(t)$, where $N(t)$ is the total number of cells in the population \cite{nozoe2017inferring}. In the case of chronological sampling, each cell line is weighted by the number of divisions that have occurred since $t=0$: each such division contributes a factor of 1/2, so the total weight assigned to a line is $2^{-m}$, where $m$ is the number of cell divisions \cite{nozoe2017inferring}.

One can now ask what is the relationship between the same probability distribution (e.g. $f_1(t, \tau)$) for $\sigma = 0$ (the case of mother-machine experiments, or equivalently, chronological sampling) and for $\sigma = 1$ (the case of batch culture, or retrospective sampling).  

For the sake of simplicity, let us restrict our attention to the original Lebowitz-Rubinow model. Then, for $\chi(t, a, \tau)$ and its marginal distributions: $\phi(t, a)$ or $f_{2}(t, \tau)$, which describe all cells in the population, this relationship  in question can be derived from the beautiful and general analogy between population dynamics models and nonequilibrium statistical mechanics \cite{garcia2019linking,genthon2020fluctuation}. However, the relationship is different for the conditional distributions: $f_0(t, \tau)$ and $f_1(t, \tau)$, which describe only newborn and mother cells, respectively. For example, we have $f_{1c}(t, \tau)/f_{1r}(t, \tau) \neq f_{2c}(t, \tau)/f_{2r}(t, \tau)$. Here we derive the relations that connect $f_{0c}(t, \tau)$ with $f_{0r}(t, \tau)$ and $f_{1c}(t, \tau)$ with $f_{1r}(t, \tau)$, which to our knowledge have not been given in the literature. These relations also make it possible to give a formula that expresses $\Lambda_r(t)$ using only chronological statistics. 

To do this, we need to consider the cell number density and all probability distributions for each generation separately, as we did in Appendix \ref{mth_generation_LR_model}. First, following Refs. \cite{nozoe2017inferring, genthon2020fluctuation} we define $\chi_c(t, a, \tau; m)$ and $\chi_r(t, a, \tau; m)$ as
\begin{equation}
\label{chi_c_definition}
\chi_c(t, a, \tau; m) = \frac{n_r(t, a, \tau; m)}{N_0 2^m}, 
\end{equation}
and 
\begin{equation}
\label{chi_r_definition}
\chi_r(t, a, \tau; m) = \frac{n_r(t, a, \tau; m)}{N_0 e^{\Omega(t)}}, 
\end{equation}
where $N_0 = N(0)$. The definition of $\chi_r(t, a, \tau, m)$ given above is consistent with the definitions of subsection \ref{Definitions_of_f_i_s} and Appendix \ref{mth_generation_LR_model}, c.f. Eqs. (\ref{P_r_FL_definition}) and (\ref{P_c_FL_definition}).

To see which equations $\chi_c(t, a, \tau, m)$ (\ref{chi_c_definition}) satisfies, let us first consider Eq. (\ref{APBE_2_m}) for the batch culture, i.e. for $\sigma=1$:
\begin{equation}
\label{APBE_2_m_BC}
n_r(t, 0, \tau; m) = 2 \int_{\tau_s}^{\tau_l} h(\tau | \tau^{\prime}) n_r(t, \tau^{\prime}, \tau^{\prime}; m-1) d \tau^{\prime}. 
\end{equation} 
If we divide (\ref{APBE_2_m_BC}) by $N_0 2^m$ and use (\ref{chi_c_definition}), we get
\begin{equation}
\label{APBE_2_m_BC_chi_c}
\chi_c(t, 0, \tau; m) = \int_{\tau_s}^{\tau_l} h(\tau | \tau^{\prime}) \chi_c(t, \tau^{\prime}, \tau^{\prime}; m-1) d \tau^{\prime}. 
\end{equation} 
$n_c(t, 0, \tau; m) = N_0 \chi_c(t, 0, \tau; m)$ satisfies the same equation. Now, if we sum both sides of (\ref{APBE_2_m_BC_chi_c}) with respect to $m$ and multiply by $N_0$, we get
\begin{equation}
\label{APBE_2_m_BC_chi_c_summed}
n_c(t, 0, \tau) = \int_{\tau_s}^{\tau_l} h(\tau | \tau^{\prime}) n_c(t, \tau^{\prime}, \tau^{\prime}) d \tau^{\prime}, 
\end{equation} 
The above equation has a form identical to (\ref{APBE_2}) with $\sigma=0$. Similarly, summing both sides of (\ref{APBE_2_m_BC}) with respect to $m$ gives (\ref{APBE_2}) with $\sigma = 1$. Next, for all $m$, both $n_c(t, 0, \tau; m)$ and $n_r(t, 0, \tau; m)$ satisfy (\ref{APBE_1}). Therefore, both $n_c(t, 0, \tau) = \sum_m n_c(t, 0, \tau; m)$ and $n_r(t, 0, \tau)  = \sum_m n_r(t, 0, \tau; m)$ satisfy  (\ref{APBE_1}). This shows that $\sigma=0$ indeed corresponds to the case of chronological sampling, while $\sigma=1$ corresponds to the case of retrospective sampling.

What is the relationship between $\chi_r(t, a, \tau; m)$ and $\chi_c(t, a, \tau; m)$? From  (\ref{chi_r_definition}) and (\ref{chi_c_definition}) or from (\ref{n_r_to_n_c_ratio}) we get
\begin{equation}
\label{chi_r_vs_chi_c}
\chi_r(t, a, \tau; m) = 2^m e^{-\Omega(t)} \chi_c(t, a, \tau; m). 
\end{equation}
The above equation also follows from the elegant formalism of fluctuation relations \cite{garcia2019linking,genthon2020fluctuation}. Using (\ref{chi_r_vs_chi_c}) we can now obtain analogous relations between  $\phi_r(t, a; m)$  and $\phi_c(t, a; m)$ as well as between $f_{2r}(t, \tau; m)$ and $f_{2c}(t, \tau; m)$ where 
\begin{equation}
\phi_{\ell(\sigma)}(t, a; m) \equiv \int_{\underline{a}}^{\tau_l} \chi_{\ell(\sigma)}(t, a, \tau; m) d \tau
\label{phi_m_as_a_marginal_of_ch_m_r}
\end{equation}
and
\begin{equation}
f_{2\ell(\sigma)}(t, \tau; m) \equiv  \int_{0}^{\tau}  \chi_{\ell(\sigma)}(t, a, \tau; m) d a,  
\label{f2_m_as_a_marginal_of_ch_m_c}
\end{equation}
and where $\ell(\sigma)$ was defined by (\ref{F_or_r_ell_od_sigma}):  $\ell(0)=c$, $\ell(1)=r$.

Now for both $\sigma = 0$ and $\sigma = 1$ we can define the probability that a randomly chosen cell belongs to the $m$th generation:
\begin{eqnarray}
\label{vartheta_PDF_definition}
\vartheta_{\ell(\sigma)}(t, m) &=& \int_{0}^{\tau_l} \int_{\underline{a}}^{\tau_l} \chi_{\ell(\sigma)}(t, a, \tau; m) d \tau d a.
\end{eqnarray}
%

We also have 
\begin{equation}
\label{f0_m_as_a_conditional_of_ch_m_r}
f_{0\ell(\sigma)}(t, \tau; m) \equiv \frac{\chi_{\ell(\sigma)}(t, 0, \tau; m)}{\phi_{\ell(\sigma)}(t, 0)} = \frac{\chi_{\ell(\sigma)}(t, 0, \tau; m)}{2^\sigma \Lambda_{\ell(\sigma)}(t)}
\end{equation}
and
\begin{equation}
\label{f1_m_as_a_conditional_of_ch_m_r}
f_{1\ell(\sigma)}(t, \tau; m) \equiv \frac{\chi_{\ell(\sigma)}(t, \tau, \tau; m)}{\Lambda_{\ell(\sigma)}(t)}.
\end{equation}
The relations between $f_{0r}(t, \tau; m)$ and $f_{0c}(t, \tau; m)$ and between $f_{1r}(t, \tau; m)$ and $f_{1c}(t, \tau; m)$ have a different form than in the case of $\chi_{\ell(\sigma)}(t, a, \tau; m)$, $\phi_{\ell(\sigma)}(t, a; m)$  or $f_{2\ell(\sigma)}(t, \tau; m)$. Instead of (\ref{chi_r_vs_chi_c}) we now have
%
%
\begin{equation}
\label{f0r_m_vs_f0c_m}
f_{0r}(t, \tau; m) = 2^m e^{-\Omega(t)} \frac{\Lambda_c(t)}{2 \Lambda_r(t)}  f_{0c}(t, \tau; m)
\end{equation}
and
\begin{equation}
\label{f1r_m_vs_f1c_m}
f_{1r}(t, \tau; m) = 2^m e^{-\Omega(t)}  \frac{\Lambda_c(t)}{\Lambda_r(t)} f_{1c}(t, \tau; m). 
\end{equation}
Note that $f_{0r}(t, \tau; 0) = 0$. If the maximum number of cell divisions in a population is $M_{\text{max}} = M$, then we must have $f_{1r}(t, \tau; M) = 0$, otherwise there will be newborn cells in a population from the $M+1$-th generation. 

Let $q_{\ell(\sigma)}(t, \mathbf{s}; m)$ be $\chi_{\ell(\sigma)}(t, a, \tau; m)$, $\phi_{\ell(\sigma)}(t, a; m)$, or $f_{i\ell(\sigma)}(t, \tau; m)$ for $i=0, 1, 2$, so $\mathbf{s} = a$, $\mathbf{s} = \tau$ or $\mathbf{s} = (a, \tau)$. We have $q_{\ell(\sigma)}(t, \mathbf{s}) = \sum_m q_{\ell(\sigma)}(t, \mathbf{s}; m)$. Following \cite{garcia2019linking} we define
\begin{equation}
\label{R_q_definition}
R_q(t, m | \mathbf{s}) \equiv \frac{q_{c}(t, \mathbf{s}; m)}{q_{c}(t, \mathbf{s})},
\end{equation}
and
\begin{equation}
\label{S_q_definition}
S_q(t, \mathbf{s}) \equiv \sum_m 2^m  R_q(t, m | \mathbf{s}).
\end{equation}
Now we are ready to express any probability distribution for the batch culture in terms of the corresponding distribution for the mother machine. For $q(t, \mathbf{s}) = \chi(t, a, \tau)$, $\phi(t, a)$ or $f_{2}(t, \tau)$ we get

%
%
\begin{equation}
\label{q_r_vs_q_c_marginal}
q_{r}(t, \mathbf{s}) = e^{-\Omega(t)} S_q(t, \mathbf{s}) q_{c}(t, \mathbf{s}).
\end{equation}
However, for $f_{0}(t, \tau)$ we have:
\begin{equation}
\label{f0_r_vs_f0_c}
f_{0r}(t, \tau) =  \frac{\Lambda_c(t)}{2 \Lambda_r(t)} e^{-\Omega(t)}  S_{f_{0}}(t, \tau) f_{0c}(t, \tau),
\end{equation}
whereas for $f_{1}(t, \tau)$ we get
\begin{equation}
\label{f1_r_vs_f1_c}
f_{1r}(t, \tau) =   \frac{\Lambda_c(t)}{\Lambda_r(t)} e^{-\Omega(t)} S_{f_{1}}(t, \tau) f_{1c}(t, \tau).
\end{equation}
We see that compared to (\ref{q_r_vs_q_c_marginal}) there is an additional factor proportional to $\Lambda_c(t)/\Lambda_r(t)$ in both (\ref{f0_r_vs_f0_c}) and (\ref{f1_r_vs_f1_c}). Now consider two definitions of the fitness landscape 
\begin{equation}
\label{H_q_alpha_definition}
H^{(\alpha)}_{q}(t, \mathbf{s}) = \frac{1}{t} \ln \left[ S_q(t, \mathbf{s}) \right],
\end{equation}
\begin{equation}
\label{H_q_beta_definition}
H^{(\beta)}_{q}(t, \mathbf{s}) = \frac{1}{t} \left\lbrace \Omega(t) + \ln \left[ \frac{q_{r}(t, \mathbf{s})}{q_{c}(t, \mathbf{s})} \right]\right\rbrace.
\end{equation}
$H^{(\alpha)}_{q}(t, \mathbf{s})$ was used in Ref. \cite{garcia2019linking} while $H^{(\beta)}_{q}(t, \mathbf{s})$ was used in Refs.\cite{genthon2020fluctuation, genthon2021universal}. For $q(t, \mathbf{s}) = \chi(t, a, \tau)$, $\phi(t, a)$ and $f_{2}(t, \tau)$ we have $H^{(\alpha)}_{q}(t, \mathbf{s}) = H^{(\beta)}_{q}(t, \mathbf{s}) = H_q(t, \mathbf{s})$, where $H^{(\beta)}_q(t, \mathbf{s})$ is a fitness landscape analyzed in Ref.  \cite{genthon2021universal} and in our section \ref{Fitness_landscapes}. However, for $q = f_0$ and $q = f_1$ we have
\begin{eqnarray}
\label{H_q_alpha_neq_H_q_beta_f0}
H^{(\beta)}_{f_0}(t, \tau) & = & H^{(\alpha)}_{f_0}(t, \tau) + \frac{1}{t} \ln \left[  \frac{\Lambda_c(t)}{2\Lambda_r(t)} \right]
\end{eqnarray}
and
\begin{eqnarray}
\label{H_q_alpha_neq_H_q_beta_f1}
H^{(\beta)}_{f_1}(t, \tau) & = & H^{(\alpha)}_{f_1}(t, \tau) + \frac{1}{t} \ln \left[  \frac{\Lambda_c(t)}{\Lambda_r(t)} \right].
\end{eqnarray}
In general, $H^{(\beta)}_{f_0}(t, \tau) \neq H^{(\alpha)}_{f_0}(t, \tau)$ and $H^{(\beta)}_{f_1}(t, \tau) \neq H^{(\alpha)}_{f_1}(t, \tau)$. As a consequence, to define fitness landscapes of $\tau$, neither $f_{0}(t, \tau)$ nor $f_{1}(t, \tau)$ should be used.


Next, following Ref. \cite{genthon2021universal} and using the non-negativity of the Kullback-Leibler divergence (\ref{KL_divergence_definition}), one can show that in the case of the Lebowitz-Rubinow model we have 
\begin{equation}
\left\langle  \ln \left[ S_q(t, \mathbf{s}) \right] \right \rangle_{q_c} \leq  \Omega(t) \leq \left\langle  \ln \left[ S_q(t, \mathbf{s}) \right] \right \rangle_{q_r}
\label{KL_D_inequality_q_r_vs_q_c_} 
\end{equation}
for $q(t, \mathbf{s}) = \chi(t, a, \tau)$, $\phi(t, a)$ or $f_{2}(t, \tau)$, where $\langle (\ldots)\rangle_{q_r} = \int (\ldots) q_r(t, \mathbf{s}) d \mathbf{s}$ and similarly for  $\langle (\ldots)\rangle_{q_c}$. But this is no longer the case for $f_{0}(t, \tau)$ and $f_{1}(t, \tau)$, for which we get
\begin{equation}
\left\langle  \ln \left[ S_{f_{1}}(t, \tau) \right] \right \rangle_{1c} \leq  \Omega(t) -  \ln \left[ \frac{\Lambda_c(t)}{\Lambda_r(t)} \right] \leq \left\langle  \ln \left[ S_{f_{1}}(t, \tau) \right] \right \rangle_{1r},
\label{KL_D_inequality_f1_r_vs_f1_c} 
\end{equation}
\begin{equation}
\left\langle  \ln \left[ S_{f_{0}}(t, \tau) \right] \right \rangle_{0c} \leq  \Omega(t) -  \ln \left[ \frac{\Lambda_c(t)}{2\Lambda_r(t)} \right] \leq \left\langle  \ln \left[ S_{f_{0}}(t, \tau) \right] \right \rangle_{0r}.
\label{KL_D_inequality_f0_r_vs_f0_c} 
\end{equation}
As for the numerical values of the expressions appearing in the double inequalities (\ref{KL_D_inequality_f1_r_vs_f1_c}) and (\ref{KL_D_inequality_f0_r_vs_f0_c}), note that the summation in the definition of  (\ref{S_q_definition}) is taken over different sets of values for $S_{f_{1}}(t, \tau)$ and $S_{f_{0}}(t, \tau)$. In the former case we have $0 \leq m\leq M-1$, while in the latter we have $1 \leq m\leq M$.

\paragraph*{Relationship between $\Lambda_c(t)$ and $\Lambda_r(t)$. Determination of $\Lambda_r(t)$ using lineage statistics.}

The equations (\ref{f0r_m_vs_f0c_m}) and (\ref{f1r_m_vs_f1c_m}) can be used to determine $\Lambda_r(t)$ using only chronological statistics, i.e., the quantities  
with $\sigma =0$, $\ell(\sigma) =c$. Integrating both sides of (\ref{f1r_m_vs_f1c_m}) with respect to $\tau$, summing over $m$ from $m=0$ to $m=M-1$, and shifting all terms with $\sigma=1$, $\ell(\sigma) = r$ to the left, we get
\begin{equation}
\label{f1r_m_vs_f1c_m_int_sum}
\Lambda_r(t) e^{\Omega(t)} =  \Lambda_c(t)  \sum_{m=0}^{M-1} 2^m  \eta_{1c}(t,  m). 
\end{equation}
where we define
\begin{equation}
\label{eta_1_PDF_definition}
 \eta_{1 \ell(\sigma)}(t,  m) = \int_{\tau_s}^{\tau_l} f_{1  \ell(\sigma)}(t, \tau; m) d \tau.  
\end{equation}
Likewise, (\ref{f0r_m_vs_f0c_m}) yields
\begin{equation}
\label{f0r_m_vs_f0c_m_int_sum}
\Lambda_r(t) e^{\Omega(t)} =  \frac{1}{2} \Lambda_c(t)   \sum_{m=1}^{M} 2^m  \eta_{0c}(t,  m),
\end{equation}
where 
\begin{equation}
\label{eta_0_PDF_definition}
 \eta_{0 \ell(\sigma)}(t,  m) = \int_{\tau_s}^{\tau_l} f_{0  \ell(\sigma)}(t, \tau; m) d \tau.  
\end{equation}
But (\ref{f1r_m_vs_f1c_m_int_sum}) and (\ref{f0r_m_vs_f0c_m_int_sum}) are actually the same equation. This is because we have 
\begin{equation}
f_0(t, \tau; m+1) \equiv  \int_{\tau_s}^{\tau_l} h(\tau |\tau^{\prime}) f_1(t, \tau^{\prime}; m) d \tau^{\prime}, 
\label{f_0_definition_2_m}
\end{equation}%
and by integrating both sides of (\ref{f_0_definition_2_m})  with respect to  $\tau$ we get
\begin{equation}
\label{f1r_m_vs_f1c_m_integrated}
 \eta_{0 \ell(\sigma)}(t,  m+1) = \eta_{1 \ell(\sigma)}(t,  m).  
\end{equation}
Equation (\ref{f1r_m_vs_f1c_m_int_sum}) should be compared with Eq. (5) or Ref. \cite{genthon2020fluctuation}, which in our notation reads 
\begin{equation}
\label{f2r_m_vs_f2c_m_int_sum}
e^{\Omega(t)} =  \sum_{m=0}^{M} 2^m  \vartheta_{c}(t,  m), 
\end{equation}
and which can be derived from (\ref{chi_r_vs_chi_c}) by integrating with respect to $a$ and $\tau$ and summing over $m$ from $m=0$ to $m=M$. The $\vartheta_{\ell(\sigma)}(t, m)$ that appears in (\ref{f2r_m_vs_f2c_m_int_sum}) is defined by (\ref{vartheta_PDF_definition}). In general, $\vartheta_{\ell(\sigma)}(t, m) \neq \eta_{1 \ell(\sigma)}(t,  m)$.

Comparing (\ref{f2r_m_vs_f2c_m_int_sum}) to (\ref{f1r_m_vs_f1c_m_int_sum}), we see that the latter formula uses only the statistics of the dividing cells, while the former takes into account all cells in the population. This is a weakness of the formula (\ref{f1r_m_vs_f1c_m_int_sum}), because at any time $t$ between $t$ and $t + \Delta t$, the number of dividing cells is much smaller than the number of all cells. Nevertheless, if one wants to express the instantaneous population growth rate $\Lambda_r(t)$ only in terms of chronological statistics, (\ref{f1r_m_vs_f1c_m_int_sum}) is an alternative to (\ref{f2r_m_vs_f2c_m_int_sum}). Note that both $\Lambda_r(t)$ and $\Lambda_c(t)$ are observable, both being proportional to the number of cells born at the observation time  $t$.

We can also combine (\ref{f2r_m_vs_f2c_m_int_sum}) and (\ref{f1r_m_vs_f1c_m_int_sum}) to get
\begin{equation}
\label{f1r_m_vs_f1c_m_int_sum_div_f2r_m_vs_f2c_m_int_sum}
\Lambda_r(t) =  \Lambda_c(t) \frac{ \sum_{m=0}^{M-1} 2^m  \eta_{1c}(t,  m)}{\sum_{m=0}^{M} 2^m  \vartheta_{c}(t,  m)}. 
\end{equation}
It may be more convenient (or more accurate) to find $\Lambda_r(t)$ using (\ref{f1r_m_vs_f1c_m_int_sum_div_f2r_m_vs_f2c_m_int_sum}) than using (\ref{f2r_m_vs_f2c_m_int_sum}) or (\ref{f1r_m_vs_f1c_m_int_sum}) alone. In addition, the equation (\ref{f1r_m_vs_f1c_m_int_sum_div_f2r_m_vs_f2c_m_int_sum}) gives a relationship between $\Lambda_r(t)$ and $\Lambda_c(t)$.

\section{Existence and uniqueness of the solution to the renewal equation (\ref{solution_for_free_propagation_t_a_tau_self_consistent_a_la_RR}) \label{Proof_of_uniqueness}} 

Here we give a simple proof that the solution to the Lebowitz-Rubinow model is unique, provided that we exclude the `pathological' forms of the distribution of inherited generation times $h(t, \tau | \tau^{\prime})$. (Here we assume that $h(t, \tau | \tau^{\prime})$ can explicitly depend on time.) The proof breaks down for $h(t, \tau | \tau^{\prime})$ containing the part proportional to the Dirac delta function, $h(t, \tau | \tau^{\prime}) \sim \delta(\tau - \tau^{\prime})$, for example for $h(t, \tau | \tau^{\prime}) = \beta \delta(\tau- \tau^{\prime}) + (1- \beta) f(\tau)$ studied in \cite{lebowitz1974theory}, if $\beta \neq 0$. In such cases the uniqueness of the solution is not guaranteed.

Following refs. \cite{rudnicki2014modele, pichor2021cell}, we define the following norm in the space of continuous real functions of two variables $\Psi : [0, T] \times [\tau_s, \tau_l] \to \mathbb{R}$: 
\begin{eqnarray}
\Big \Vert \Psi(t, \tau) \Big \Vert_{\omega} = \max \left\lbrace e^{-\omega t}|\Psi(t, \tau)|: t \in [0, T], \tau \in [\tau_s, \tau_l] \right\rbrace, \nonumber \\
\label{Norm_omega_or_lambda_or_rielecki}
\end{eqnarray}
where 
\begin{eqnarray}
\label{definition_of_omega_proof}
\omega = 2 \max \left\lbrace 2^{\sigma} h(t, \tau | \tau^{\prime}): t \in [0, T]; \tau, \tau^{\prime} \in [\tau_s, \tau_l] \right\rbrace. 
\end{eqnarray}
Using the renewal equation (\ref{solution_for_free_propagation_t_a_tau_self_consistent_a_la_RR}) for a given initial condition $\Phi(t,\tau)$, we define
\begin{eqnarray}
\label{solution_for_free_propagation_t_a_tau_self_consistent_a_la_RR_opeartor_S}
S[\Psi(t, \tau)] &=& 2^{\sigma} \Theta(t-\tau_s) \int_{\tau_s}^{\overline{t} } h(t, \tau | \xi) \Psi(t-\xi, \xi) d \xi \nonumber \\ &+& 2^{\sigma} \Theta(\tau_l - t) \int_{\underline{t}}^{\tau_l} h(t, \tau | \xi) \Phi(\xi - t, \xi) d \xi. \nonumber \\ 
\end{eqnarray}
The uniqueness of the solution to (\ref{solution_for_free_propagation_t_a_tau_self_consistent_a_la_RR}) may be proved by invoking the Banach contraction principle applied to the operation $S$ defined by (\ref{solution_for_free_propagation_t_a_tau_self_consistent_a_la_RR_opeartor_S}). Namely, we will show that for two solutions of (\ref{solution_for_free_propagation_t_a_tau_self_consistent_a_la_RR}), $\Psi(t,\tau)$ and $\Psi^{\prime}(t,\tau)$ with the same initial condition ($\Phi(t,\tau) = \Phi^{\prime}(t,\tau)$), we have
%
%
\begin{eqnarray}
\Vert S\Psi - S\Psi^{\prime} \Vert_{\omega} &=& \max \left\lbrace e^{-\omega t} \left | S\Psi(t, \tau)- S\Psi^{\prime}(t, \tau) \right |: t \in [0, T], \tau \in [\tau_s, \tau_l] \right\rbrace \leq \frac{1}{2}\Vert \Psi - \Psi^{\prime} \Vert_{\omega}. 
\label{proof_ranach_1}
\end{eqnarray}
Indeed, 
\begin{eqnarray}
\left | S\Psi(t, \tau)- S\Psi^{\prime}(t, \tau) \right | &=& \left | \int_{\tau_s}^{\overline{t}} 2^{\sigma} h(t, \tau | \xi) \left[ \Psi(t-\xi, \xi) - \Psi^{\prime}(t-\xi, \xi) \right] d \xi \right | \nonumber \\ & \leq & \int_{\tau_s}^{\overline{t}} \left | 2^{\sigma} h(t, \tau | \xi) \right | \left | \Psi(t-\xi, \xi) - \Psi^{\prime}(t-\xi, \xi) \right | d \xi \nonumber \\ & \leq & \frac{\omega}{2} \int_{\tau_s}^{\overline{t}} \left | \Psi(t-\xi, \xi) - \Psi^{\prime}(t-\xi, \xi) \right | d \xi.
\label{proof_ranach_2}
\end{eqnarray}
and consequently
\begin{eqnarray}
\Vert S\Psi - S\Psi^{\prime} \Vert_{\omega} &\leq & \max \left\lbrace \frac{\omega e^{-\omega t}}{2} \int_{\tau_s}^{\overline{t}} \left | \Psi(t-\xi, \xi) - \Psi^{\prime}(t-\xi, \xi) \right | d \xi: t \in [0, T], \tau \in [\tau_s, \tau_l] \right\rbrace \nonumber \\ & \leq & \frac{\omega}{2} \max \left\lbrace e^{-\omega t} \int_{\tau_s}^{\overline{t}} e^{\omega(t-\xi)} \Vert \Psi - \Psi^{\prime} \Vert_{\omega} d \xi: t \in [0, T], \tau \in [\tau_s, \tau_l] \right\rbrace \nonumber \\ & \leq & \frac{\omega}{2} \Vert \Psi - \Psi^{\prime} \Vert_{\omega} \int_0^\infty e^{-\omega \xi} d\xi = \frac{1}{2}\Vert \Psi - \Psi^{\prime} \Vert_{\omega}. 
\label{proof_ranach_3}
\end{eqnarray}
%

In the transition from the second to the third line of (\ref{proof_ranach_2}) we used (\ref{definition_of_omega_proof}), whereas in (\ref{proof_ranach_3}) we used the following inequality: 
\begin{equation}
\forall \in [0, T], \tau \in [\tau_s, \tau_l] : \left | \Psi(\tilde{t}, \tau) \right | \leq \exp(\omega \tilde{t}) \Vert \Psi \Vert_{\omega},
\end{equation}
which is a consequence of the definition (\ref{Norm_omega_or_lambda_or_rielecki}) of $\Vert \Psi \Vert_{\omega}$.

\section{More general description of the generation time inheritance \label{correlations_and_inheritance}} 

In the Lebowitz-Rubinow model (\ref{APBE_1})--(\ref{APBE_3}), generation-time inheritance and mother-daughter generation-time correlations are described by the parameterized probability distribution $h(\tau | \tau^{\prime})$. This quantity appears in Eq. (\ref{APBE_2}): $h(\tau | \tau^{\prime}) d\tau$ is the probability that the generation time of both daughter cells is $\tau$, given that the generation time of their mother was $\tau^{\prime}$.

Thus, it is implicitly assumed that each of the two daughter cells (denoted $+$ and $-$ from now on) inherits the same generation time upon division: $\tau_{+} = \tau_{-} = \tau$. Moreover, the common value of $\tau$ inherited by both daughters is assumed to depend only on the generation time $\tau^{\prime}$ of their mother, but not on the generation times of their more distant ancestors.

Both assumptions can be relaxed. In this appendix, we propose a more general model that explicitly takes into account generation-time correlations between sisters, as well as between the cell of interest and the cells of $G \geq 1$ previous generations in the lineage. We show that, under certain simplifying assumptions, such a model can be reduced to the effective model of the form analyzed in the main text.

\subsection{Elimination of cell sister generation time \label{Sisters}}

\subsubsection{Heuristic justification}

The simplifying assumption $\tau_{+} = \tau_{-} = \tau$ can be justified by the following argument. First, consider a general situation where one of the sister cells inherits the generation time $\tau_{+}$ upon cell division, and the other inherits ${\tau}_{-}$ (${\tau}_{+}$ need not equal $\tau_{-}$), provided that their mother's generation time was $\tau^{\prime}$. Such an event is denoted by $(\tau_{+}, \tau_{-}|\tau^{\prime})$. Let the probability of $(\tau_{+}, \tau_{-}|\tau^{\prime})$ be $\mathcal{P}(\tau_{+}, \tau_{-}|\tau^{\prime})$ and assume that it is equal to the probability of the situation where $\tau_{+}$ and $\tau_{-}$ are swapped between the daughters: $\mathcal{P}(\tau_{+}, \tau_{-}|\tau^{\prime}) =\mathcal{P}(\tau_{-}, \tau_{+}|\tau^{\prime})$. In such a case, the average number of $(\tau_{+}, \tau_{-}|\tau^{\prime})$ cell divisions between $t$ and $t+dt$ is equal to the number of $(\tau_{-}, \tau_{+}|\tau^{\prime})$ divisions. Now we can cut and rearrange the lineage tree so that instead of the two `asymmetric' divisions: $(\tau_{+}, \tau_{-}|\tau^{\prime})$ and $(\tau_{-}, \tau_{+}|\tau^{\prime})$, we have two `symmetric' ones, $(\tau_{+}, \tau_{+}|\tau^{\prime})$ and $(\tau_{-}, \tau_{-}|\tau^{\prime})$. Such a rearrangement changes the generation time correlation between the sister cells, but does not affect the number of cells born with a given value of generation time $\tau$ at observation time $t$. Thus, both $\Lambda(t)$ and $n(t, 0, \tau)= \Psi(t, \tau)$ remain unchanged. As long as we are not interested in the generation time correlation between sisters, all model predictions remain the same after this ``reshuffling'' of the lineage tree. However, both daughters of a given mother now inherit the same generation time.

\subsubsection{Formal justification}

Now we will try to make the arguments given above a little more rigorous. First, consider the case of a population in a batch culture. At each cell division, we can distinguish between the old-pole and the new-pole daughter cells. The former will be called `red' and marked with a plus ($+$); the latter will be called `blue' and marked with a minus ($-$). It has been shown experimentally that the new pole cells in \textit{E. coli} grow faster than the old pole cells \cite{stewart2005aging}. However, we ignore the effects of aging here: We assume that it is just as likely that one daughter inherits the time of generation $\tau_{+}$ and the other $\tau_{-}$ as it is that $\tau_{+}$ and $\tau_{-}$ are interchanged.

Furthermore, we assume that each cell is characterized not only by its age $a$ and the inherited generation time $\tau$, but also by the generation time of its sister, $\tilde{\tau}$, and that of its mother ($\tau_1$), grandmother ($\tau_2$), and by the generation times of the more distant ancestors: $\tau_3$, $\ldots$, $\tau_G$. Therefore, instead of $n(t, a, \tau)$, e.g, in (\ref{APBE_1}) and (\ref{APBE_2}), we now have to introduce the following cell number densities: $n_{+}(t, a, \tau, \tilde{\tau}, \tau_1, \ldots \tau_G)$ for the `red' cells and $n_{-}(t, a, \tau, \tilde{\tau}, \tau_1, \ldots \tau_G)$ for the `blue' ones. We have
\begin{equation}
n_{+}(t, 0, \tau, \tilde{\tau}, \vec{\tau}) = n_{-}(t, 0, \tilde{\tau}, \tau, \vec{\tau})
\label{n_plus_swap_n_minus}
\end{equation}
where $\vec{\tau} = (\tau_1, \tau_2, \ldots \tau_G)$. We also define 
\begin{equation}
n_s(t, a, \tau, \tilde{\tau}, \vec{\tau}) = n_{+}(t, a, \tau, \tilde{\tau}, \vec{\tau}) + n_{-}(t, a, \tau, \tilde{\tau}, \vec{\tau})
\label{n_plus_plus_n_minus_equals_n_s}
\end{equation}
and 
\begin{equation}
n(t, a, \tau, \vec{\tau}) \equiv \int_{\tau_s}^{\tau_l} n_s(t, a, \tau, \tilde{\tau}, \vec{\tau}) d \tilde{\tau}.
\label{n_s_integrated_equals_n}
\end{equation}
The time-evolution equation for both $n_{+}(t, a, \tau, \tilde{\tau}, \vec{\tau})$ and $n_{-}(t, a, \tau, \tilde{\tau}, \vec{\tau})$ is identical to (\ref{APBE_1}), i.e, 
\begin{eqnarray}
\label{APBE_GC_Sisters_plus}
\frac{\partial n_{+}(t, a, \tau, \tilde{\tau},\vec{\tau})}{\partial t} + \frac{\partial n_{+}(t, a, \tau, \tilde{\tau},\vec{\tau})}{\partial a} = 0, 
\end{eqnarray}
\begin{eqnarray}
\label{APBE_GC_Sisters_minus}
\frac{\partial n_{-}(t, a, \tau, \tilde{\tau}, \vec{\tau})}{\partial t} + \frac{\partial n_{-}(t, a, \tau, \tilde{\tau}, \vec{\tau})}{\partial a} = 0. 
\end{eqnarray}
From (\ref{n_plus_plus_n_minus_equals_n_s}) and (\ref{n_s_integrated_equals_n}) it also follows that both $n_{s}(t, a, \tau, \tilde{\tau}, \vec{\tau})$ and $n(t, a, \tau, \vec{\tau})$ obey (\ref{APBE_1}), too:  
\begin{eqnarray}
\label{APBE_GC_Sisters_n_s}
\frac{\partial n_{s}(t, a, \tau, \tilde{\tau}, \vec{\tau})}{\partial t} + \frac{\partial n_{s}(t, a, \tau, \tilde{\tau}, \vec{\tau})}{\partial a}= 0,
\end{eqnarray}
\begin{eqnarray}
\label{APBE_GC_Sisters_n}
\frac{\partial n(t, a, \tau, \vec{\tau})}{\partial t} + \frac{\partial n(t, a, \tau, \vec{\tau})}{\partial a}= 0. 
\end{eqnarray}
(The operator ${\partial}/{\partial t} + {\partial}/{\partial a}$ does not depend on $\tilde{\tau}$ or the components $\tau_1, \tau_2, \tau_3$, $\ldots$, $\tau_G$ of $\vec{\tau}$).

We should add the appropriate initial conditions to the time evolution equations (\ref{APBE_GC_Sisters_plus})-(\ref{APBE_GC_Sisters_n}): 
\begin{eqnarray}
\label{APBE_GC_Sisters_initial_condition_plus}
\Phi_{+}(a, \tau, \tilde{\tau},\vec{\tau}) = n_{+}(0, a, \tau, \tilde{\tau},\vec{\tau}) \\
\label{APBE_GC_Sisters_initial_condition_minus}
\Phi_{-}(a, \tau, \tilde{\tau},\vec{\tau}) = n_{-}(0, a, \tau, \tilde{\tau},\vec{\tau}) \\
\label{APBE_GC_Sisters_initial_condition_s}
\Phi_{s}(a, \tau, \tilde{\tau},\vec{\tau}) = \Phi_{+}(a, \tau, \tilde{\tau},\vec{\tau}) + \Phi_{-}(a, \tau, \tilde{\tau},\vec{\tau}) \\
\label{APBE_GC_Sisters_initial_condition}
\Phi(a, \tau, \vec{\tau}) = \int_{\tau_s}^{\tau_l} \Phi_{s}(a, \tau, \tilde{\tau};\vec{\tau}) d \tilde{\tau}.
\end{eqnarray}

Now consider the boundary condition, i.e. the influx of newborn cells due to cell division. Let us focus on the `red' newborns. Such cells can be daughters of either `red' or `blue' mothers. The probability of these two situations is proportional to $h_{++}(\tau, \tilde{\tau} | \vec{\tau})$ and $h_{+-}(\tau, \tilde{\tau} | \vec{\tau})$, respectively. $h_{++}(\tau, \tilde{\tau} | \vec{\tau}) d\tau d \tilde{\tau}$ is the probability that a `red' cell inherits the generation time $\tau$ and its `blue' sister inherits the generation time $\tilde{\tau}$, provided that their mother is `red' and the generation times of $G$ consecutive common ancestors of the two daughters are the components of $\vec{\tau}$. $h_{+-}(\tau, \tilde{\tau} | \vec{\tau})$ has an analogous interpretation. So we have
%
\begin{eqnarray}
\label{APBE_GC_2_Sisters_plus}
n_{+}(t, 0, \tau, \tilde{\tau}, \vec{\tau}) &=& \int_{\tau_s}^{\tau_l} \int_{\tau_s}^{\tau_l} h_{++}(\tau, \tilde{\tau} | \vec{\tau}) n_{+}(t, \tau_1, \tau_1, \tilde{\tau}_1, \vec{\tau}^{\prime}) d\tilde{\tau}_1 d \tau_{G+1} \nonumber \\ &+& \int_{\tau_s}^{\tau_l} \int_{\tau_s}^{\tau_l} h_{+-}(\tau, \tilde{\tau} | \vec{\tau}) n_{-}(t, \tau_1, \tau_1, \tilde{\tau}_1, \vec{\tau}^{\prime}) d\tilde{\tau}_1 d \tau_{G+1}. 
\end{eqnarray}
%
%
In the above, $\vec{\tau}^{\prime} = (\tau_2, \tau_3, \ldots \tau_{G+1})$, {i.e., the primed quantities refer to mothers.} We have assumed here that the generation time $\tau$ inherited by the `red' cell depends on the generation times of its mother, grandmother, etc., up to the $G$-th generation, but it does not depend on $\tau_{G+1}$, nor on the generation times of the ancestors' siblings. %

Now we make another simplifying assumption: The `red' cell can be equally likely a daughter of a `red' or a `blue' mother,
\begin{equation}
\label{simplification_of_h_plus_1_symmetry}
h_{++}(\tau, \tilde{\tau} | \vec{\tau}) = h_{+-}(\tau, \tilde{\tau} | \vec{\tau}) = h_{+}(\tau, \tilde{\tau} | \vec{\tau})
\end{equation}
Using (\ref{n_plus_plus_n_minus_equals_n_s}), (\ref{n_s_integrated_equals_n}) and (\ref{simplification_of_h_plus_1_symmetry}), we rewrite (\ref{APBE_GC_2_Sisters_plus}) as 
\begin{eqnarray}
\label{APBE_GC_2_Sisters_plus_integrated}
n_{+}(t, 0, \tau, \tilde{\tau}, \vec{\tau}) &=& \int_{\tau_s}^{\tau_l} h_{+}(\tau, \tilde{\tau} | \vec{\tau}) n(t, \tau_1, \tau_1, \vec{\tau}^{\prime}) d \tau_{G+1}. \nonumber \\ 
\end{eqnarray}
%
%
For the ``blue'' cells, we have analogous equation:
\begin{eqnarray}
\label{APBE_GC_2_Sisters_minus_integrated}
n_{-}(t, 0, \tau, \tilde{\tau}, \vec{\tau}) &=& \int_{\tau_s}^{\tau_l} h_{-}(\tau, \tilde{\tau} | \vec{\tau}) n(t, \tau_1, \tau_1, \vec{\tau}^{\prime}) d \tau_{G+1}, \nonumber \\ 
\end{eqnarray}
where $h_{-}(\tau, \tilde{\tau} | \vec{\tau}) = h_{+-}(\tau, \tilde{\tau} | \vec{\tau}) = h_{--}(\tau, \tilde{\tau} | \vec{\tau})$, analogous to the case of the 'red' cells. We also have
\begin{equation}
\label{swap_relations_between_h_plus_h_minus}
h_{+}(\tau, \tilde{\tau} | \vec{\tau}) = h_{-}(\tilde{\tau}, \tau | \vec{\tau}).
\end{equation}
Next, we make another assumption about the symmetry between the `blue' and `red' cells: The probability that a `red' cell inherits the generation time ${\tau}$ and its `blue' sister inherits the generation time $\tilde{\tau}$ is equal to the probability of the situation where the generation time values are swapped between the sisters, 
\begin{equation}
\label{simplification_of_h_plus}
h_{+}(\tau, \tilde{\tau} | \vec{\tau}) = h_{-}(\tau, \tilde{\tau} | \vec{\tau}) \equiv h(\tau, \tilde{\tau} | \vec{\tau}).
\end{equation}
Instead of defining $h(\tau, \tilde{\tau} | \vec{\tau})$ as in (\ref{simplification_of_h_plus}), we can add (\ref{APBE_GC_2_Sisters_plus_integrated}) and (\ref{APBE_GC_2_Sisters_minus_integrated}) to get
\begin{eqnarray}
\label{APBE_GC_2_Sisters_s_integrated}
n_{s}(t, 0, \tau, \tilde{\tau}, \vec{\tau}) &=& 2 \int_{\tau_s}^{\tau_l} h(\tau, \tilde{\tau} | \vec{\tau}) n(t, \tau_1, \tau_1, \vec{\tau}^{\prime}) d \tau_{G+1}, \nonumber 
\end{eqnarray}
where
\begin{equation}
\label{simplification_of_h_plus_symmetric}
h(\tau, \tilde{\tau} | \vec{\tau}) \equiv \frac{h_{+}(\tau, \tilde{\tau} | \vec{\tau}) + h_{-}(\tau, \tilde{\tau} | \vec{\tau})}{2}.
\end{equation}
The intuitive interpretation of (\ref{simplification_of_h_plus_symmetric}) is as follows: One of the cells inherits the generation time $\tau$ upon division, but we have no information whether this is a red or a blue cell. Now $\tilde{\tau}$ can be integrated out, and we finally get 
\begin{eqnarray}
\label{APBE_GC_2_Sisters_n_integrated_again}
n(t, 0, \tau, \vec{\tau}) &=& 2 \int_{\tau_s}^{\tau_l} h(\tau | \vec{\tau}) n(t, \tau_1, \tau_1, \vec{\tau}^{\prime}) d \tau_{G+1}, \nonumber \\ 
\end{eqnarray}
where
\begin{equation}
\label{simplification_of_h_bis_marginalisation}
h(\tau | \vec{\tau}) \equiv \int_{\tau_s}^{\tau_l} h(\tau, \tilde{\tau} | \vec{\tau}) d \tilde{\tau}.
\end{equation}
So far we have considered the batch culture. To get analogous results for the mother machine, we have to consider only the 'red' cells and ignore the 'blue' ones. As a result, the factor 2 in the boundary condition disappears. 

\subsection{Elimination of generation times of the cell's grandmother and more distant ancestors \label{Grandmas}}

In the previous subsection, we showed how to eliminate the generation time of the cell's sister from the model description, provided that certain simplifying conditions are satisfied. We derived the equations describing the time evolution of the cell density $n(t, a, \tau, \vec{\tau})$, where $\vec{\tau} = (\tau_1, \tau_2, \ldots \tau_G)$ and $\tau_1$ is the generation time of the mother, $\tau_2$ is the generation time of the grandmother, and so on.

Now our task is to keep the dependence of the inherited generation time distribution $h(\tau | \ldots)$ on the mother's generation time $\tau_1$, but to get rid of the generation times of more distant ancestors. We are going to obtain the equations (\ref{APBE_1})--(\ref{APBE_3}) of the Lebowitz-Rubinow model, i.e. the time evolution equation with initial and boundary conditions for the cell number density, 
\begin{equation}
\label{GC_Consistency_1}
n(t, a, \tau) = \int_{\tau_s}^{\tau_l} \ldots \int_{\tau_s}^{\tau_l} n(t, a, \tau, \vec{\tau}) d {\tau_1} \ldots d {\tau_G}. 
\end{equation}
Note that we no longer require that $a <\tau_i$ for $i = 1, 2, \ldots, G$. 

Our starting point for the following analysis are now the equations (\ref{APBE_GC_Sisters_n}), (\ref{APBE_GC_Sisters_initial_condition}), and (\ref{APBE_GC_2_Sisters_n_integrated_again}). We also define 
\begin{eqnarray} 
\label{APBE_GC_3_Sisters}
 n(t, 0, \tau, \vec{\tau}) &=& \Psi(t, \tau, \vec{\tau}).
\end{eqnarray}
The inherited value of the generation time $\tau=\tau_0$ depends on $\tau_i$, $i= 1, 2, \ldots, G$ as described by $h(\tau | \vec{\tau})$ (\ref{simplification_of_h_bis_marginalisation}). We assume that the correlations between more distant generations ($i > G$) are vanishing and that the environmental conditions are constant. 

Integrating (\ref{APBE_GC_Sisters_n}) with respect to the components of $\vec{\tau}$, we obtain (\ref{APBE_1}). Similarly, the initial conditions (\ref{APBE_GC_Sisters_initial_condition}) reduce to (\ref{APBE_3}). However, to obtain the boundary condition (\ref{APBE_2}) from (\ref{APBE_GC_2_Sisters_n_integrated_again}), we have to define an effective distribution of the inherited generation times:
\begin{eqnarray}
\label{effective_h_MD}
h_e(t, \tau | \tau_1) &\equiv & \frac{\int_{\tau_s}^{\tau_l} \ldots \int_{\tau_s}^{\tau_l} h(\tau | \tau_1, \tau_2, \ldots, \tau_G) n(t, \tau_1, \tau_1, \tau_2, \ldots, \tau_G, \tau_{G+1})d \tau_2 \ldots d {\tau_G} d \tau_{G+1}}{\int_{\tau_s}^{\tau_l} \ldots \int_{\tau_s}^{\tau_l} n(t, \tau_1, \tau_1, \tau_2, \ldots, \tau_G, \tau_{G+1})d \tau_2 \ldots d {\tau_G} d \tau_{G+1}} \nonumber \\ &=& \frac{\int_{\tau_s}^{\tau_l} \ldots \int_{\tau_s}^{\tau_l} h(\tau | \tau_1, \tau_2, \ldots, \tau_G) f_1(t, \tau_1, \tau_2, \ldots, \tau_G, \tau_{G+1})d \tau_2 \ldots d {\tau_G} d \tau_{G+1}}{\int_{\tau_s}^{\tau_l} \ldots \int_{\tau_s}^{\tau_l} f_1(t, \tau_1, \tau_2, \ldots, \tau_G, \tau_{G+1})d \tau_2 \ldots d {\tau_G} d \tau_{G+1}} \nonumber \\ 
& = & \frac{\mathcal{P}(t, \tau, \tau_1)}{\mathcal{P}(t, \tau_1)} = \mathcal{P}(t, \tau | \tau_1),
\end{eqnarray}
where 
\begin{equation}
\label{generation_time_for_mothers_PDF_GC}
f_1(t, \tau_1, \tau_2, \ldots, \tau_G, \tau_{G+1}) = \frac{n(t, \tau_1, \tau_1, \tau_2, \ldots, \tau_G, \tau_{G+1})}{N(t)\Lambda(t)},
\end{equation}
is a generalization of the mother age distribution, $f_1(t, \tau)$, considered in the main text. We have 
\begin{eqnarray}
\label{GC_Consistency_1_f_1}
\int_{\tau_s}^{\tau_l} \ldots \int_{\tau_s}^{\tau_l} f_1(t, \tau, \vec{\tau}) d {\tau_1} \ldots d {\tau_G} &=& f_1(t, \tau)
\end{eqnarray}
where $f_1(t, \tau)$ is given by (\ref{carrier_PDF_time_dependent_NA}) and (\ref{t_dependent_f_1_definition}). In (\ref{effective_h_MD}) we have also defined
%
\begin{equation}
\label{Mathcal_P_PDF_GC_marginal_tau_tau_1a}
\mathcal{P}(t, \tau, \tau_1) = \int_{\tau_s}^{\tau_l} \int_{\tau_s}^{\tau_l} \ldots \int_{\tau_s}^{\tau_l} \mathcal{P}(t, \tau, \tau_1, \tau_2, \ldots, \tau_G, \tau_{G+1}) d \tau_2 \ldots d {\tau_G} d {\tau_{G+1}},
\end{equation}
and
\begin{equation}
\label{Mathcal_P_PDF_GC_marginal_tau_tau_1}
\mathcal{P}(t, \tau_1) = \int_{\tau_s}^{\tau_l} \int_{\tau_s}^{\tau_l} \ldots \int_{\tau_s}^{\tau_l} \int_{\tau_s}^{\tau_l} \mathcal{P}(t, \tau, \tau_1, \tau_2, \ldots, \tau_G, \tau_{G+1}) d \tau d \tau_2 \ldots d {\tau_G} d {\tau_{G+1}},
\end{equation}
where
\begin{equation}
\label{Mathcal_P_PDF_GC_joint}
\mathcal{P}(t, \tau, \tau_1, \tau_2, \ldots, \tau_G, \tau_{G+1}) = h(\tau | \tau_1, \tau_2, \ldots, \tau_G) f_1(t, \tau_1, \tau_1, \tau_2, \ldots, \tau_G, \tau_{G+1}).
\end{equation}
%
%
Note that {$\mathcal{P}(t, \tau, \tau_1, \tau_2, \ldots, \tau_G, \tau_{G+1})$ defined above is properly normalized.} Importantly, the time dependence of $\mathcal{P}(t, \tau, \tau_1)$ and $\mathcal{P}(t, \tau_1)$ need not cancel in (\ref{effective_h_MD}) and $h_e(t, \tau | \tau_1)$ can depend on the observation time $t$ even if $h(\tau | \tau_1, \tau_2, \ldots, \tau_G)$ does not.

Finally, invoking (\ref{effective_h_MD}), we integrate (\ref{APBE_GC_2_Sisters_n_integrated_again}) with respect to the components of $\vec{\tau}$ and we arrive at (\ref{APBE_2}):
%
\begin{eqnarray}
\label{GC_Consistency_2}
n(t, 0, \tau) &=& \int_{0}^{\tau_l} \int_{0}^{\tau_l} \ldots \int_{0}^{\tau_l} n(t, 0, \tau, \vec{\tau}) d \tau_1 d \tau_2 \ldots d \tau_{G} = 2^{\sigma} \int_{0}^{\tau_l} \int_{0}^{\tau_l} \ldots \int_{0}^{\tau_l} h(\tau |\tau_1, \vec{\tau}^{\prime}) n(t, \tau_1, \tau_1, \vec{\tau}^{\prime}) d \tau_{G+1} \ldots d \tau_2 d \tau_1 \nonumber \\ &=& 2^{\sigma} \int_{0}^{\infty} h_e(t, \tau | \tau_1) n(t, \tau_1, \tau_1) d \tau_1. 
\end{eqnarray}
In this way we have reduced the model defined by (\ref{APBE_GC_Sisters_plus}), (\ref{APBE_GC_Sisters_minus}) and (\ref{APBE_GC_2_Sisters_plus}) to the simpler one given by (\ref{APBE_1})--(\ref{APBE_3}).

\section{Derivation of the Rubinow model from the Lebowitz-Rubinow model \label{Rubinow}}

Starting from the Lebowitz-Rubinow model (\ref{APBE_1})--(\ref{APBE_3}), we derive here a model that is formally identical to the one proposed by Rubinow in 1968 \cite{rubinow1968maturity}. In the latter model, there is only a single state variable $x\in[0, 1]$, called maturity. $x$ increases with cell age $a$ (or observation time $t$) from $x = 0$ at the beginning of the cell cycle to $x = 1$ at cell division. The time evolution of $x$ for each cell is assumed to be deterministic and is given by the maturation velocity function $\tilde{g}(t, x)$: 
\begin{equation}
\frac{dx}{dt} = \tilde{g}(t, x). 
\label{deterministic_evolution_of_x_origial_RM}
\end{equation}
Let $u(t, x)dx$ be the number of cells with maturity $x$ at time $t$. The equations of the Rubinow model are
\begin{eqnarray}
\label{Rubinow_x_defining_Eqs_1}
\frac{\partial}{\partial t} u(t, x) + \frac{\partial}{\partial x} \left[\tilde{g}(t, x) u(t, x) \right] = 0, \\
\label{Rubinow_x_defining_Eqs_2}
\tilde{g}(t, 0) u(t, 0) = 2^{\sigma} \tilde{g}(t, 1) u(t, 1),\\
\label{Rubinow_x_defining_Eqs_3}
u(0, x) = u_0(x).
\end{eqnarray}
We consider here a slightly different set of model equations than the one originally proposed by Rubinow \cite{rubinow1968maturity} or analyzed in \cite{rudnicki2014modele}. First, we allow $\tilde{g}(t, x)$, which appears in (\ref{deterministic_evolution_of_x_origial_RM}), (\ref{Rubinow_x_defining_Eqs_2}) and (\ref{Rubinow_x_defining_Eqs_1}), to depend on the observation time $t$. Second, in analogy to the case of the Lebowitz-Rubinow model, we introduce the parameter $\sigma$ in the boundary condition (\ref{Rubinow_x_defining_Eqs_2}): $\sigma=1$ for the batch culture and $\sigma=0$ for the mother machine. However, unlike the original model, we ignore cell death \cite{rubinow1968maturity}.


\subsection{The maturity representation of the Lebowitz-Rubinow model}

No unique, precise definition of maturity $x$ was given in ref. \cite{rubinow1968maturity}. In this reference we find the following passage `By level of maturity is meant the various stages in the growth of the cell such as birth, onset of DNA synthesis, onset of mitosis, etc. These may or may not be readily observable. In fact, it is difficult to say in what manner the maturity level of a cell should be determined. For bacterial cells such as \textit{E. coli} in which DNA synthesis continues from the moment of birth, the amount of DNA in the cell could be utilized as a measure of cell maturity. Or $x$ could simply be considered to represent the amount of DNA in the cell. However, for many cells in which DNA synthesis is only a portion of the life cycle, such a measure is not completely satisfactory. Thus, at the present time even the dimensions of $x$ must be left unspecified. Another possibility is to let $x$ represent cell volume.'. In the above quote, we have changed the original notation of maturity from $\mu$ to $x$. Here, our definition of maturity is simply
\begin{equation}
x = \frac{a}{\tau}. 
\label{deterministic_evolution_of_x_a_tau_model_heuristic}
\end{equation}
If, instead of the cell age $a$, we use $x$ as defined in Eq. (\ref{deterministic_evolution_of_x_a_tau_model_heuristic}) as the independent variable of the Lebowitz-Rubinow model, then (\ref{APBE_1}) becomes 
\begin{eqnarray}
\label{Rubinow_APBE_x_tau_1}
\frac{\partial \tilde{n}(t, x, \tau)}{\partial t} &+& \frac{1}{\tau} \frac{\partial \tilde{n}(t, x, \tau)}{\partial x} = 0,
\end{eqnarray}
where 
\begin{equation}
\tilde{n}(t, x, \tau) = n(t, x \tau, \tau) \tau
\label{n_Rubinow_definition_of}
\end{equation} 
and $n(t, a, \tau)$ is the solution of (\ref{APBE_1})--(\ref{APBE_3}). Where there is a risk of confusion, we use tilde to distinguish the quantities of the Rubinow model (maturity representation) from those of the Lebowitz-Rubinow model (age--generation time representation). Next, using (\ref{n_Rubinow_definition_of}), we define 
\begin{equation}
\tilde{\chi}(t, x, \tau) = \frac{\tilde{n}(t, x, \tau)}{N(t)} = \chi(t, x \tau, \tau) \tau,
\label{chi_Rubinow_definition_of}
\end{equation}
where $\chi(t, a, \tau)$ is given by (\ref{chi_definition_of}). From (\ref{Rubinow_APBE_x_tau_1}) and (\ref{chi_Rubinow_definition_of}) we obtain the time evolution equation for $\tilde{\chi}(t, x, \tau)$
\begin{eqnarray}
\label{chi_Rubinow_time_evolution_PDE}
\frac{\partial \tilde{\chi}(t, x, \tau)}{\partial t} &+& \frac{1}{\tau} \frac{\partial \tilde{\chi}(t, x, \tau)}{\partial x} + \sigma \Lambda(t) \tilde{\chi}(t, x, \tau) = 0. \nonumber \\
\end{eqnarray}
We also define
\begin{equation}
u(t, x) = \int_{\tau_{s}}^{\tau_{l}} \tilde{n}(t, x, \tau) d \tau
\label{u_definition_Rubinow}
\end{equation}
and
\begin{equation}
\tilde{\varphi}(t, x) = \frac{u(t, x)}{N(t)} = \int_{\tau_{s}}^{\tau_{l}} \tilde{\chi}(t, x, \tau) d \tau.
\label{varphi_definition_Rubinow}
\end{equation}
Within the Rubinow model, one can show that the total number of cells in the population,
\begin{equation}
N(t) = \int_{0}^{1} \int_{\tau_{s}}^{\tau_{l}} \tilde{n}(t, x, \tau) d \tau d x,
\label{Total_cell_number_Rubinow}
\end{equation}
is given by (\ref{Number_of_cells_solution}), as it should be. Now, by integrating both sides of (\ref{Rubinow_APBE_x_tau_1}) with respect to $\tau$ and using (\ref{chi_Rubinow_definition_of}), (\ref{u_definition_Rubinow}), and (\ref{varphi_definition_Rubinow}), we get (\ref{Rubinow_x_defining_Eqs_1}), provided that
\begin{equation}
\tilde{g}(t, x) u(t, x) = \int_{\tau_{s}}^{\tau_{l}} \frac{\tilde{n}(t, x, \tau)}{\tau} d \tau. 
\label{g_r_effective_definition_1}
\end{equation}
As a result,
\begin{eqnarray}
\tilde{g}(t, x) &=& \int_{\tau_{s}}^{\tau_{l}} \frac{1}{\tau} \frac{\tilde{n}(t, x, \tau)}{u(t, x)} d \tau = \int_{\tau_{s}}^{\tau_{l}} \frac{1}{\tau} \frac{\tilde{\chi}(t, x, \tau)}{\tilde{\varphi}(t, x)} d \tau \nonumber \\ &=& \int_{\tau_{s}}^{\tau_{l}} \tau^{-1} \tilde{\chi}(\tau |x, t) d \tau \equiv \Big \langle \frac{1}{\tau} \Big \rangle. 
\label{g_r_effective_definition_2}
\end{eqnarray}
We can now derive (\ref{Rubinow_x_defining_Eqs_2}) from (\ref{APBE_2}). After changing the variables from $a$ to $x$, Eq. (\ref{APBE_2}) reads
\begin{eqnarray}
\label{APBE_2_in_x_for_Rubinow_model}
\frac{\tilde{n}(t, 0, \tau)}{\tau} &=& 2^{\sigma} \int_{\tau_{s}}^{\tau_{l}} h(\tau | \xi, t) \tilde{n}(t, 1, \xi) \frac{d \xi}{\xi}. 
\end{eqnarray}
Using (\ref{g_r_effective_definition_1}), we actually get (\ref{Rubinow_x_defining_Eqs_2}) from (\ref{u_definition_Rubinow}) and (\ref{APBE_2_in_x_for_Rubinow_model}). Finally, using (\ref{n_Rubinow_definition_of}) and (\ref{varphi_definition_Rubinow}), we obtain the initial condition (\ref{Rubinow_x_defining_Eqs_3}) from the initial condition $n(0, a, \tau) = \Phi(a, \tau)$ (\ref{APBE_3}) of the Lebowitz-Rubinow model. 

Note that although $h(\tau | \xi, t)$, which describes the inheritance of generation times, does not appear explicitly in the Rubinow model, it affects the form of the boundary condition (\ref{Rubinow_x_defining_Eqs_2}). We see that (\ref{Rubinow_x_defining_Eqs_1})--(\ref{Rubinow_x_defining_Eqs_3}) can be regarded as an effective model with $\tilde{g}(t, x)$ equal to $\tau^{-1} $ averaged over all cell cycle lengths for given values of $t$ and $x$.

\subsection{Stationary solution for the Rubinow model}

\subsubsection{Steady exponential growth in batch culture} 

For $\sigma=1$, in the steady-state limit, $\tilde{\chi}(t, x, \tau)=\tilde{\chi}_{r}(x, \tau)$, $\Lambda(t)=\Lambda_r$ and $N(t) = N_0 \exp (\Lambda_r t)$. Invoking (\ref{solution_for_chi_sigma_1}) and changing the variable $a$ to $x$, we obtain
\begin{equation}
\tilde{\chi}_{r}(x, \tau) = 2 \Lambda_r f_{0r}(\tau) \tau \exp\left(-\Lambda_r x \tau\right).
\label{chi_solution_exponential_population_growth_Rubinow}
\end{equation}
Eq. (\ref{chi_solution_exponential_population_growth_Rubinow}) can also be derived directly from (\ref{chi_Rubinow_time_evolution_PDE}), which in this case is: 
\begin{eqnarray}
\label{chi_Rubinow_time_evolution_PDE_ss}
\frac{d \tilde{\chi}_{r}(x, \tau)}{d x} + \tau \Lambda_r \tilde{\chi}_{r}(x, \tau) = 0.
\end{eqnarray}
Integrating (\ref{chi_solution_exponential_population_growth_Rubinow}) with respect to $x$, we obtain
\begin{equation}
\label{f_2_RM}
\int_{0}^{1} \tilde{\chi}_{r}(x, \tau) d x = f_{2r}(\tau) = 2 f_{0r}(\tau)(1 - e^{-\Lambda_r \tau}),
\end{equation}
as expected. For the maturity distribution $ \tilde{\varphi}_{r}(x)$ defined by (\ref{varphi_definition_Rubinow}), we have 
%
%
\begin{eqnarray}
\tilde{\varphi}_{r}(x) &=& \int_{\tau_{s}}^{\tau_{l}} \tilde{\chi}_{r}(x, \tau) d \tau = \int_{\tau_{s}}^{\tau_{l}} 2 \Lambda_r f_{0r}(\tau) \tau e^{-\Lambda_r x \tau} d \tau \nonumber \\ &=& - 2 \Lambda_r \hat{f}^{\prime}_{0r}(z)\vert_{ z = \Lambda_r x},
\label{varphi_exponential_population_growth_Rubinow}
\end{eqnarray}
%
%
where $\hat{f}_{0r}(z)$ is the Laplace transform of $f_{0r}(\tau)$, and $\hat{f}^{\prime}_{0r}(z)$ is the derivative of $\hat{f}_{0r}(z)$ with respect to $z$. (We can extend the integration limits in (\ref{varphi_exponential_population_growth_Rubinow}) by replacing $\tau_{s}$ with $0$ and $\tau_{l}$ with $\infty$, since $f_{0r}(\tau) = 0$ for $\tau < \tau_{s}$ and $\tau > \tau_{l}$). If the generation time distribution of newborns is the gamma distribution, 
\begin{equation}
f_{0r}(\tau) = \frac{ \tau^{\alpha-1} e^{- \tau/\beta} }{\beta^{\alpha}\Gamma(\alpha)}, ~~~ \hat{f}_{0r}(z) = \frac{1}{(1 + \beta z)^{\alpha}},
\label{gamma_distribution_of_generation_times_f_0_alpha_beta}
\end{equation}
then we obtain from (\ref{varphi_exponential_population_growth_Rubinow}):
\begin{equation}
\tilde{\varphi}_{r}(x) = \frac{2 \alpha \beta \Lambda_r}{(1 + \beta \Lambda_r x)^{\alpha + 1}}.
\label{varphi_exponential_population_growth_Rubinow_gamma}
\end{equation}
The normalization condition for $\tilde{\varphi}_{r}(x)$ (\ref{varphi_exponential_population_growth_Rubinow_gamma}) is the Euler-Lotka equation, which now reads 
\begin{equation}
1 = \frac{2}{(1 + \beta \Lambda_r)^{\alpha}}.
\label{Euler_Lotka_ss_Rubinow}
\end{equation}
We rewrite (\ref{varphi_exponential_population_growth_Rubinow_gamma}) using (\ref{Euler_Lotka_ss_Rubinow}) as
\begin{equation}
\tilde{\varphi}_{r}(x) = \frac{2 \alpha (2^{\frac{1}{\alpha}} - 1)}{[1 + (2^{\frac{1}{\alpha}} - 1) x]^{\alpha + 1}}.
\label{varphi_exponential_population_growth_Rubinow_gamma_final}
\end{equation}
From (\ref{g_r_effective_definition_2}), we also find the explicit form of the maturation velocity:
\begin{equation}
\tilde{g}_r(x) = - \frac{\hat{f}_{0r}(z)\vert_{ z = \Lambda_r x}}{\hat{f}^{\prime}_{0r}(z)\vert_{ z = \Lambda_r x}}. 
\label{g_r_effective_exponential_population_growth_general}
\end{equation}
In particular, for $f_{0r}(\tau)$, which is given by (\ref{gamma_distribution_of_generation_times_f_0_alpha_beta}), we get
\begin{equation}
\tilde{g}_r(x) = \frac{(1 + \beta \Lambda_r x)}{\alpha \beta} = \frac{[1 + (2^{\frac{1}{\alpha}} - 1) x]}{\alpha \beta}. 
\label{g_r_effective_exponential_population_growth_gamma}
\end{equation}
We rewrite equation (\ref{g_r_effective_exponential_population_growth_general}) as
\begin{equation}
\frac{\hat{f}^{\prime}_{0r}(z)}{\hat{f}_{0r}(z)} = -\frac{1}{g_{rB}\left(\Lambda_r^{-1} z \right)}, 
\label{g_r_effective_exponential_population_growth_general_bis}
\end{equation}
and therefore
\begin{equation}
\hat{f}_{0r}(\Lambda_r x) = \exp \left( - \int_{0}^{x} \frac{\Lambda_r}{\tilde{g}_r(\tilde{x})} d \tilde{x} \right). 
\label{g_r_effective_exponential_population_growth_general_bis_solution}
\end{equation}\\

\subsubsection{Steady state in the mother machine experiment} 

For the mother machine experiment ($\sigma = 0$) in steady state, Eq. (\ref{chi_Rubinow_time_evolution_PDE}) reduces to $d\tilde{\chi}_{c}(x, \tau)/dx = 0$; hence $\tilde{\chi}_{c}(x, \tau)$ is a constant function of $x$, so it depends in a nontrivial way only on $\tau$. (In this and the following equations, $c$ stands for `chronological'). To find the explicit form of $\tilde{\chi}_{c}(x, \tau)$, it is most convenient to use (\ref{chi_MM}),
\begin{equation}
\chi_c(a, \tau) = \frac{f_{0c}(\tau)}{\int_{0}^{\infty} \tau^{\prime} f_{0c}(\tau^{\prime}) d \tau^{\prime}},
\label{chi_MM_RM}
\end{equation}
and (\ref{chi_Rubinow_definition_of}), from which, after changing the variables, we get
\begin{equation}
\tilde{\chi}_{c}(x, \tau) = \frac{\tau f_{0c}(\tau)}{\int_{0}^{\infty} \tau^{\prime} f_{0c}(\tau^{\prime}) d \tau^{\prime}}.
\label{chi_solution_mother_machine_Rubinow}
\end{equation}
We see that $\tilde{\chi}_{c}(x, \tau)$ is identical to $f_{2c}(\tau)$ given by (\ref{f_2_MM_SS_first}). As a consequence,
\begin{equation}
\tilde{\varphi}_{c}(x) = \int_{\tau_{s}}^{\tau_{l}} \tilde{\chi}_{c}(x, \tau) d \tau = 1.
\label{varphi_mother_machine_Rubinow}
\end{equation}
(More precisely, $\tilde{\varphi}_{c}(x) = \Theta (x)\Theta (1-x)$, where $\Theta (x)$ is the Heaviside step function). Finally, we get the maturation velocity function for the present case: %
\begin{equation}
\tilde{g}_c(x) = \frac{1}{\int_{\tau_{s}}^{\tau_{l}} \tau f_{0c}(\tau) d \tau} = \frac{1}{\langle \tau \rangle_{0c}} =  \Lambda_c. 
\label{g_r_effective_mother_machine_gamma}
\end{equation}

\end{widetext}


\bibliography{bibliography53a}

\begin{thebibliography}{39}%
\makeatletter
\providecommand \@ifxundefined [1]{%
 \@ifx{#1\undefined}
}%
\providecommand \@ifnum [1]{%
 \ifnum #1\expandafter \@firstoftwo
 \else \expandafter \@secondoftwo
 \fi
}%
\providecommand \@ifx [1]{%
 \ifx #1\expandafter \@firstoftwo
 \else \expandafter \@secondoftwo
 \fi
}%
\providecommand \natexlab [1]{#1}%
\providecommand \enquote  [1]{``#1''}%
\providecommand \bibnamefont  [1]{#1}%
\providecommand \bibfnamefont [1]{#1}%
\providecommand \citenamefont [1]{#1}%
\providecommand \href@noop [0]{\@secondoftwo}%
\providecommand \href [0]{\begingroup \@sanitize@url \@href}%
\providecommand \@href[1]{\@@startlink{#1}\@@href}%
\providecommand \@@href[1]{\endgroup#1\@@endlink}%
\providecommand \@sanitize@url [0]{\catcode `\\12\catcode `\$12\catcode
  `\&12\catcode `\#12\catcode `\^12\catcode `\_12\catcode `\%12\relax}%
\providecommand \@@startlink[1]{}%
\providecommand \@@endlink[0]{}%
\providecommand \url  [0]{\begingroup\@sanitize@url \@url }%
\providecommand \@url [1]{\endgroup\@href {#1}{\urlprefix }}%
\providecommand \urlprefix  [0]{URL }%
\providecommand \Eprint [0]{\href }%
\providecommand \doibase [0]{http://dx.doi.org/}%
\providecommand \selectlanguage [0]{\@gobble}%
\providecommand \bibinfo  [0]{\@secondoftwo}%
\providecommand \bibfield  [0]{\@secondoftwo}%
\providecommand \translation [1]{[#1]}%
\providecommand \BibitemOpen [0]{}%
\providecommand \bibitemStop [0]{}%
\providecommand \bibitemNoStop [0]{.\EOS\space}%
\providecommand \EOS [0]{\spacefactor3000\relax}%
\providecommand \BibitemShut  [1]{\csname bibitem#1\endcsname}%
\let\auto@bib@innerbib\@empty
\bibitem [{\citenamefont {Powell}(1956)}]{powell1956growth}%
  \BibitemOpen
  \bibfield  {author} {\bibinfo {author} {\bibfnamefont {E.~O.}\ \bibnamefont
  {Powell}},\ }\href@noop {} {\bibfield  {journal} {\bibinfo  {journal}
  {Journal of General Microbiology}\ }\textbf {\bibinfo {volume} {15}},\
  \bibinfo {pages} {492} (\bibinfo {year} {1956})}\BibitemShut {NoStop}%
\bibitem [{\citenamefont {Powell}(1964)}]{powell1964note}%
  \BibitemOpen
  \bibfield  {author} {\bibinfo {author} {\bibfnamefont {E.}~\bibnamefont
  {Powell}},\ }\href@noop {} {\bibfield  {journal} {\bibinfo  {journal}
  {Microbiology}\ }\textbf {\bibinfo {volume} {37}},\ \bibinfo {pages} {231}
  (\bibinfo {year} {1964})}\BibitemShut {NoStop}%
\bibitem [{\citenamefont {Lebowitz}\ and\ \citenamefont
  {Rubinow}(1974)}]{lebowitz1974theory}%
  \BibitemOpen
  \bibfield  {author} {\bibinfo {author} {\bibfnamefont {J.~L.}\ \bibnamefont
  {Lebowitz}}\ and\ \bibinfo {author} {\bibfnamefont {S.}~\bibnamefont
  {Rubinow}},\ }\href@noop {} {\bibfield  {journal} {\bibinfo  {journal}
  {Journal of Mathematical Biology}\ }\textbf {\bibinfo {volume} {1}},\
  \bibinfo {pages} {17} (\bibinfo {year} {1974})}\BibitemShut {NoStop}%
\bibitem [{\citenamefont {Hashimoto}\ \emph {et~al.}(2016)\citenamefont
  {Hashimoto}, \citenamefont {Nozoe}, \citenamefont {Nakaoka}, \citenamefont
  {Okura}, \citenamefont {Akiyoshi}, \citenamefont {Kaneko}, \citenamefont
  {Kussell},\ and\ \citenamefont {Wakamoto}}]{hashimoto2016noise}%
  \BibitemOpen
  \bibfield  {author} {\bibinfo {author} {\bibfnamefont {M.}~\bibnamefont
  {Hashimoto}}, \bibinfo {author} {\bibfnamefont {T.}~\bibnamefont {Nozoe}},
  \bibinfo {author} {\bibfnamefont {H.}~\bibnamefont {Nakaoka}}, \bibinfo
  {author} {\bibfnamefont {R.}~\bibnamefont {Okura}}, \bibinfo {author}
  {\bibfnamefont {S.}~\bibnamefont {Akiyoshi}}, \bibinfo {author}
  {\bibfnamefont {K.}~\bibnamefont {Kaneko}}, \bibinfo {author} {\bibfnamefont
  {E.}~\bibnamefont {Kussell}}, \ and\ \bibinfo {author} {\bibfnamefont
  {Y.}~\bibnamefont {Wakamoto}},\ }\href@noop {} {\bibfield  {journal}
  {\bibinfo  {journal} {Proceedings of the National Academy of Sciences}\
  }\textbf {\bibinfo {volume} {113}},\ \bibinfo {pages} {3251} (\bibinfo {year}
  {2016})}\BibitemShut {NoStop}%
\bibitem [{\citenamefont {Lin}\ and\ \citenamefont
  {Amir}(2017)}]{lin2017effects}%
  \BibitemOpen
  \bibfield  {author} {\bibinfo {author} {\bibfnamefont {J.}~\bibnamefont
  {Lin}}\ and\ \bibinfo {author} {\bibfnamefont {A.}~\bibnamefont {Amir}},\
  }\href@noop {} {\bibfield  {journal} {\bibinfo  {journal} {Cell Systems}\
  }\textbf {\bibinfo {volume} {5}},\ \bibinfo {pages} {358} (\bibinfo {year}
  {2017})}\BibitemShut {NoStop}%
\bibitem [{\citenamefont {Quedeville}\ \emph {et~al.}(2019)\citenamefont
  {Quedeville}, \citenamefont {Morchain}, \citenamefont {Villedieu},\ and\
  \citenamefont {Fox}}]{quedeville2019critical}%
  \BibitemOpen
  \bibfield  {author} {\bibinfo {author} {\bibfnamefont {V.}~\bibnamefont
  {Quedeville}}, \bibinfo {author} {\bibfnamefont {J.}~\bibnamefont
  {Morchain}}, \bibinfo {author} {\bibfnamefont {P.}~\bibnamefont {Villedieu}},
  \ and\ \bibinfo {author} {\bibfnamefont {R.~O.}\ \bibnamefont {Fox}},\
  }\href@noop {} {\bibfield  {journal} {\bibinfo  {journal} {Scientific
  Reports}\ }\textbf {\bibinfo {volume} {9}},\ \bibinfo {pages} {1} (\bibinfo
  {year} {2019})}\BibitemShut {NoStop}%
\bibitem [{\citenamefont {Levien}\ \emph {et~al.}(2020)\citenamefont {Levien},
  \citenamefont {Kondev},\ and\ \citenamefont {Amir}}]{levien2020interplay}%
  \BibitemOpen
  \bibfield  {author} {\bibinfo {author} {\bibfnamefont {E.}~\bibnamefont
  {Levien}}, \bibinfo {author} {\bibfnamefont {J.}~\bibnamefont {Kondev}}, \
  and\ \bibinfo {author} {\bibfnamefont {A.}~\bibnamefont {Amir}},\ }\href@noop
  {} {\bibfield  {journal} {\bibinfo  {journal} {Journal of the Royal Society
  interface}\ }\textbf {\bibinfo {volume} {17}},\ \bibinfo {pages} {20190827}
  (\bibinfo {year} {2020})}\BibitemShut {NoStop}%
\bibitem [{\citenamefont {Maruyama}\ and\ \citenamefont
  {Yanagita}(1956)}]{Maruyama1956}%
  \BibitemOpen
  \bibfield  {author} {\bibinfo {author} {\bibfnamefont {Y.}~\bibnamefont
  {Maruyama}}\ and\ \bibinfo {author} {\bibfnamefont {T.}~\bibnamefont
  {Yanagita}},\ }\href@noop {} {\bibfield  {journal} {\bibinfo  {journal}
  {Journal of Bacteriology}\ }\textbf {\bibinfo {volume} {71}},\ \bibinfo
  {pages} {542} (\bibinfo {year} {1956})}\BibitemShut {NoStop}%
\bibitem [{\citenamefont {Newton}\ and\ \citenamefont
  {Wildy}(1959)}]{Newton1959}%
  \BibitemOpen
  \bibfield  {author} {\bibinfo {author} {\bibfnamefont {A.}~\bibnamefont
  {Newton}}\ and\ \bibinfo {author} {\bibfnamefont {P.}~\bibnamefont {Wildy}},\
  }\href@noop {} {\bibfield  {journal} {\bibinfo  {journal} {Experimental Cell
  Research}\ }\textbf {\bibinfo {volume} {16}},\ \bibinfo {pages} {624}
  (\bibinfo {year} {1959})}\BibitemShut {NoStop}%
\bibitem [{\citenamefont {Shehata}\ and\ \citenamefont
  {Marr}(1970)}]{Shehata1970}%
  \BibitemOpen
  \bibfield  {author} {\bibinfo {author} {\bibfnamefont {T.~E.}\ \bibnamefont
  {Shehata}}\ and\ \bibinfo {author} {\bibfnamefont {A.~G.}\ \bibnamefont
  {Marr}},\ }\href@noop {} {\bibfield  {journal} {\bibinfo  {journal} {Journal
  of Bacteriology}\ }\textbf {\bibinfo {volume} {103}},\ \bibinfo {pages} {789}
  (\bibinfo {year} {1970})}\BibitemShut {NoStop}%
\bibitem [{\citenamefont {Brown}(1940)}]{Brown1940}%
  \BibitemOpen
  \bibfield  {author} {\bibinfo {author} {\bibfnamefont {A.}~\bibnamefont
  {Brown}},\ }\href@noop {} {\bibfield  {journal} {\bibinfo  {journal} {The
  Annals of Mathematical Statistics}\ }\textbf {\bibinfo {volume} {11}},\
  \bibinfo {pages} {448} (\bibinfo {year} {1940})}\BibitemShut {NoStop}%
\bibitem [{\citenamefont {Kendall}(1948)}]{Kendall1948}%
  \BibitemOpen
  \bibfield  {author} {\bibinfo {author} {\bibfnamefont {D.~G.}\ \bibnamefont
  {Kendall}},\ }\href@noop {} {\bibfield  {journal} {\bibinfo  {journal}
  {Biometrika}\ }\textbf {\bibinfo {volume} {35}},\ \bibinfo {pages} {316}
  (\bibinfo {year} {1948})}\BibitemShut {NoStop}%
\bibitem [{\citenamefont {Hirsch}\ and\ \citenamefont
  {Engelberg}(1966)}]{Hirsch1966}%
  \BibitemOpen
  \bibfield  {author} {\bibinfo {author} {\bibfnamefont {H.~R.}\ \bibnamefont
  {Hirsch}}\ and\ \bibinfo {author} {\bibfnamefont {J.}~\bibnamefont
  {Engelberg}},\ }\href@noop {} {\bibfield  {journal} {\bibinfo  {journal} {The
  Bulletin of Mathematical Biophysics}\ }\textbf {\bibinfo {volume} {28}},\
  \bibinfo {pages} {391} (\bibinfo {year} {1966})}\BibitemShut {NoStop}%
\bibitem [{\citenamefont {Burnett-Hall}\ and\ \citenamefont
  {Waugh}(1967)}]{Burnett-Hall1967}%
  \BibitemOpen
  \bibfield  {author} {\bibinfo {author} {\bibfnamefont {D.~G.}\ \bibnamefont
  {Burnett-Hall}}\ and\ \bibinfo {author} {\bibfnamefont {W.~A.~O.}\
  \bibnamefont {Waugh}},\ }\href@noop {} {\bibfield  {journal} {\bibinfo
  {journal} {Biometrics}\ }\textbf {\bibinfo {volume} {23}},\ \bibinfo {pages}
  {693} (\bibinfo {year} {1967})}\BibitemShut {NoStop}%
\bibitem [{\citenamefont {Bell}\ and\ \citenamefont
  {Anderson}(1967)}]{bell1967cell}%
  \BibitemOpen
  \bibfield  {author} {\bibinfo {author} {\bibfnamefont {G.~I.}\ \bibnamefont
  {Bell}}\ and\ \bibinfo {author} {\bibfnamefont {E.~C.}\ \bibnamefont
  {Anderson}},\ }\href@noop {} {\bibfield  {journal} {\bibinfo  {journal}
  {Biophysical Journal}\ }\textbf {\bibinfo {volume} {7}},\ \bibinfo {pages}
  {329} (\bibinfo {year} {1967})}\BibitemShut {NoStop}%
\bibitem [{\citenamefont {Anderson}\ and\ \citenamefont
  {Petersen}(1967)}]{anderson1967cell}%
  \BibitemOpen
  \bibfield  {author} {\bibinfo {author} {\bibfnamefont {E.~C.}\ \bibnamefont
  {Anderson}}\ and\ \bibinfo {author} {\bibfnamefont {D.~F.}\ \bibnamefont
  {Petersen}},\ }\href@noop {} {\bibfield  {journal} {\bibinfo  {journal}
  {Biophysical journal}\ }\textbf {\bibinfo {volume} {7}},\ \bibinfo {pages}
  {353} (\bibinfo {year} {1967})}\BibitemShut {NoStop}%
\bibitem [{\citenamefont {Anderson}\ \emph {et~al.}(1969)\citenamefont
  {Anderson}, \citenamefont {Bell}, \citenamefont {Petersen},\ and\
  \citenamefont {Tobey}}]{anderson1969cell}%
  \BibitemOpen
  \bibfield  {author} {\bibinfo {author} {\bibfnamefont {E.}~\bibnamefont
  {Anderson}}, \bibinfo {author} {\bibfnamefont {G.}~\bibnamefont {Bell}},
  \bibinfo {author} {\bibfnamefont {D.}~\bibnamefont {Petersen}}, \ and\
  \bibinfo {author} {\bibfnamefont {R.}~\bibnamefont {Tobey}},\ }\href@noop {}
  {\bibfield  {journal} {\bibinfo  {journal} {Biophysical journal}\ }\textbf
  {\bibinfo {volume} {9}},\ \bibinfo {pages} {246} (\bibinfo {year}
  {1969})}\BibitemShut {NoStop}%
\bibitem [{\citenamefont {Hagander}(1980)}]{Hagander1980}%
  \BibitemOpen
  \bibfield  {author} {\bibinfo {author} {\bibfnamefont {P.}~\bibnamefont
  {Hagander}},\ }\href@noop {} {\bibfield  {journal} {\bibinfo  {journal}
  {Mathematical Biosciences}\ }\textbf {\bibinfo {volume} {48}},\ \bibinfo
  {pages} {241} (\bibinfo {year} {1980})}\BibitemShut {NoStop}%
\bibitem [{\citenamefont {Chiorino}\ \emph {et~al.}(2001)\citenamefont
  {Chiorino}, \citenamefont {Metz}, \citenamefont {Tomasoni},\ and\
  \citenamefont {Ubezio}}]{chiorino2001desynchronization}%
  \BibitemOpen
  \bibfield  {author} {\bibinfo {author} {\bibfnamefont {G.}~\bibnamefont
  {Chiorino}}, \bibinfo {author} {\bibfnamefont {J.}~\bibnamefont {Metz}},
  \bibinfo {author} {\bibfnamefont {D.}~\bibnamefont {Tomasoni}}, \ and\
  \bibinfo {author} {\bibfnamefont {P.}~\bibnamefont {Ubezio}},\ }\href@noop {}
  {\bibfield  {journal} {\bibinfo  {journal} {Journal of Theoretical Biology}\
  }\textbf {\bibinfo {volume} {208}},\ \bibinfo {pages} {185} (\bibinfo {year}
  {2001})}\BibitemShut {NoStop}%
\bibitem [{\citenamefont {Gavagnin}\ \emph {et~al.}(2021)\citenamefont
  {Gavagnin}, \citenamefont {Vittadello}, \citenamefont {Gunasingh},
  \citenamefont {Haass}, \citenamefont {Simpson}, \citenamefont {Rogers},\ and\
  \citenamefont {Yates}}]{gavagnin2021synchronized}%
  \BibitemOpen
  \bibfield  {author} {\bibinfo {author} {\bibfnamefont {E.}~\bibnamefont
  {Gavagnin}}, \bibinfo {author} {\bibfnamefont {S.~T.}\ \bibnamefont
  {Vittadello}}, \bibinfo {author} {\bibfnamefont {G.}~\bibnamefont
  {Gunasingh}}, \bibinfo {author} {\bibfnamefont {N.~K.}\ \bibnamefont
  {Haass}}, \bibinfo {author} {\bibfnamefont {M.~J.}\ \bibnamefont {Simpson}},
  \bibinfo {author} {\bibfnamefont {T.}~\bibnamefont {Rogers}}, \ and\ \bibinfo
  {author} {\bibfnamefont {C.~A.}\ \bibnamefont {Yates}},\ }\href@noop {}
  {\bibfield  {journal} {\bibinfo  {journal} {Biophysical Journal}\ } (\bibinfo
  {year} {2021})}\BibitemShut {NoStop}%
\bibitem [{\citenamefont {Jafarpour}\ \emph {et~al.}(2018)\citenamefont
  {Jafarpour}, \citenamefont {Wright}, \citenamefont {Gudjonson}, \citenamefont
  {Riebling}, \citenamefont {Dawson}, \citenamefont {Lo}, \citenamefont
  {Fiebig}, \citenamefont {Crosson}, \citenamefont {Dinner},\ and\
  \citenamefont {Iyer-Biswas}}]{jafarpour2018bridging}%
  \BibitemOpen
  \bibfield  {author} {\bibinfo {author} {\bibfnamefont {F.}~\bibnamefont
  {Jafarpour}}, \bibinfo {author} {\bibfnamefont {C.~S.}\ \bibnamefont
  {Wright}}, \bibinfo {author} {\bibfnamefont {H.}~\bibnamefont {Gudjonson}},
  \bibinfo {author} {\bibfnamefont {J.}~\bibnamefont {Riebling}}, \bibinfo
  {author} {\bibfnamefont {E.}~\bibnamefont {Dawson}}, \bibinfo {author}
  {\bibfnamefont {K.}~\bibnamefont {Lo}}, \bibinfo {author} {\bibfnamefont
  {A.}~\bibnamefont {Fiebig}}, \bibinfo {author} {\bibfnamefont
  {S.}~\bibnamefont {Crosson}}, \bibinfo {author} {\bibfnamefont {A.~R.}\
  \bibnamefont {Dinner}}, \ and\ \bibinfo {author} {\bibfnamefont
  {S.}~\bibnamefont {Iyer-Biswas}},\ }\href@noop {} {\bibfield  {journal}
  {\bibinfo  {journal} {Physical Review X}\ }\textbf {\bibinfo {volume} {8}},\
  \bibinfo {pages} {021007} (\bibinfo {year} {2018})}\BibitemShut {NoStop}%
\bibitem [{\citenamefont {Jafarpour}(2019)}]{jafarpour2019cell}%
  \BibitemOpen
  \bibfield  {author} {\bibinfo {author} {\bibfnamefont {F.}~\bibnamefont
  {Jafarpour}},\ }\href@noop {} {\bibfield  {journal} {\bibinfo  {journal}
  {Physical Review Letters}\ }\textbf {\bibinfo {volume} {122}},\ \bibinfo
  {pages} {118101} (\bibinfo {year} {2019})}\BibitemShut {NoStop}%
\bibitem [{\citenamefont {Hein}\ and\ \citenamefont
  {Jafarpour}(2022)}]{hein2022asymptotic}%
  \BibitemOpen
  \bibfield  {author} {\bibinfo {author} {\bibfnamefont {Y.}~\bibnamefont
  {Hein}}\ and\ \bibinfo {author} {\bibfnamefont {F.}~\bibnamefont
  {Jafarpour}},\ }\href@noop {} {\bibfield  {journal} {\bibinfo  {journal}
  {arXiv preprint arXiv:2209.14683}\ } (\bibinfo {year} {2022})}\BibitemShut
  {NoStop}%
\bibitem [{\citenamefont {Thomas}(2017)}]{thomas2017making}%
  \BibitemOpen
  \bibfield  {author} {\bibinfo {author} {\bibfnamefont {P.}~\bibnamefont
  {Thomas}},\ }\href@noop {} {\bibfield  {journal} {\bibinfo  {journal}
  {Journal of The Royal Society Interface}\ }\textbf {\bibinfo {volume} {14}},\
  \bibinfo {pages} {20170467} (\bibinfo {year} {2017})}\BibitemShut {NoStop}%
\bibitem [{\citenamefont {Nozoe}\ \emph {et~al.}(2017)\citenamefont {Nozoe},
  \citenamefont {Kussell},\ and\ \citenamefont
  {Wakamoto}}]{nozoe2017inferring}%
  \BibitemOpen
  \bibfield  {author} {\bibinfo {author} {\bibfnamefont {T.}~\bibnamefont
  {Nozoe}}, \bibinfo {author} {\bibfnamefont {E.}~\bibnamefont {Kussell}}, \
  and\ \bibinfo {author} {\bibfnamefont {Y.}~\bibnamefont {Wakamoto}},\
  }\href@noop {} {\bibfield  {journal} {\bibinfo  {journal} {PLoS genetics}\
  }\textbf {\bibinfo {volume} {13}},\ \bibinfo {pages} {e1006653} (\bibinfo
  {year} {2017})}\BibitemShut {NoStop}%
\bibitem [{\citenamefont {Genthon}\ and\ \citenamefont
  {Lacoste}(2020)}]{genthon2020fluctuation}%
  \BibitemOpen
  \bibfield  {author} {\bibinfo {author} {\bibfnamefont {A.}~\bibnamefont
  {Genthon}}\ and\ \bibinfo {author} {\bibfnamefont {D.}~\bibnamefont
  {Lacoste}},\ }\href@noop {} {\bibfield  {journal} {\bibinfo  {journal}
  {Scientific Reports}\ }\textbf {\bibinfo {volume} {10}},\ \bibinfo {pages}
  {1} (\bibinfo {year} {2020})}\BibitemShut {NoStop}%
\bibitem [{\citenamefont {Genthon}\ and\ \citenamefont
  {Lacoste}(2021)}]{genthon2021universal}%
  \BibitemOpen
  \bibfield  {author} {\bibinfo {author} {\bibfnamefont {A.}~\bibnamefont
  {Genthon}}\ and\ \bibinfo {author} {\bibfnamefont {D.}~\bibnamefont
  {Lacoste}},\ }\href@noop {} {\bibfield  {journal} {\bibinfo  {journal}
  {Physical Review Research}\ }\textbf {\bibinfo {volume} {3}},\ \bibinfo
  {pages} {023187} (\bibinfo {year} {2021})}\BibitemShut {NoStop}%
\bibitem [{\citenamefont {M'Kendrick}(1925)}]{MKendrick1925}%
  \BibitemOpen
  \bibfield  {author} {\bibinfo {author} {\bibfnamefont {A.~G.}\ \bibnamefont
  {M'Kendrick}},\ }\href@noop {} {\bibfield  {journal} {\bibinfo  {journal}
  {Proceedings of the Edinburgh Mathematical Society}\ }\textbf {\bibinfo
  {volume} {44}},\ \bibinfo {pages} {98} (\bibinfo {year} {1925})}\BibitemShut
  {NoStop}%
\bibitem [{\citenamefont {Von~Foerster}\ and\ \citenamefont
  {Stohlman}(1959)}]{von1959kinetics}%
  \BibitemOpen
  \bibfield  {author} {\bibinfo {author} {\bibfnamefont {H.}~\bibnamefont
  {Von~Foerster}}\ and\ \bibinfo {author} {\bibfnamefont {F.}~\bibnamefont
  {Stohlman}},\ }\href@noop {} {\bibfield  {journal} {\bibinfo  {journal}
  {Grune \& Stratton}\ }\textbf {\bibinfo {volume} {3}} (\bibinfo {year}
  {1959})}\BibitemShut {NoStop}%
\bibitem [{\citenamefont
  {Trucco}(1965{\natexlab{a}})}]{trucco1965mathematical1}%
  \BibitemOpen
  \bibfield  {author} {\bibinfo {author} {\bibfnamefont {E.}~\bibnamefont
  {Trucco}},\ }\href@noop {} {\bibfield  {journal} {\bibinfo  {journal} {The
  Bulletin of Mathematical Biophysics}\ }\textbf {\bibinfo {volume} {27}},\
  \bibinfo {pages} {285} (\bibinfo {year} {1965}{\natexlab{a}})}\BibitemShut
  {NoStop}%
\bibitem [{\citenamefont
  {Trucco}(1965{\natexlab{b}})}]{trucco1965mathematical2}%
  \BibitemOpen
  \bibfield  {author} {\bibinfo {author} {\bibfnamefont {E.}~\bibnamefont
  {Trucco}},\ }\href@noop {} {\bibfield  {journal} {\bibinfo  {journal} {The
  Bulletin of Mathematical Biophysics}\ }\textbf {\bibinfo {volume} {27}},\
  \bibinfo {pages} {449} (\bibinfo {year} {1965}{\natexlab{b}})}\BibitemShut
  {NoStop}%
\bibitem [{\citenamefont {Rudnicki}(2014)}]{rudnicki2014modele}%
  \BibitemOpen
  \bibfield  {author} {\bibinfo {author} {\bibfnamefont {R.}~\bibnamefont
  {Rudnicki}},\ }\href@noop {} {\emph {\bibinfo {title} {Modele i metody
  biologii matematycznej: Modele deterministyczne}}}\ (\bibinfo  {publisher}
  {Instytut Matematyczny Polskiej Akademii Nauk},\ \bibinfo {year}
  {2014})\BibitemShut {NoStop}%
\bibitem [{\citenamefont {Taheri-Araghi}\ \emph {et~al.}(2015)\citenamefont
  {Taheri-Araghi}, \citenamefont {Bradde}, \citenamefont {Sauls}, \citenamefont
  {Hill}, \citenamefont {Levin}, \citenamefont {Paulsson}, \citenamefont
  {Vergassola},\ and\ \citenamefont {Jun}}]{taheri2015cell}%
  \BibitemOpen
  \bibfield  {author} {\bibinfo {author} {\bibfnamefont {S.}~\bibnamefont
  {Taheri-Araghi}}, \bibinfo {author} {\bibfnamefont {S.}~\bibnamefont
  {Bradde}}, \bibinfo {author} {\bibfnamefont {J.~T.}\ \bibnamefont {Sauls}},
  \bibinfo {author} {\bibfnamefont {N.~S.}\ \bibnamefont {Hill}}, \bibinfo
  {author} {\bibfnamefont {P.~A.}\ \bibnamefont {Levin}}, \bibinfo {author}
  {\bibfnamefont {J.}~\bibnamefont {Paulsson}}, \bibinfo {author}
  {\bibfnamefont {M.}~\bibnamefont {Vergassola}}, \ and\ \bibinfo {author}
  {\bibfnamefont {S.}~\bibnamefont {Jun}},\ }\href@noop {} {\bibfield
  {journal} {\bibinfo  {journal} {Current Biology}\ }\textbf {\bibinfo {volume}
  {25}},\ \bibinfo {pages} {385} (\bibinfo {year} {2015})}\BibitemShut
  {NoStop}%
\bibitem [{\citenamefont {Rubinow}(1968)}]{rubinow1968maturity}%
  \BibitemOpen
  \bibfield  {author} {\bibinfo {author} {\bibfnamefont {S.}~\bibnamefont
  {Rubinow}},\ }\href@noop {} {\bibfield  {journal} {\bibinfo  {journal}
  {Biophysical Journal}\ }\textbf {\bibinfo {volume} {8}},\ \bibinfo {pages}
  {1055} (\bibinfo {year} {1968})}\BibitemShut {NoStop}%
\bibitem [{\citenamefont {Garc{\'\i}a-Garc{\'\i}a}\ \emph
  {et~al.}(2019)\citenamefont {Garc{\'\i}a-Garc{\'\i}a}, \citenamefont
  {Genthon},\ and\ \citenamefont {Lacoste}}]{garcia2019linking}%
  \BibitemOpen
  \bibfield  {author} {\bibinfo {author} {\bibfnamefont {R.}~\bibnamefont
  {Garc{\'\i}a-Garc{\'\i}a}}, \bibinfo {author} {\bibfnamefont
  {A.}~\bibnamefont {Genthon}}, \ and\ \bibinfo {author} {\bibfnamefont
  {D.}~\bibnamefont {Lacoste}},\ }\href@noop {} {\bibfield  {journal} {\bibinfo
   {journal} {Physical Review E}\ }\textbf {\bibinfo {volume} {99}},\ \bibinfo
  {pages} {042413} (\bibinfo {year} {2019})}\BibitemShut {NoStop}%
\bibitem [{\citenamefont {Genthon}\ \emph {et~al.}(2023)\citenamefont
  {Genthon}, \citenamefont {Nozoe}, \citenamefont {Peliti},\ and\ \citenamefont
  {Lacoste}}]{genthon2023cell}%
  \BibitemOpen
  \bibfield  {author} {\bibinfo {author} {\bibfnamefont {A.}~\bibnamefont
  {Genthon}}, \bibinfo {author} {\bibfnamefont {T.}~\bibnamefont {Nozoe}},
  \bibinfo {author} {\bibfnamefont {L.}~\bibnamefont {Peliti}}, \ and\ \bibinfo
  {author} {\bibfnamefont {D.}~\bibnamefont {Lacoste}},\ }\href@noop {}
  {\bibfield  {journal} {\bibinfo  {journal} {arXiv preprint arXiv:2305.05406}\
  } (\bibinfo {year} {2023})}\BibitemShut {NoStop}%
\bibitem [{\citenamefont {Bell}(1968)}]{bell1968cell}%
  \BibitemOpen
  \bibfield  {author} {\bibinfo {author} {\bibfnamefont {G.~I.}\ \bibnamefont
  {Bell}},\ }\href@noop {} {\bibfield  {journal} {\bibinfo  {journal}
  {Biophysical journal}\ }\textbf {\bibinfo {volume} {8}},\ \bibinfo {pages}
  {431} (\bibinfo {year} {1968})}\BibitemShut {NoStop}%
\bibitem [{\citenamefont {Pich{\'o}r}\ and\ \citenamefont
  {Rudnicki}(2021)}]{pichor2021cell}%
  \BibitemOpen
  \bibfield  {author} {\bibinfo {author} {\bibfnamefont {K.}~\bibnamefont
  {Pich{\'o}r}}\ and\ \bibinfo {author} {\bibfnamefont {R.}~\bibnamefont
  {Rudnicki}},\ }\href@noop {} {\bibfield  {journal} {\bibinfo  {journal}
  {arXiv preprint arXiv:2104.01442}\ } (\bibinfo {year} {2021})}\BibitemShut
  {NoStop}%
\bibitem [{\citenamefont {Stewart}\ \emph {et~al.}(2005)\citenamefont
  {Stewart}, \citenamefont {Madden}, \citenamefont {Paul},\ and\ \citenamefont
  {Taddei}}]{stewart2005aging}%
  \BibitemOpen
  \bibfield  {author} {\bibinfo {author} {\bibfnamefont {E.~J.}\ \bibnamefont
  {Stewart}}, \bibinfo {author} {\bibfnamefont {R.}~\bibnamefont {Madden}},
  \bibinfo {author} {\bibfnamefont {G.}~\bibnamefont {Paul}}, \ and\ \bibinfo
  {author} {\bibfnamefont {F.}~\bibnamefont {Taddei}},\ }\href@noop {}
  {\bibfield  {journal} {\bibinfo  {journal} {PLoS Biol}\ }\textbf {\bibinfo
  {volume} {3}},\ \bibinfo {pages} {e45} (\bibinfo {year} {2005})}\BibitemShut
  {NoStop}%
\end{thebibliography}%


\end{document}